\documentclass[a4paper,12pt]{article}
\pdfoutput=1
\usepackage{epsfig,graphicx,xcolor,amsbsy,amssymb,latexsym,amsfonts,amsmath,setspace}
\usepackage{pstricks}
\usepackage{color}
\usepackage{soul}
\usepackage[numbers,square,comma, compress]{natbib}
\usepackage{placeins}
\usepackage{wrapfig}

\usepackage{eurosym}
\usepackage[a4paper]{geometry}
\usepackage{graphicx}
\usepackage{tikz}
\usepackage{xcolor}
\usepackage{amsmath,amssymb,amsfonts,pstricks,setspace}
\usepackage[enableskew]{youngtab}

\numberwithin{equation}{section}

\newcommand{\bea}{\begin{eqnarray}\displaystyle}
\newcommand{\eea}{\end{eqnarray}}

\newcommand{\figref}[1]{Fig.~\protect\ref{#1}}

\newcommand{\pf}[2]{\mathcal{Z}_{#1,#2}}

\title{
\vspace{-3cm}
\begin{flushright}
{\small LYCEN 2017-08}\\[40pt]
\end{flushright}
\bf{Triality in Little String Theories}\\[15pt]}
\author{\large \textsc{Brice Bastian\footnote{\tt b.bastian@ipnl.in2p3.fr}},~~\textsc{Stefan~Hohenegger\footnote{\tt s.hohenegger@ipnl.in2p3.fr},~ Amer Iqbal\footnote{\tt  amer@alum.mit.edu},~ Soo-Jong Rey\,\footnote{\tt sjrey@snu.ac.kr}}}
\date{}

\begin{document}

\maketitle

\begin{center}
\renewcommand{\thefootnote}{\fnsymbol{footnote}}\vspace{-0.5cm}
${}^{\footnotemark[1]\footnotemark[2]}$ Universit\'e de Lyon\\
UMR 5822, CNRS/IN2P3, Institut de Physique Nucl\'eaire de Lyon\\ 4 rue Enrico Fermi, 69622 Villeurbanne Cedex, \rm FRANCE\\[0.4cm]
${}^{\footnotemark[3]}$ Abdus Salam School of Mathematical Sciences \\ Government College University, Lahore, PAKISTAN\\[0.4cm]
${}^{\footnotemark[3]}$ Center for Theoretical Physics, Lahore, PAKISTAN\\[0.4cm]
${}^{\footnotemark[4]}$ School of Physics and Astronomy \& Center for Theoretical Physics\\
Seoul National University, Seoul 08826 \rm KOREA\\[1cm]
\end{center}

\begin{abstract}
We study a class of eight-supercharge little string theories (LSTs) on the world-volume of $N$ M5-branes with transverse space $\mathbb{S}^{1}\times (\mathbb{C}^2/\mathbb{Z}_M)$. These M-brane configurations compactified on a circle are dual to $M$ D5-branes intersecting $N$ NS5-branes on $\mathbb{T}^2\times \mathbb{R}^{7,1}$ as well as to F-theory compactified on a toric Calabi-Yau threefold $X_{N,M}$. We argue that the K\"ahler cone of $X_{N,M}$ admits three regions associated with weakly coupled quiver gauge theories of gauge groups $[U(N)]^M, [U(M)]^N$ and $[U(\frac{NM}{k})]^k$ where $k=\mbox{gcd}(N,M)$. These provide low-energy descriptions of different LSTs.
The duality between the first two gauge theories is well known and is a consequence of the S-duality between D5- and NS5-branes or the T-duality of the LSTs. The triality involving the third gauge theory is new and we demonstrate it using several examples. We also discuss implications of this triality for the W-algebras associated with the Alday-Gaiotto-Tachikawa dual theories. 

 \end{abstract}

\newpage

\tableofcontents

\onehalfspacing

\vskip1cm

%%%%%%%%%%%%%%%%%%%%%%%%%%%%%%%%%%%%%%
\section{Introduction}
Throughout the years, dualities have been a driving tool in the exploration and development of statistical physics, quantum field theory and (non)perturbative string theory. Either as conceptual means to analyse  the (mathematical) structure of the theory or as computational tools to reformulate specific questions in a more tractable framework, dualities were proven extremely useful and, in many cases, have been at the forefront of new discoveries. While new dualities (or associated symmetries) are typically difficult to find (or prove) in full-fledge string theories, they typically also leave imprints on other physical system that are engineered by string (or M-) theory. One such example is that of Little String Theories (LSTs) \cite{Seiberg:1996vs,Berkooz:1997cq,Blum:1997mm,Seiberg:1997zk,Losev:1997hx,Intriligator:1997dh,Aharony:1998ub} (see \cite{Aharony:1999ks,Kutasov:2001uf} for a review): using various brane constructions (in string or M-theory) and their dual geometric description in F-theory \cite{Bhardwaj:2015oru}, different realisations of LSTs can be constructed and duality relations among them can be studied \cite{Hohenegger:2016yuv, Hohenegger:2016eqy}, including heterotic LSTs \cite{Haghighat:2014pva, Choi:2017vtd, Choi:2017luj}. In this paper, we aim to further analyse the web of dualities connecting different LSTs as well as their low energy descriptions in terms of gauge field theories.

LSTs refer to a class of interacting, ultraviolet-complete quantum theories in six dimensions whose nonlocal dynamics is governed by self-dual noncritical strings. While being easier to handle due to the fact that gravity is decoupled (\emph{i.e.} the spectrum does not contain a massless spin-two field), these theories still share many features in common with the critical string theory in ten dimensions. In fact, LSTs is operationally definable from Type II string theory (or its dual avatars) through a particular decoupling limit that sends the string coupling constant to zero ($g_{\text{st}}\to 0$) while at the same time keeps the string length $\ell_{\text{st}}$ finite. Depending on details of the original Type II setup, one can construct various different LSTs with $(2,0)$ or $(1,0)$ supersymmetry.

The construction of LSTs which will be relevant for us is in terms of F-theory compactifications \cite{Morrison:1996na,Morrison:1996pp,Bershadsky:1996nu,Aspinwall:1996vc,Bershadsky:1997sb, Esole:2015xfa} on a class of toric, non-compact Calabi-Yau threefolds called $X_{N,M}$: the latter are equipped with the structure of a double elliptic fibration, in which one elliptic fibration has a singularity of type $I_{N-1}$ and the other one of $I_{M-1}$ \cite{Hohenegger:2015btj, Kanazawa:2016tnt}. These LSTs of type $(N,M)$ can also be engineered using $N$ parallel M5-branes with a transverse orbifold of type $A_{M-1}$ \cite{Blum:1997fw,Cecotti:2013mba,Haghighat:2013gba, Haghighat:2013tka}. Moreover, using the refined topological vertex formalism \cite{Iqbal:2007ii}, the topological string partition function $\pf{N}{M}(\mathbf{T},\mathbf{t},m,\epsilon_1,\epsilon_2)$ of $X_{N,M}$ has been computed in \cite{Haghighat:2013tka,Hohenegger:2013ala,Hohenegger:2016eqy} at a particular region of the K\"aher moduli space: in this region $\mathbf{T}=\{T_1,\ldots,T_M\}$ and $\mathbf{t}=\{t_1,\ldots,t_N\}$ are two sets of K\"ahler parameters related to the two elliptic fibrations of $X_{N,M}$, respectively (and $m$ is a further K\"ahler modulus), while $\epsilon_{1,2}$ are related to the topological string coupling and the beta deformation due to the $\Omega$-background \cite{Nekrasov:2002qd,Losev:2003py,Hollowood:2003cv}. $\pf{N}{M}(\mathbf{T},\mathbf{t},m,\epsilon,-\epsilon)$ have also been studied for quantized values of the parameters ${\bf t}$ in units of $\epsilon$ and shown to be given by highest weight representations of an affine algebra \cite{Bastian:2017jje}.
It was proposed in \cite{Hohenegger:2016eqy} that the topological string partition function $\pf{N}{M}$ captures the partition functions of a class of LST with 8 supercharges. 
Compactification of LSTs of type $(N,M)$ on a circle then relates these with LSTs of type $(M,N)$ by T-duality. This T-duality relation is reflected in the Calabi-Yau $X_{N,M}$ by the exchange of the two fibers \cite{Bhardwaj:2015oru,Hohenegger:2015btj}.  See \cite{Klemm:2012sx, Huang:2015sta} for discussion of topological string partition functions on elliptically fibered Calabi-Yau threefolds.

Furthermore, the low energy limits of the compactified LSTs of type $(N,M)$ and $(M,N)$ are the (circular) quiver gauge theories with gauge group $[U(N)]^M$ and $[U(M)]^N$, respectively. In these two descriptions, the parameters $\mathbf{T}$ and $\mathbf{t}$ are interpreted as either the gauge coupling constants or the Coulomb-branch parameters (while $m$ is interpreted as the mass scale of the bi-fundamental hypermultiplets). Therefore, the T-duality relation above leads to a duality relation between these two gauge theories in which the coupling constants are exchanged with the Coulomb branch parameters and vice versa. 
\begin{figure}[htb]
\begin{center}
\scalebox{0.75}{\parbox{14.5cm}{\begin{tikzpicture}[scale = 1.1]
%horizontal lines%%%%%%%%%%
%first layer
\draw[ultra thick, green!50!black] (-1,0) -- (0,0);
\draw[ultra thick, green!50!black] (1,1) -- (2,1);
\draw[ultra thick, green!50!black] (3,2) -- (4,2);
\node at (4.5,2) {\Large $\cdots$};
\draw[ultra thick, green!50!black] (5,2) -- (6,2);
\draw[ultra thick, green!50!black] (7,3) -- (8,3);
%second layer
\draw[ultra thick, green!50!black] (0,2) -- (1,2);
\draw[ultra thick, green!50!black] (2,3) -- (3,3);
\draw[ultra thick, green!50!black] (4,4) -- (5,4);
\node at (5.5,4) {\Large $\cdots$};
\draw[ultra thick, green!50!black] (6,4) -- (7,4);
\draw[ultra thick, green!50!black] (8,5) -- (9,5);
%third layer
\draw[ultra thick, green!50!black] (1,6) -- (2,6);
\draw[ultra thick, green!50!black] (3,7) -- (4,7);
\draw[ultra thick, green!50!black] (5,8) -- (6,8);
\node at (6.5,8) {\Large $\cdots$};
\draw[ultra thick, green!50!black] (7,8) -- (8,8);
\draw[ultra thick, green!50!black] (9,9) -- (10,9);
%vertical lines%%%%%%%%%%
%first layer
\draw[ultra thick, red] (0,0) -- (0,-1);
\draw[ultra thick, red] (2,1) -- (2,0);
\draw[ultra thick, red] (6,2) -- (6,1);
%second layer
\draw[ultra thick, red] (1,1) -- (1,2);
\draw[ultra thick, red] (3,2) -- (3,3);
\draw[ultra thick, red] (7,3) -- (7,4);
%third layer
\draw[ultra thick, red] (2,3) -- (2,4);
\draw[ultra thick, red] (4,4) -- (4,5);
\draw[ultra thick, red] (8,5) -- (8,6);
%dots
\node[rotate=90] at (2,4.5) {\Large $\cdots$};
\node[rotate=90] at (4,5.5) {\Large $\cdots$};
\node[rotate=90] at (8,6.5) {\Large $\cdots$};
%fourth layer
\draw[ultra thick, red] (2,5) -- (2,6);
\draw[ultra thick, red] (4,6) -- (4,7);
\draw[ultra thick, red] (8,7) -- (8,8);
%fifth layer
\draw[ultra thick, red] (3,7) -- (3,8);
\draw[ultra thick, red] (5,8) -- (5,9);
\draw[ultra thick, red] (9,9) -- (9,10);
%diagonal lines%%%%%%%%%%
%first layer
\draw[ultra thick, blue] (0,0) -- (1,1);
\draw[ultra thick, blue] (2,1) -- (3,2);
\draw[ultra thick, blue] (6,2) -- (7,3);
%second layer
\draw[ultra thick, blue] (1,2) -- (2,3);
\draw[ultra thick, blue] (3,3) -- (4,4);
\draw[ultra thick, blue] (7,4) -- (8,5);
%third layer
\draw[ultra thick, blue] (2,6) -- (3,7);
\draw[ultra thick, blue] (4,7) -- (5,8);
\draw[ultra thick, blue] (8,8) -- (9,9);
%%%%%%%%%%%%%%%%%%%%%%
%ID tags vertical
\node at (0,-1.25) {{\small \bf 1}};
\node at (2,-0.25) {{\small \bf 2}};
\node at (6,0.75) {{\small \bf N}};
\node at (3,8.25) {{\small \bf 1}};
\node at (5,9.25) {{\small \bf 2}};
\node at (9,10.25) {{\small \bf N}};
%ID tags horizontal
\node at (-1.3,0) {{\small $\mathbf{a_1}$}};
\node at (-0.3,2) {{\small $\mathbf{a_2}$}};
\node at (0.6,6) {{\small $\mathbf{a_M}$}};
\node at (8.3,3) {{\small $\mathbf{a_1}$}};
\node at (9.3,5) {{\small $\mathbf{a_2}$}};
\node at (10.35,9) {{\small $\mathbf{a_M}$}};
%%%%%%%%%%%%%%%%%%%%%%%
%vertical labels
\node at (-0.25,-0.5) {{\small \bf $v_{1}$}};
\node at (1.75,0.5) {{\small \bf $v_{2}$}};
\node at (5.7,1.5) {{\small \bf $v_{N}$}};
\node at (0.5,1.5) {{\small \bf $v_{N+1}$}};
\node at (2.5,2.5) {{\small \bf $v_{N+2}$}};
\node at (6.65,3.5) {{\small \bf $v_{2N}$}};
\node at (1.4,3.5) {{\small \bf $v_{2N+1}$}};
\node at (3.4,4.5) {{\small \bf $v_{2N+2}$}};
\node at (7.65,5.5) {{\small \bf $v_{3N}$}};
\node[rotate=45] at (1.25,5) {{\small \bf $v_{(M-1)N+1}$}};
\node[rotate=45] at (3.25,6) {{\small \bf $v_{(M-1)N+2}$}};
\node at (7.6,7.5) {{\small \bf $v_{MN}$}};
\node at (2.75,7.5) {{\small \bf $v_1$}};
\node at (4.75,8.5) {{\small \bf $v_2$}};
\node at (8.7,9.5) {{\small \bf $v_N$}};
%horizontal labels
\node at (-0.5,0.3) {{\small \bf $h_{N}$}};
\node at (1.5,1.3) {{\small \bf $h_{1}$}};
\node at (3.5,2.3) {{\small \bf $h_{2}$}};
\node at (5.5,2.3) {{\small \bf $h_{N-1}$}};
\node at (7.5,3.3) {{\small \bf $h_{N}$}};
\node at (0.5,2.3) {{\small \bf $h_{2N}$}};
\node at (2.5,3.3) {{\small \bf $h_{N+1}$}};
\node at (4.5,4.3) {{\small \bf $h_{N+2}$}};
\node at (6.5,4.3) {{\small \bf $h_{2N-1}$}};
\node at (8.5,5.3) {{\small \bf $h_{2N}$}};
\node at (1.5,6.3) {{\small \bf $h_{MN}$}};
\node[rotate=90] at (3.5,8) {{\small \bf $h_{(M-1)N+1}$}};
\node[rotate=90] at (5.5,9) {{\small \bf $h_{(M-1)N+2}$}};
\node at (7.5,8.3) {{\small \bf $h_{MN-1}$}};
\node at (9.5,9.3) {{\small \bf $h_{MN}$}};
%diagonal labels
\node at (0.75,0.25) {{\small \bf $m_{1}$}};
\node at (2.75,1.25) {{\small \bf $m_{2}$}};
\node at (6.75,2.25) {{\small \bf $m_{N}$}};
\node at (1.85,2.2) {{\small \bf $m_{N+1}$}};
\node at (3.85,3.2) {{\small \bf $m_{N+2}$}};
\node at (7.85,4.3) {{\small \bf $m_{2N}$}};
\node[rotate=-45] at (1.65,7.2) {{\small \bf $m_{(M-1)N+1}$}};
\node[rotate=-45] at (5.1,6.65) {{\small \bf $m_{(M-1)N+2}$}};
\node at (8.95,8.3) {{\small \bf $m_{MN}$}};
%rho and tau
\draw[<->] (-1,-1.7) -- (6,-1.7);
\node at (2.5,-2) {{\small $\rho$}};
\draw[<->] (11,2) -- (11,10);
\node at (11.3,6) {{\small $\tau$}};
\end{tikzpicture}}}
\end{center}
\caption{\sl The 5-brane web corresponding to $X_{N,M}$ with a generic parametrisation of all line segments. Not all variables $\mathbf{h}=(h_1,\ldots,h_{MN})$, $\mathbf{v}=(v_1,\ldots,v_{MN})$ and $\mathbf{m}=(m_1,\ldots,m_{MN})$ are independent, but are subject to $2NM-2$ consistency conditions.}
\label{Fig:WebToric}
\end{figure}
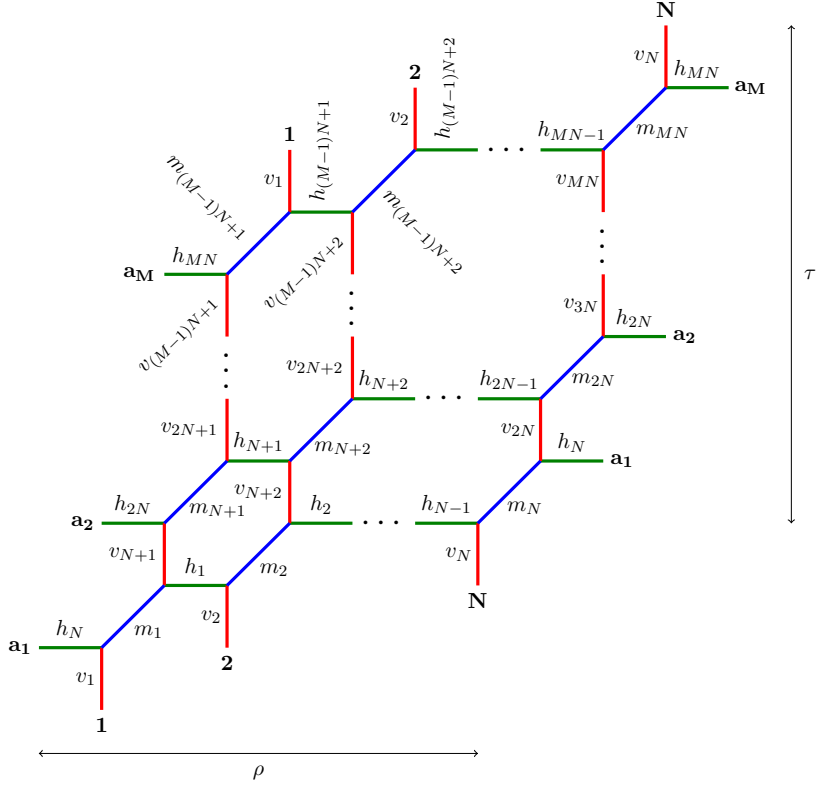
%%%%%%%%%%%%%%%%%%%%%%%%%%%%%%%%%%
%%%%%%%%%%%%%%%%%%%%%%%%%%%%%%%%%%

It should be noted that the corresponding instanton partition functions can be obtained by suitable expansions of the topological string partition function $\pf{N}{M}$. Indeed, in \cite{Hohenegger:2015btj}, these two gauge theories have also been referred to as \emph{vertical}- and \emph{horizontal description} respectively, reflecting two possible choices of a preferred direction in the toric web diagram of $X_{N,M}$ for calculating $\pf{N}{M}$ using the (refined) topological vertex: the web diagram is shown \figref{Fig:WebToric}, with a generic labelling of the various line segments, which represent rational curves in $X_{N,M}$. Since each trivalent vertex contains a horizontal (green) leg and a vertical (red) leg,  we can choose either of them as the preferred direction, which gives rise to a representation of $\pf{N}{M}$ as a series expansion in $e^{-h_{1,\ldots,MN}}$ or $e^{-v_{1,\ldots,MN}}$, respectively. At the particular region in the moduli space space considered in \cite{Hohenegger:2015btj}, these in turn could readily be identified with the instanton expansions of the quiver gauge theories mentioned above. 

However, inspecting \figref{Fig:WebToric}, we see that all vertices also contain a diagonal (blue) leg, which can equally be chosen as the preferred direction. Hereafter, we refer to this as {\sl diagonal description}. In fact, in \cite{Bastian:2017ing}, diagonal expansion of $\pf{N}{M}$ in the form of a series in powers of $e^{-m_{1,\ldots,NM}}$ have recently been studied. While well-defined from the perspective of the topological string, it is an interesting question if this presentation of $\pf{N}{M}$ (in some region of the parameter space) can also be interpreted as the instanton partition function of a new (quiver) gauge theory that can be engineered from $X_{N,M}$. If this were so, it would extend the T-duality relation of LSTs to a \emph{triality}. In this paper, we present strong evidence that this is indeed the case and that the %\emph
{diagonal description} gives a (circular) quiver gauge theory with gauge group $[U(NM/k)]^k$ (where $k=\text{gcd}(N,M)$), which is  the weak coupling limit of a new LST. Working at a generic point in the K\"ahler moduli space of $X_{N,M}$, we analyse in detail the horizontal, vertical and diagonal gauge theories, whose gauge group is
\begin{align}
&G_{\text{hor}}=[U(M)]^N\,,&&G_{\text{vert}}=[U(N)]^M\,,&&G_{\text{diag}}=[U(\tfrac{NM}{k})]^k\,,\nonumber
\end{align}
respectively. We propose explicit parametrisations of the K\"ahler cone of $X_{N,M}$ which makes these three gauge symmetries manifest by allowing to extract the corresponding instanton partition functions $Z^{(N,M)}_{\text{hor}}$, $Z^{(N,M)}_{\text{vert}}$ and $Z^{(N,M)}_{\text{diag}}$ from $\pf{N}{M}$. 

The Alday-Gaiotto-Tachikawa (AGT) relation for 
four-dimensional ${\cal N}=2$ supersymmetric gauge theories \cite{AGT, AGT2} has been extended to five-dimensional  gauge theories \cite{Awata:2009ur,Schiappa:2009cc, Awata:2010yy} and to five-dimensional quiver gauge theories \cite{Kimura:2015rgi}.  The triality we newly proposed in this paper has interesting implications for this AGT relation. 
It is known that there is a relation between the five-dimensional $[U(N)]^M$ quiver gauge theory and an $W_N$ algebra \cite{Awata:2010yy}. By duality, there is also a relation between the same quiver gauge theory and an $W_{M}$ algebra \cite{Kimura:2015rgi}. The newly proposed triality
implies a further relation between the same quiver gauge theory and an $W_{\frac{NM}{k}}$ algebra.

Finally, based on our previous work \cite{Bastian:2017ing}, we remark that in the extended K\"ahler moduli space of $X_{N,M}$ (extended through flop transformations) there are yet more LSTs. By analysing the associated toric diagrams, it was shown in \cite{Hohenegger:2016yuv} that $X_{N,M}\sim X_{N',M'}$ for $MN=M'N'$ and $\text{gcd}(M,N)=k=\text{gcd}(M',N')$, \emph{i.e.} the two Calabi-Yau threefolds lie in the same extended moduli space and can be related by a combination of symmetry and flop transforms. For example, the K\"ahler cone of $X_{N',M'}$ again affords different parametrisations, which suggest the appearance of a triad of quiver gauge theories, with gauge groups $[U(M')]^{N'}$, $[U(N')]^{M'}$ and $[U(N'M'/k)]^k$, which are generically different from the theories mentioned above. These, along with other dual quiver gauge theories and  associated $W$-algebras, will be discussed in detail in the forthcoming publication \cite{paper3}.

This paper is organised as follows. In section 2, we discuss LSTs of type $(N,M)$ from the viewpoints of branes in M-theory, $(p,q)$ 5-brane webs in type IIB theory, and dual Calabi-Yau threefolds $X_{N,M}$. We also discuss the K\"ahler cone of $X_{N,M}$ and the theory data of the dual gauge theories. In section 3, we analyze three different limits of the LST partition function leading to distinct (but dual) weakly coupled affine A-type quiver gauge theories. In section 4, we make some general comments about the implication of triality for the W-algebras associated with these gauge theories. In section 5, we present our conclusions and directions for future work. In appendix, we relegate special situations of $(N, N)$ and $(nN, N)$ webs. We also elaborate the residual dualities that survive in the noncompact limit that the LSTs descend to superconformal field theories.

%%%%%%%%%%%%%%%%%%%%%%%%%%%%%%%%%
\section{Little Strings, Gauge Theories and Brane Webs}
In this section, we briefly review how six-dimensional LSTs with $\mathcal{N}=(2,0)$ and $\mathcal{N}=(1,0)$ supersymmetry and world-volume $\mathbb{R}^{4}\times \mathbb{T}^2$ can be constructed from brane configurations in M-theory, along with their parameter space. 
These M5-brane configurations are U-dual to a class of $(p,q)$ 5-brane webs in type IIB string theory and can give rise to, in the limit of decoupling little strings, four-dimensional superconformal field theories (SCFTs) with eight supercharges in the infrared. We will also discuss in this section the gauge group and the coupling constants of the quiver gauge theories dual to LSTs on $\mathbb{R}^{4}\times \mathbb{T}^2$ and the K\"ahler cone of the associated Calabi-Yau threefold $X_{N,M}$ which is dual to the $(p,q)$ 5-brane web.
 
 \subsection{Little String Theories from M-Theory}\label{Sect:NonMaximal}
 In this subsection, we discuss how ${\cal N} = (1,0)$ LSTs arise from M-theory along with their parameter spaces. We begin with a brief review of the M-brane configurations which engineer LSTs. For more details on this construction, see \cite{Cecotti:2013mba, Haghighat:2013gba, Haghighat:2013tka, Hohenegger:2015btj}.

%%%%%%%%%%%%%%%%%%%%%%%%% 
 \subsubsection{M-brane Configurations}
A stack of $N$ coincident M5-branes probing a transverse $\mathbb{R}^5$ space at low energies is described by a six-dimensional ${\cal N} = (2,0)$ SCFT of $A_{N-1}$ type~\cite{Witten:1995zh,Witten:1995gx,Strominger:1995ac,Ganor:1996mu,Seiberg:1996qx}. We can move away from the conformal point by separating the M5-branes along a $\mathbb{R}_{\text{trans}}\subset\mathbb{R}^5$ transverse to the branes: indeed, in this case massive states appear in the form of M2-branes stretched between the individual M5-branes. If $\mathbb{R}_{\text{trans}}$ is furthermore compactified to a circle $\mathbb{S}^1_{\text{trans}}$ of circumference $\rho$, the dynamics on the $N$ M5-branes, whether coincident or not, defines a maximally supersymmetric LST of $A$-type. The positions of the M5-branes on the transverse circle parametrise a $(\mathbb{S}^1)^N/S_N$ moduli subspace of the tensor branch of the $N$ M5-branes on the (partially) compactified space. The circumference $\rho$ sets a defining length scale of the LST, which appears since an M2-brane can wrap $\mathbb{S}^1_{\text{trans}}$ while ending on separate M5-branes, thus giving rise to a string configuration whose tension $T \sim \rho$ (in M theory unit) is proportional to $\rho$. If the $\mathbb{S}^1_{\text{trans}}\times \mathbb{R}^4$ transverse to the M5-branes is orbifolded by $\mathbb{Z}_{M}$ (\emph{i.e.} replaced by $\mathbb{S}_{\text{trans}}\times \mathbb{R}^{4}/\mathbb{Z}_{M}$), we obtain a LST with $\mathcal{N}=(1,0)$ supersymmetry. The spectrum of the latter is consists of towers of tensor multiplets, vector multiplets, and hypermultiplets.

The configurations we are mostly concerned with in this work are obtained by compactifying not only the time direction on $\mathbb{S}^1_0$ (whose radius is chosen to be $1$), but also a further direction parallel to the M5-branes on $\mathbb{S}^1_{||}$ (with circumference $\tau$), such that the world-volume of the M5-branes becomes $\mathbb{S}^1_0\times \mathbb{S}^1_{||}\times \mathbb{R}^{4}$. As has been discussed in detail in \cite{Haghighat:2013gba}, these M-brane configurations are U-dual to webs of $M$ parallel D5-branes intersecting $N$ parallel NS5-branes in type II string theory. Upon resolving the intersections of these branes, the web can be presented in the form shown in \figref{Fig:WebToric}. This description in turn is dual to F-theory compactifications on a toric Calabi-Yau manifold $X_{N,M}$, whose toric web diagram takes the same form as \figref{Fig:WebToric}. For more details on these dual descriptions, we refer the reader to \cite{Haghighat:2013gba, Haghighat:2013tka,Hohenegger:2013ala} and references therein.

\subsubsection{Little Strings and Their Parameter Space}\label{Sect:KaehlerParameters}
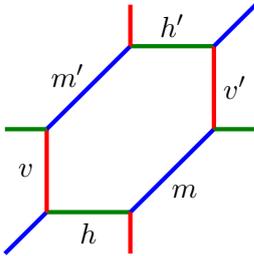
\begin{wrapfigure}{L}{0.3\textwidth}
\centering
\scalebox{1}{\parbox{3.5cm}{\begin{tikzpicture}[scale = 1.1]
\draw[ultra thick, blue] (-0.5,-0.5) -- (0,0);
\draw[ultra thick, green!50!black] (0,0) -- (1,0);
\draw[ultra thick, red] (0,0) -- (0,1);
\draw[ultra thick, blue] (1,0) -- (2,1);
\draw[ultra thick, blue] (0,1) -- (1,2);
\draw[ultra thick, green!50!black] (1,2) -- (2,2);
\draw[ultra thick, red] (2,1) -- (2,2);
\draw[ultra thick, blue] (2,2) -- (2.5,2.5);
\draw[ultra thick, red] (1,0) -- (1,-0.5);
\draw[ultra thick, red] (1,2) -- (1,2.5);
\draw[ultra thick, green!50!black] (0,1) -- (-0.5,1);
\draw[ultra thick, green!50!black] (2,1) -- (2.5,1);
\node at (0.5,-0.25) {{\small $h$}};
\node at (1.5,2.25) {{\small $h'$}};
\node at (-0.25,0.5) {{\small $v$}};
\node at (2.25,1.5) {{\small $v'$}};
\node at (1.65,0.25) {{\small $m$}};
\node at (0.25,1.65) {{\small $m'$}};
\end{tikzpicture}}}
\caption{\sl Imposing consistency conditions on a hexagon appearing in \figref{Fig:WebToric}.}
\label{Fig:ConsistencyCondition}
\end{wrapfigure}

The brane web shown in \figref{Fig:WebToric} is dual to the Calabi-Yau threefold $X_{N,M}$ with sizes of the various line segments corresponding to the K\"ahler parameters of $X_{N,M}$ \cite{Leung:1997tw}. The $3NM$ parameters labelling the web, as shown in \figref{Fig:WebToric}, (hereafter we collectively denote them as $({\bf h},{\bf v},{\bf m})$) are not all independent. There are only $(NM+2)$ independent K\"ahler parameters (see \cite{Haghighat:2013tka}), as they are constrained by the condition that all horizontal, vertical and diagonal lines must be parallel (\emph{i.e.} oriented along $(1,0)$, $(0,1)$ and $(1,1)$, respectively). Considering a hexagon web in \figref{Fig:WebToric} consisting of two horizontal, two vertical and two diagonal lines (see \figref{Fig:ConsistencyCondition} with some 

\noindent 
generic labelling of the areas associated with these line segments), the condition that the horizontal, vertical and diagonal lines are pairwise parallel, leads to the following constraints
\begin{align}
&h+m=h'+m'\, &&\text{and} &&v+m'=m+v'\,.\label{ConsistencyCondSchem}
\end{align}
Imposition of these conditions for every hexagon appearing in \figref{Fig:WebToric} leads to a system of linear equations whose solution contains $(MN+2)$ independent parameters (see \cite{Hohenegger:2016yuv} for more details). How to choose these independent variables is a priori not fixed and, as we shall see in following section, different choices allow us to explore different  gauge theories engineered by the web diagrams. 

We also remark that, in many works in the literature (see \cite{Haghighat:2013tka,Hohenegger:2013ala,Hohenegger:2015btj,Hohenegger:2016eqy,Ahmed:2017hfr}), a non-maximal set of parameters $(T_1,\ldots,T_M, t_1,\ldots,t_N,m)$ is  often considered, which (with respect to the labelling of  \figref{Fig:WebToric}) is related to our parametrization as 
{\allowdisplaybreaks
\begin{align}
&T_i=m_{(i-1)N+s}+v_{iN+s}\,,&&\forall s=1,\ldots,N\qquad \text{ and }i=1,\ldots,M\,,\nonumber\\
&t_a=m_{a+rN}+h_{a+rN}\,,&&\forall r=0,\ldots,M-1\text{ and }a=1,\ldots,N\,,\nonumber\\
&m=m_k\,,&&\forall k=1,\ldots, NM\,. \label{RestrictedRegionModuli}
\end{align}
}
Indeed, the set of $(N+M+1)$ variables $(T_1,\ldots,T_M, t_1,\ldots,t_N,m)$ is a solution of the consistency conditions of the type (\ref{ConsistencyCondSchem}).

The partition function of the LSTs engineered by the M-brane configurations corresponding to the web diagram in \figref{Fig:WebToric} is computed by the refined topological string partition function $\pf{N}{M}$ of  the Calabi-Yau threefold $X_{N,M}$. This partition function depends on the K\"ahler parameters $(\mathbf{h},\mathbf{v},\mathbf{m})$ (subject to the consistency conditions mentioned above). In addition, it depends on the two refinement parameters $\epsilon_1$ and $\epsilon_2$ required by the refined topological string:
\bea
\pf{N}{M}({\bf h},{\bf v},{\bf m},\epsilon_{1},\epsilon_2)\,\label{PartitionFunction}
\eea
From the perspective of the gauge theories engineered by $X_{N,M}$, $\epsilon_{1,2}$ parametrise the $\Omega$ background (see \cite{Moore:1997dj,Lossev:1997bz,Nekrasov:2002qd}), suitable for regularising the instanton contribution. Moreover, 
$\pf{N}{M}$ can be computed with the help of the refined topological vertex, which requires picking a preferred direction in the web diagram. Each choice of the preferred direction corresponds to a possibly distinct gauge theory partition function \cite{Iqbal:2007ii,Haghighat:2013gba,Bastian:2017ing}.

%%%%%%%%%%%%%%%%%%%%%%%%%%%%%%%
\subsection{Dual Gauge Theories and Series Expansions}

\subsubsection{Theory Data}\label{Sect:TheoryData}

At low energies below the compactification scale, the LSTs associated with the Calabi-Yau threefolds $X_{N,M}$ discussed above are described by five-dimensional quiver gauge theories of $\widehat{A}_{r-1}$ type with a unitary gauge group at each node $U(s)_{a}$\,($a=1,\cdots,r)$. For a general low-energy description of M-theory compactified on a Calabi-Yau threefold, see \cite{Witten:1996qb,Ferrara:1996wv,Chou:1997ba}. The field content of these theories includes gauge vector multiplets $(\varphi_a, A_a)$, transforming in the adjoint of $U(s)_{a}$, and matter hypermultiplets $(H^{a}, \widetilde{H}^{a})$, transforming in the bi-fundamental representations of $U(s)_{a}\times U(s)_{a+1}$. From the perspective of the gauge theories, at a generic point of the Coulomb branch of the theory with gauge group $G$, the dynamics is described by the prepotential ${\cal F}$,
\bea
{\cal F} = {1 \over 2!} t_{ij} \varphi^i \varphi^j
+ {1 \over 3!} c^{0}_{ijk} \varphi^{i} \varphi^j \varphi^k +\frac{1}{12}\Big(\sum_{R}|R\cdot \varphi|^3-\sum_{f}\sum_{w\in W_{f}}|w\cdot \varphi+m_{f}|^3\Big)\,.
\eea
Here, $\varphi$'s are local coordinates in an open patch of the Coulomb branch moduli space ${\cal M}$ and are the vacuum expectation values of the scalars in the vector multiplet with $i=1,\cdots, \mbox{rank}(G)$. $R$ is the set of weights of the adjoint representation, $W_{f}$ are the sets of weights in the representation of the hypermultiplets with mass $m_{f}$ which for us is the bi-fundamental representation and the effective gauge couplings are given by,
\bea
\Big(g^{-2}\Big)_{ij} = \partial_i \partial_j {\cal F}=t_{ij}+\cdots\,.
\nonumber
\eea
The five-dimensional Lagrangian contains the Chern-Simons term,
\bea\label{CSterm}
{\cal L}=\frac{1}{24\pi^2}\,c_{ijk}\,A^{i}\wedge F^{j}\wedge F^{k} \quad
\mbox{where} \quad c_{ijk}=\frac{\partial^{3}{\cal F}}{\partial \varphi_{i}\partial \varphi_{j}\partial \varphi_{k}}=c^{0}_{ijk}+\cdots,
\eea
and $F^{i}$ is the field strength two-form associated with gauge potential one-form $A^{i}$. The presence of this term spoils the gauge invariance of the path integral at quantum level unless the coefficients $c_{ijk}$ are integrally quantized (in suitable units) and therefore cannot change under continuous symmetry transformations of the theory.

Recall that, if the five-dimensional theory were obtained by compactification of M-theory on a Calabi-Yau threefold, then the prepotential of the theory is related to the triple intersection of the divisors. For instance, consider a five-dimensional $SU(N)$ gauge theory obtained from a Calabi-Yau threefold $Y_{N}$ which is an $A_{N-1}$ singularity fibered over $\mathbb{P}^1$. Resolving the $A_{N-1}$ singularity gives a chain of $(N-1)$ $\mathbb{P}^1$'s in the fiber. The compact divisors of $Y_{N}$ are the 4-cycles $\{D_{1},D_{2},\cdots,D_{N-1}\}$, which are the $i$-th fiber $\mathbb{P}^1$ fibered over the base $\mathbb{P}^1$. Each of the divisors $D_{i}$ is a Hirzebruch surface. The size of the $\mathbb{P}^{1}$'s in the fiber is $t_{i}=(\varphi_{i+1}-\varphi_{i})$\,($i=1,\cdots,N$) where $\varphi_{i}$'s are the scalars introduced above. The cubic part of the prepotential, which gives the Chern-Simons coefficient (\ref{CSterm}), of the five-dimensional theory is then given by \cite{Intriligator:1997pq}
\bea
{\cal F}|_{\text{cubic}}=\frac{1}{6}\Big(\sum_{i,j=1}^{N-1}(A^{-1})^{ij}t_{i}D_{j}\Big)^3\,,
\eea
where $(A^{-1})^{ij}$ is the inverse Cartan matrix of $A_{N-1}$ and $D_{i}D_{j}D_{k}$ are the triple intersection numbers of the divisors in $Y_{N}$. From this, we can see that the coefficients $c_{ijk}$ in Eq.(\ref{CSterm}) depend on the triple intersection numbers of the divisors in the Calabi-Yau threefold $Y_N$. When the five-dimensional theory is considered on $\mathbb{R}^4\times \mathbb{S}^1$, the M2-branes wrapping the holomorphic curves in $Y_N$ and the $\mathbb{S}^1$ (with momentum along the $\mathbb{S}^1$) contribute to the prepotential as well. In the type IIA string description,  these are the bound-states of D2-branes and the D0-branes. The triple intersection numbers now get contributions from these bound-states such that 
\bea
D_{i}D_{j}D_{k}\quad \mapsto \quad D_{i}D_{j}D_{k}+\sum_{\beta\in H_{2}(Y_{N},\mathbb{Z})}\,N^{\beta}_{ijk}\,e^{-\int_{\beta}\omega},
\eea 
where $N^{\beta}_{ijk}$ is the contribution from the holomorphic curves in the class $\beta$ which intersect the divisors $D_{i},D_{j}$ and $D_{k}$. These contributions are precisely captured by the genus-zero topological string amplitude. The full partition function of the gauge theory on $\mathbb{R}^{4}\times \mathbb{S}^1$ is thus given by the topological string partition function of the corresponding Calabi-Yau threefold used to engineer the gauge theory itself \cite{Iqbal:2003ix,Iqbal:2003zz}.

We can relate this gauge theory description to the Calabi-Yau description discussed in the previous section: the K\"ahler parameters of $X_{N,M}$ take the role of either the gauge coupling constants, Coulomb branch parameters or hypermultiplet masses. As already alluded to above, the precise correspondence is not unique, \emph{i.e.} the $(NM+2)$ independent K\"ahler parameters can be assigned in different ways to these three sets of gauge theory parameters. This means that in general different gauge theories can be associated with a single $X_{N,M}$ which, however, are dual to each other. To clarify more precisely what we mean by `different but dual', we need to provide more details on the characteristic quantities that are necessary to distinguish two gauge theories engineered from the same $X_{N,M}$. They are as follows:
\begin{itemize}
\item gauge group $G$\\
The gauge theories discussed above that are associated with $X_{N,M}$ are of $\widehat{A}$ quiver type with gauge group
\begin{align}
&G=[U(s)]^r\,,&&\text{with} &&rs=NM\,.\label{GaugeGroup}
\end{align}
Here the condition $rs=NM$ can be understood from the fact that these gauge theories are low-energy limits of LSTs of A-type, which can be obtained as a particular decoupling limits of type IIB string theory: conservation of D-brane charges in the latter imposes that the rank of the gauge group, $|G|$, is the same for all theories associated with $X_{N,M}$.
%%%
\item gauge coupling constants\\
To each of the $r$ factors of $U(s)$ in the gauge group $G$ in (\ref{GaugeGroup}), we can associate an independent inverse coupling constant $g_i^{-2}$. The coupling constants can be varied continuously and the weak coupling limit is $g_i^{-2}\to \infty$ for $i=1,\ldots,r$. 
\item moduli space\\
The $(NM+2-r)$ remaining independent K\"ahler parameters of $X_{N,M}$ parametrise the  hypermultiplet masses and Coulomb branch parameters of the gauge theory. The latter can be varied continuously and we assume to always work at a point in the moduli space where the latter are positive definite.
%%%
\item Chern-Simons terms\\
As we just discussed, the five-dimensional gauge theories contain Chern-Simons couplings for the vector multiplet gauge fields Eq.(\ref{CSterm}).
The coefficients $c_{ijk}$ in this term are given by the triple intersection number of the Calabi-Yau threefold. They appear in the topological string partition function of the corresponding Calabi-Yau threefolds  \cite{Bershadsky:1993cx} (see also \cite{Tachikawa:2004ur}). Recall that in general the topological string partition function of a Calabi-Yau threefold $X$ is given by
\bea
Z_{X}=\text{exp}(F)\,,
\eea
where $F=\sum_{g=0}^{\infty}\lambda^{2g-2}F_{g}$ is the free energy given by the topological string coupling $\lambda$ and the genus-$g$ topological string amplitudes $F_g$ which are given by integrating topological string measure over the moduli space of genus-$g$ curves. If we denote the K\"ahler class of the Calabi-Yau threefold $X$ by $\omega$ \cite{Gopakumar:1998ii, Gopakumar:1998jq},
\bea
F&=&{1 \over \lambda^2}\int_{X} {1 \over 3!} \omega\wedge \omega\wedge \omega-
\int_{X}{c_{2}(X) \over 24} \wedge \omega+{\chi(X) \over 2} \ \text{ln} \ M(e^{i\lambda})\\\nonumber
&&+\sum_{0\neq \beta\in H_{2}(X,\mathbb{Z})}\sum_{g\geq 0}n^{g}(\beta)\,\sum_{n=1}^{\infty}{1 \over n}\,\Big(2 \ \text{sin}(\tfrac{n\lambda}{2})\Big)^{2g-2}e^{-n\int_{\beta}\omega}\,,
\eea
where $c_{2}(X)$ is the second Chern class of the tangent bundle of $X$, $\chi(X)$ is the Euler characteristic of $X$ and $M(q)$ is the MacMahon function,
\bea
M(q)=\prod_{k=1}^{\infty}(1-q^k)^{-k}\,.
\eea
The $n^{g}(\beta)$ are the genus-$g$ Gopakumar-Vafa invariants \cite{Gopakumar:1998ii,Gopakumar:1998jq}  of the curve class $\beta$ and are conjectured to be integers since they are related to the dimensions of the various homology groups of the moduli space of D2-branes wrapping the holomorphic curve in class $\beta$. The genus-zero contribution to the free energy contains the classical contribution $\int_{X}\omega \wedge\omega \wedge \omega$ which gives the triple intersection numbers $(\omega=\sum_{i=1}^{\text{dim}H^{1,1}(X,\mathbb{C})}x^{i}\,\omega_{i}$ where $\omega_{i}\in H^{1,1}(X,\mathbb{C})$),
\bea
\int_{X}\omega \wedge\omega \wedge \omega=c_{ijk}x^{i}x^{j}x^{k}\, \quad \mbox{where} \quad c_{ijk}=\int_{X}\omega_{i}\wedge\omega_{j}\wedge \omega_{k}\,.
\eea
From this, we see that the triple intersection numbers and hence the Chern-Simons coefficients appear in the definition of $F_{0}$ (the genus-zero amplitude) and hence in the partition function. However, this is not the only place the Chern-Simons terms appear. Since the geometry of the Calabi-Yau threefold is related to the triple intersection numbers, the Gopakumar-Vafa invariants $n^{g}(\beta)$ change as the triple intersection numbers change. One can see this very clearly when the topological string partition function is expressed in the form of Nekrasov's instanton partition function. 
 
In the case of $SU(N)$ gauge theory we discussed above, the Chern-Simons coefficient can take values from $-N$ to $+N$, where theories of opposite sign Chern-Simons coefficients are related by charge-parity conjugation. So, there are $(N + 1)$ distinct gauge theories.This is reflected in the geometry as well since the fibration of the $A_{N-1}$ singularity over $\mathbb{P}^1$ is not unique either and there are actually $(N+1)$ distinct fibrations corresponding to distinct five-dimensional gauge theories. We denote by $Y_{N, k}$ the Calabi-Yau threefolds which engineer $SU(N)$ gauge theory with Chern-Simons coefficient $k$. 

The topological string partition function of $Y_{N, k}$ were calculated in \cite{Iqbal:2003zz} (for the refined case, see \cite{Taki}) and it was found that partition functions for different $k$ were related to each other. Define the topological string partition function of $Y_{N,k}$ as
\bea
Z_{Y_{N,k}}=\sum_{\nu_{1}\cdots \nu_{N}}Q_{B}^{|\nu_{1}|+\cdots+|\nu_{N}|}\,Z^{k}_{\nu_{1}\cdots\nu_{N}}({\bf t},\epsilon)\,,
\eea
where $-\mbox{ln} \ Q_{B}$ is the K\"ahler parameter of the base $\mathbb{P}^1$. Then,  
\bea
Z^{k}_{\vec{\nu}}({\bf t},\epsilon)&=&f^{k}_{\vec{\nu}}({\bf t},\epsilon)\,Z^{0}_{\vec{\nu}}({\bf t},\epsilon)\,, 
\eea
where the $k$-dependent part reads
\bea
f^{k}_{\vec{\nu}}({\bf t},\epsilon)&=&\Big(\prod_{i=1}^{\lfloor\frac{N+k-1}{2}\rfloor}Q_{t_{i}}^{(N+k-2i)(|\nu_{1}|+\cdots+|\nu_{i}|)}\! \prod_{i=\lfloor\frac{N+ k+1}{2}\rfloor}^{N-1}Q_{t_{i}}^{-(N+ k -2i)(|\nu_{i+1}|+\cdots +|\nu_{N}|)}\Big)\,q^{{1 \over 2} k\sum_{i=1}^{N}{\kappa(\nu_{i})}}\,,
\nonumber
\eea
where $\kappa(\nu)=\sum_{(i,j)\in \nu}(j-i)$. Thus, we see that the effect of the Chern-Simons coefficient is to modify the prefactor once the topological string partition function is expressed in the form of the Nekrasov partition function.
 
\end{itemize}

\indent
The non-perturbative gauge theory partition function for a theory with the gauge group $G$ in Eq.(\ref{GaugeGroup}) can be obtained by writing $\pf{N}{M}$ as a series expansion in powers of $\text{exp}[-1/g_a^2]$. In practice, we will identify  sets of parameters, the decoupling parameters $d_{a}$, such that in the limit $d_{a}\mapsto \infty$ the $(N,M)$ web breaks up into pieces whose partition function can be identified with the perturbative part of gauge theory partition function. The parameters $d_{a}$ are related to the gauge theory coupling constants $\frac{1}{g_{a}^2}$ such that
\bea\label{DecouplingDef}
d_{a}\mapsto \infty \qquad \implies \qquad \frac{1}{g_{a}^2}\mapsto \infty\,.
\eea
The $d_a$'s differ from $\frac{1}{g_{a}^2}$ by a combination of Coulomb branch parameters and masses of the bi-fundamental hypermultiplets. Thus, we are naturally led to studying different types of series expansions of Eq.(\ref{PartitionFunction}), as we shall do in the following.

%%%%%%%%%%%%%%%%%%%%%%%%
\subsubsection{Series Expansions and Building Blocks}
\begin{figure}[htb]
\begin{center}
%\hspace*{-2cm}%
\scalebox{1.1}{\parbox{4cm}{\begin{tikzpicture}[scale = 0.6]
%horizontal lines
\draw[ultra thick,green!50!black] (-6,0) -- (-5,0);
\draw[ultra thick,green!50!black] (-4,1) -- (-3,1);
\draw[ultra thick,green!50!black] (-5,2) -- (-4,2);
\draw[ultra thick,green!50!black] (-3,3) -- (-2,3);
\draw[ultra thick,green!50!black] (-2,8) -- (-1,8);
\draw[ultra thick,green!50!black] (-4,7) -- (-3,7);
%diagonal lines
\draw[ultra thick,blue] (-5,0) -- (-4,1);
\draw[ultra thick,blue] (-4,2) -- (-3,3);
\draw[ultra thick,blue] (-3,7) -- (-2,8);
%vertical lines
\draw[ultra thick,red] (-5,-1) -- (-5,0);
\draw[ultra thick,red] (-4,1) -- (-4,2);
\draw[ultra thick,red] (-3,3) -- (-3,4);
\draw[ultra thick,red] (-2,8) -- (-2,9);
\draw[ultra thick,red] (-3,6) -- (-3,7);
%dots
\node[rotate=90] at (-3,5) {\Large $\cdots$};
%nodes
\node at (-5,-0.8) {{\footnotesize $=$}};
\node at (-2,8.8) {{\footnotesize $=$}};
%partitions
\node at (-6.4,0) {{\tiny$\beta^t_M$}};
\node at (-5.65,2) {{\tiny$\beta^t_{M-1}$}};
\node at (-4.4,7) {{\tiny$\beta^t_{1}$}};
\node at (-2.4,1) {{\tiny$\alpha_M$}};
\node at (-1.3,3) {{\tiny$\alpha_{M-1}$}};
\node at (-0.5,8) {{\tiny$\alpha_{1}$}};
%labels v
\node at (-2.3,8.5) {{\tiny$v_{1}$}};
\node at (-3.3,6.5) {{\tiny$v_{2}$}};
\node at (-3.65,3.5) {{\tiny$v_{M-1}$}};
\node at (-4.35,1.5) {{\tiny$v_{M}$}};
\node at (-5.3,-0.5) {{\tiny$v_{1}$}};
%labels m
\node at (-2.1,7.3) {{\tiny$m_{1}$}};
\node at (-2.75,2.3) {{\tiny$m_{M-1}$}};
\node at (-3.9,0.3) {{\tiny$m_{M}$}};
%caption
\node at (-3.5,-2) {\small (a)}; 
\end{tikzpicture}
}}
%%%%%%%%%%%%%%%%%%%%%%%%%%
\hspace{0.2cm}
%%%%%%%%%%%%%%%%%%%%%%%%%%
\scalebox{1.1}{\parbox{6cm}{\begin{tikzpicture}[scale = 0.6]
%horizontal lines
\draw[ultra thick,green!50!black] (-1,0) -- (0,0);
\draw[ultra thick,green!50!black] (1,1) -- (2,1);
\draw[ultra thick,green!50!black] (3,2) -- (4,2);
\draw[ultra thick,green!50!black] (6,2) -- (7,2);
\draw[ultra thick,green!50!black] (8,3) -- (9,3);
%diagonal lines
\draw[ultra thick,blue] (0,0) -- (1,1);
\draw[ultra thick,blue] (2,1) -- (3,2);
\draw[ultra thick,blue] (7,2) -- (8,3);
%vertical lines
\draw[ultra thick,red] (0,0) -- (0,-1);
\draw[ultra thick,red] (2,1) -- (2,0);
\draw[ultra thick,red] (7,2) -- (7,1);
\draw[ultra thick,red] (1,1) -- (1,2);
\draw[ultra thick,red] (3,2) -- (3,3);
\draw[ultra thick,red] (8,3) -- (8,4);
%dots
\node at (5,2) {\Large $\cdots$};
%nodes
\node[rotate=90] at (-0.7,0) {{\footnotesize $=$}};
\node[rotate=90] at (8.8,3) {{\footnotesize $=$}};
%partitions
\node at (0,-1.3) {{\tiny$\beta^t_1$}};
\node at (2,-0.3) {{\tiny$\beta^t_2$}};
\node at (7,0.7) {{\tiny$\beta^t_M$}};
\node at (1,2.3) {{\tiny$\alpha_1$}};
\node at (3,3.3) {{\tiny$\alpha_2$}};
\node at (8,4.3) {{\tiny$\alpha_N$}};
%labels h
\node at (-0.3,0.3) {{\tiny$h_1$}};
\node at (1.5,1.3) {{\tiny$h_2$}};
\node at (3.5,2.3) {{\tiny$h_3$}};
\node at (6.5,2.3) {{\tiny$h_N$}};
\node at (8.3,2.7) {{\tiny$h_1$}};
%labels m
\node at (0.9,0.3) {{\tiny$m_1$}};
\node at (2.9,1.3) {{\tiny$m_2$}};
\node at (7.9,2.3) {{\tiny$m_N$}};
%caption
\node at (4,-4) {\small (b)}; 
\node at (4,7) {\small ${}$}; 
\end{tikzpicture}
}}
%%%%%%%%%%%%%%%%%%%%%%%%%%
\hspace{0.4cm}
%%%%%%%%%%%%%%%%%%%%%%%%%%
\scalebox{1.1}{\parbox{4.3cm}{\begin{tikzpicture}[scale = 0.6]
%horizontal lines
\draw[ultra thick,green!50!black] (-1,0) -- (0,0);
\draw[ultra thick,green!50!black] (0,-1) -- (1,-1);
\draw[ultra thick,green!50!black] (1,-2) -- (2,-2);
\draw[ultra thick,green!50!black] (4,-2) -- (5,-2);
\draw[ultra thick,green!50!black] (5,-3) -- (6,-3);
%vertical lines
\draw[ultra thick,red] (0,0) -- (0,-1);
\draw[ultra thick,red] (1,-1) -- (1,-2);
\draw[ultra thick,red] (5,-2) -- (5,-3);
%diagonal lines
\draw[ultra thick,blue] (0,0) -- (0.5,0.5);
\draw[ultra thick,blue] (1,-1) -- (1.5,-0.5);
\draw[ultra thick,blue] (5,-2) -- (5.5,-1.5);
\draw[ultra thick,blue] (0,-1) -- (-0.5,-1.5);
\draw[ultra thick,blue] (1,-2) -- (0.5,-2.5);
\draw[ultra thick,blue] (5,-3) -- (4.5,-3.5);
%dots
\node at (3,-2) {\Large $\cdots$};
%nodes
\node[rotate=90] at (-0.7,0) {{\footnotesize $=$}};
\node[rotate=90] at (5.8,-3) {{\footnotesize $=$}};
%partitions
\node at (0.7,0.7) {{\tiny$\alpha_1$}};
\node at (1.7,-0.3) {{\tiny$\alpha_2$}};
\node at (5.7,-1.3) {{\tiny$\alpha_{\frac{NM}{k}}$}};
\node at (-0.8,-1.8) {{\tiny$\beta^t_1$}};
\node at (0.2,-2.8) {{\tiny$\beta^t_2$}};
\node at (4.2,-3.8) {{\tiny$\beta^t_{\frac{NM}{k}}$}};
%labels h
\node at (-0.3,0.3) {{\tiny$h_1$}};
\node at (0.5,-0.7) {{\tiny$h_2$}};
\node at (1.5,-1.7) {{\tiny$h_3$}};
\node at (4.5,-1.6) {{\tiny$h_{\frac{NM}{k}}$}};
\node at (5.4,-2.7) {{\tiny$h_1$}};
%labels v
\node at (-0.3,-0.5) {{\tiny$v_1$}};
\node at (0.7,-1.5) {{\tiny$v_2$}};
\node at (4.4,-2.5) {{\tiny$v_{\frac{NM}{k}}$}};
%caption
\node at (2,-7) {\small (c)}; 
\node at (2,3.75) {\small ${}$}; 
\end{tikzpicture}
}}
%%%%%%%%%%%%%%%%%%%%%%%%%%
\end{center}
\caption{\sl Basic building blocks for decomposition of the web diagram~\figref{Fig:WebToric}: (a) strip of length $M$ representing the building block suitable for horizontal preferred direction (b) strip of length $N$ representing the building block suitable for vertical preferred direction (c) strip of length $\frac{NM}{k}$ (with $k=\text{gcd}(N,M)$) representing the building block suitable for diagonal preferred direction.}
\label{Fig:StripVertical}
\end{figure}
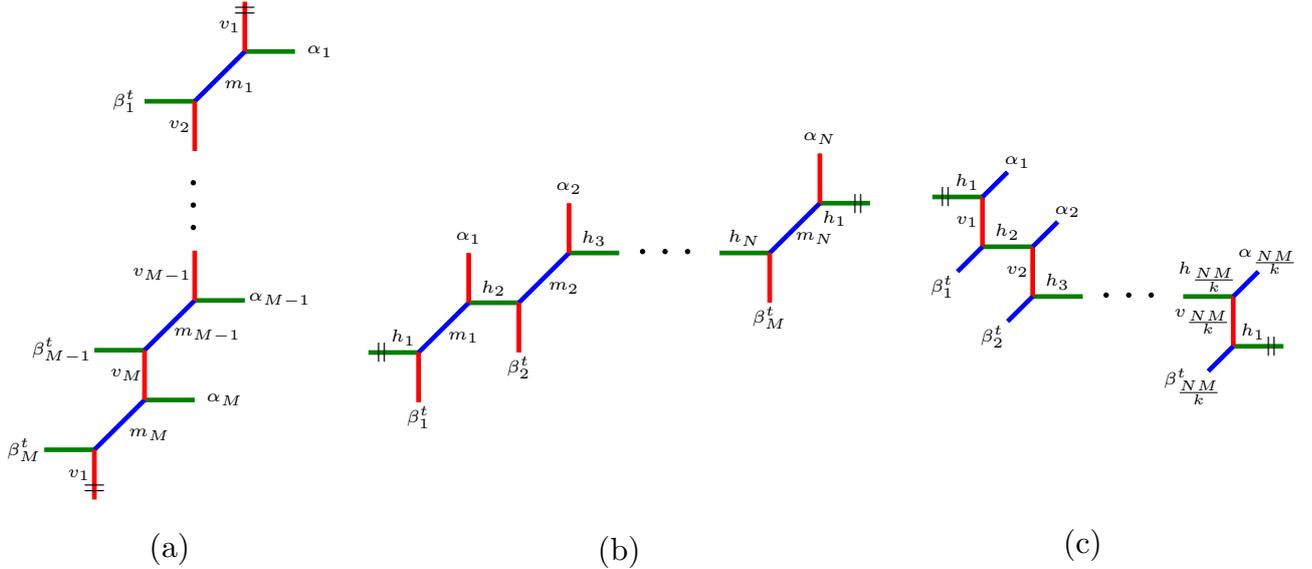

The topological string partition function $\pf{N}{M}$ appears naturally in the form of an infinite series when computed using the refined topological vertex. Indeed, gluing the trivalent vertices together according to the web diagram of $X_{N,M}$ requires to choose a preferred direction common to all vertices. Thus, the preferred direction is a feature of the entire web diagram and not only of an individual vertex. A given brane web, however, may allow for several different choices, which lead to different (but equivalent) representations of the topological string partition function. For web diagrams of the form shown in \figref{Fig:WebToric}, we have three different choices of the preferred direction, namely, horizontal along $(1,0)$,  vertical along $(0,1)$ and diagonal along $(1,1)$. For each of the three, the web-diagram can be cut into strips as shown in~\figref{Fig:StripVertical}, which represent the basic building blocks for computing $\pf{N}{M}$.

\noindent
Here, the external legs of each of the strips are labelled in terms of two sets of integer partitions $\{\alpha_{1,\ldots,L}\}$ and $\{\beta_{1,\ldots,L}\}$ (where $L=M$ or $L=N$ or $L=\frac{NM}{k}$ for (a), (b) or (c), respectively, and $\beta^t$ denotes the transposed partition of $\beta$). 

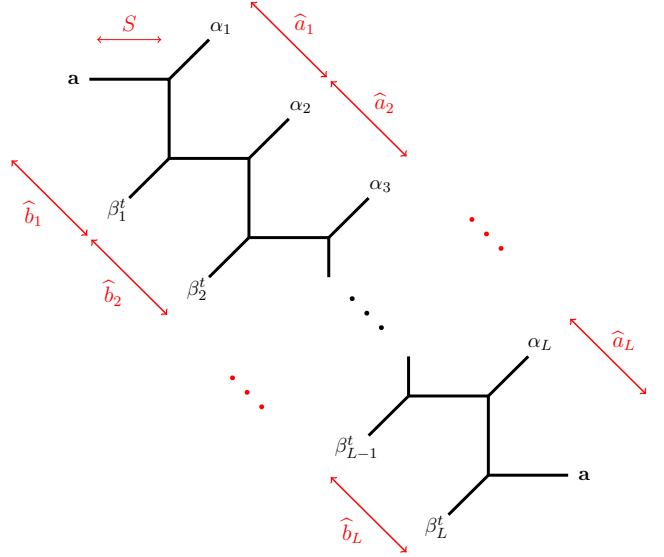
\begin{wrapfigure}[12]{r}{0.5\textwidth}
\centering
%\hspace*{-2cm}%
\scalebox{0.7}{\parbox{12cm}{\begin{tikzpicture}[scale = 1.50]
\draw[ultra thick] (0,0) -- (1,0) -- (1,-1) -- (2,-1) -- (2,-2) -- (3,-2) -- (3,-2.5) ;
%\draw[ultra thick,dashed] (3.5,-3) -- (4.5,-4) ;
\node[rotate=315] at (3.5,-3) {\Huge $\cdots$};
%diagonals up
\draw[ultra thick] (1,0) -- (1.5,0.5);
\draw[ultra thick] (2,-1) -- (2.5,-0.5);
\draw[ultra thick] (3,-2) -- (3.5,-1.5);
%diagonals down
\draw[ultra thick] (1,-1) -- (0.5,-1.5);
\draw[ultra thick] (2,-2) -- (1.5,-2.5);
%labels up
\node at (1.65,0.65) {{\small \bf $\alpha_1$}};
\node at (2.65,-0.35) {{\small \bf $\alpha_2$}};
\node at (3.65,-1.35) {{\small \bf $\alpha_3$}};
%labels down
\node at (0.35,-1.65) {{\small \bf $\beta^t_1$}};
\node at (1.35,-2.65) {{\small \bf $\beta^t_2$}};
%labels horizontal
\node at (-0.2,0) {{\small $\mathbf{a}$}};
%lower part
\begin{scope}[xshift=-1cm,yshift=1cm]
\draw[ultra thick] (5,-4.5) -- (5,-5) -- (6,-5) -- (6,-6) -- (7,-6);
\node at (4.35,-5.65) {{\small \bf $\beta^t_{L-1}$}};
\node at (5.35,-6.65) {{\small \bf $\beta^t_L$}};
\node at (6.65,-4.35) {{\small \bf $\alpha_L$}};
\node at (7.2,-6) {{\small $\mathbf{a}$}};
\draw[ultra thick] (6,-5) -- (6.5,-4.5);
\draw[ultra thick] (5,-5) -- (4.5,-5.5);
\draw[ultra thick] (6,-6) -- (5.5,-6.5);
\end{scope}
%roots top
\draw[thick,<->,red] (2.025,0.975) -- (2.975,0.025);
\node[red] at (2.7,0.7) {$\widehat{a}_1$};
\draw[thick,<->,red] (3.025,-0.025) -- (3.975,-0.975);
\node[red] at (3.7,-0.3) {$\widehat{a}_2$};
\node[red,rotate=315] at (5,-2) {\Huge$\cdots$};
\draw[thick,<->,red] (6.025,-3.025) -- (6.975,-3.975);
\node[red] at (6.7,-3.3) {$\widehat{a}_L$};
%roots bottom
\draw[thick,<->,red] (-0.975,-1.025) -- (-0.025,-1.975);
\node[red] at (-0.7,-1.7) {$\widehat{b}_1$};
\draw[thick,<->,red] (0.025,-2.025) -- (0.975,-2.975);
\node[red] at (0.3,-2.7) {$\widehat{b}_2$};
\node[red,rotate=315] at (2,-4) {\Huge$\cdots$};
\draw[thick,<->,red] (3.025,-5.025) -- (3.975,-5.975);
\node[red] at (3.3,-5.7) {$\widehat{b}_L$};
%Sline
\draw[<->,red] (0.1,0.5) -- (0.9,0.5);
\node[red] at (0.5,0.7) {$S$};
\end{tikzpicture}}}
\caption{\sl Generic strip of length $L$ with a labelling of the various parameters suitable for the computation of the building block in Eq.(\ref{BuildingBlockPF}). The external legs are labelled by integer partition $\{\alpha_{1,\ldots,L}\}$ and $\{\beta_1,\ldots,\beta_L\}$ (where $\beta_i^t$ denotes the transposed partition).}
\label{Fig:GenericStrip}
\end{wrapfigure}

Since the three strips can be transformed  into one another with the help of an $SL(2,\mathbb{Z})$-symmetry (under which the partition function is invariant), they can all be computed in the same fashion (using the refined topological vertex). Indeed, the common generic expression has been calculated explicitly in \cite{Bastian:2017ing} for a generic parametrisation. In order to find the strips in \figref{Fig:StripVertical}, we simply need to adapt the K\"ahler parameters accordingly. To be precise, in \cite{Bastian:2017ing}, the so-called `staircase diagram' shown in \figref{Fig:GenericStrip} was considered, which is parametrised by $(\widehat{a}_{1,\ldots,L},\widehat{b}_{1,\ldots,L},S)$ (where $\sum_{i=1}^L(\widehat{a}_i-\widehat{b}_i)=0$). The latter can directly be adapted to fit each of the three strips (a), (b) and (c) in \figref{Fig:StripVertical}. Specifically, for $i=1,\ldots,L$, we have the following 

\noindent
correspondence\vspace{0.2cm}

\begin{tabular}{c|c|c|c}
& {\bf strip (a)} & {\bf strip (b)} & {\bf strip (c)} \\\hline
$\widehat{a}_i$ & $v_{i+1}+m_i$ & $h_{i}+m_i$ & $v_i+h_{i+1}$ \\
$\widehat{b}_i$ & $v_{i}+m_{i}$ & $h_{i}+m_{i-1}$ & $h_i+v_i$ \\
$S$ & $v_1$ & $m_N$ & $h_1$\\
$L$ & $M$ & $N$ & $\frac{NM}{k}$\\[4pt]
\end{tabular}

\noindent
where we identified $h_{L+1}=h_1$, $v_{L+1}=v_1$ and $m_{0}=m_L$. Defining $q=e^{2\pi i\epsilon_1}$, $t=e^{-2\pi i\epsilon_2}$ and
\begin{align}
&Q_{a_i}=e^{-\widehat{a}_i}\,,&&Q_{b_i}=e^{-\widehat{b}_i}\,,&&Q_S=e^{-S}\,,
\end{align}
as well as the following quantities
\begin{align}
&\widehat{Q}_{i,j} = Q_S\,\prod_{r=1}^i(Q_{a_r}Q_{b_r}^{-1})\, \prod_{k=1}^{j-1}Q_{a_{i-k}}\,,\nonumber\\
&\overline{Q}_{i,j} = \left\{\begin{array}{ll} 1 & \quad \text{if } j=L \\
 \prod_{k=1}^j Q_{a_{i-k}} & \quad \text{if }j\neq L \\
\end{array}\right. \nonumber\\		
&\dot{Q}_{i,j} = \prod_{k=1}^jQ_{b_{i+k}}	\,.		
\end{align}

\noindent
We can write the generic building block associated with the strip in \figref{Fig:GenericStrip} in the following fashion \cite{Bastian:2017ing} (see also \cite{Haghighat:2013gba,Hohenegger:2013ala} for earlier results working in the particular region (\ref{RestrictedRegionModuli}) of the moduli space of $X_{N,M}$)
\begin{align}
W^{\alpha_1 \dots \alpha_L}_{\beta_1 \dots \beta_L} = W_L(\emptyset) \cdot \hat{Z}\cdot \prod_{i,j=1}^L \frac{\mathcal{J}_{\alpha_i \beta_{j}}(\widehat{Q}_{i,i-j};q,t)\mathcal{J}_{\beta_{j}\alpha_i}((\widehat{Q}_{i,i-j})^{-1}Q_\rho;q,t)}{\mathcal{J}_{\alpha_i \alpha_{j}}(\overline{Q}_{i,i-j}\sqrt{q/t};q,t)\mathcal{J}_{\beta_{j}\beta_i}(\dot{Q}_{i,j-i}\sqrt{t/q};q,t)} \, ,  \label{BuildingBlockPF}
\end{align}
where the prefactors are given by (with $Q_\rho=\prod_{i=1}^L Q_{a_i}$)
\begin{align}
&W_L(\emptyset)=\prod_{i,j=1}^L \prod_{k,r,s=1}^{\infty} \frac{(1-\widehat{Q}_{i,j}\,Q_\rho^{k-1} q^{r-\frac{1}{2}}t^{s-\frac{1}{2}})(1-\widehat{Q}^{-1}_{i,j}Q_\rho^{k}q^{s-\frac{1}{2}}t^{r-\frac{1}{2}})}{(1-\overline{Q}_{i,j}Q_\rho^{k-1} q^{r}t^{s-1})(1-\dot{Q}_{i,j}Q_\rho^{k-1} q^{s-1}t^r)}\,,\nonumber\\
&\hat{Z}= \prod_{i=1}^L t^{\frac{||\alpha_k||^2}{2}} q^{\frac{||\alpha_k^t||^2}{2}} \tilde{Z}_{\alpha_k}(q,t) \tilde{Z}_{\alpha_k^t}(t,q)\,,\hspace{1cm} \tilde{Z}_\nu(t,q)=\prod_{(i,j)\in\nu}\left(1-t^{\nu_j^t-i+1}q^{\nu_i-j}\right)^{-1}\,,
\end{align}
and the function $\mathcal{J}_{\mu\nu}$ (for two partitions $\mu$ and $\nu$) is defined as
{\allowdisplaybreaks
\begin{align}
\mathcal{J}_{\mu\nu}(x;t,q)&=\prod_{k=1}^\infty J_{\mu\nu}(Q_\rho^{k-1}x;t,q)\,,\nonumber\\
J_{\mu\nu}(x;t,q)=&\prod_{(i,j)\in\mu}\left(1-x\,t^{\nu^t_j-i+\frac{1}{2}}q^{\mu_i-j+\frac{1}{2}}\right)\times\prod_{(i,j)\in\nu}\left(1-x\,t^{-\mu^t_j+i-\frac{1}{2}}q^{-\nu_i+j-\frac{1}{2}}\right)\,.\nonumber
\end{align}
}
Gluing several of strips together (see \cite{Bastian:2017ing} for more details), we get,
\bea\label{PartitionFunctionXNM}
{\cal Z}_{N,M}=\sum_{{\bf \alpha}}\Big(\prod_{i=1,j=1}^{M,N}e^{-u_{ij}\,|\alpha^{i}_{j}|}\Big)\,\prod_{j=1}^{N}W^{\alpha^{1}_{j}\cdots \alpha^{M}_{j}}_{\alpha^{1}_{j+1}\cdots \alpha^{M}_{j+1}}
\eea
where $e^{-u_{ij}}$ are the parameters used to glue the strips together and we are using the fact that the compactification of the web gives $\alpha^{i}_{N+1}=\alpha^{i}_{1}$. We can obtain three different series representations of $\pf{N}{M}$ which we write in the following form
{\allowdisplaybreaks
\begin{align}
\pf{N}{M}({\bf h},{\bf v},{\bf m},\epsilon_{1,2})&=Z_p({\bf v},{\bf m})\sum_{\vec{k}}\,e^{-\vec{k}\cdot \mathbf{h}}\,Z_{\vec{k}}({\bf v},{\bf m})=Z^{(N,M)}_{\text{hor}}\nonumber\\
&=Z_p({\bf h},{\bf m})\sum_{\vec{k}}\,e^{-\vec{k}\cdot \mathbf{v}}\,Z_{\vec{k}}({\bf h},{\bf m})=Z^{(N,M)}_{\text{vert}}\nonumber\\
&=Z_p({\bf h},{\bf v})\sum_{\vec{k}}\,e^{-\vec{k}\cdot \mathbf{m}}\,Z_{\vec{k}}({\bf h},{\bf v})=Z^{(N,M)}_{\text{diag}}\,,\label{TrialitySchemat}
\end{align}
}
\noindent
where $Z_p$ denotes the perturbative part in each of the three descriptions. The first two of these expansions ($Z^{(N,M)}_{\text{hor}}$ and $Z^{(N,M)}_{\text{vert}}$) have been extensively studied in the literature~\cite{Haghighat:2013gba,Hohenegger:2013ala} at a point in the K\"ahler moduli space of $X_{N,M}$ where $m_i=m_j$ (for $i,j=1,\ldots,NM$). The diagonal expansion $Z^{(N,M)}_{\text{diag}}$ has recently~\cite{Bastian:2017ing} been used to prove the duality \cite{Hohenegger:2016yuv} $X_{N,M}\sim X_{N',M'}$ (with $N'M'=NM$ and $\text{gcd}(N,M)=k=\text{gcd}(N',M')$) explicitly for $k=1$ and give further evidence for $k>1$. However, its gauge theory interpretation has not been studied so far. 

The main proposition of this paper is to interpret the three different expansions in (\ref{TrialitySchemat}) as instanton expansions in field theory. Therefore we associate a \emph{triality} of gauge theories to a given Calabi-Yau threefold $X_{N,M}$. However, identifying $Z^{(N,M)}_{\text{hor}}$, $Z^{(N,M)}_{\text{vert}}$ or $Z^{(N,M)}_{\text{diag}}$ in Eq.(\ref{TrialitySchemat}) with the instanton series of a gauge theory is, however, a priori delicate: indeed, due to the consistency conditions discussed in section~\ref{Sect:KaehlerParameters}, in general, the parameters $(\mathbf{h},\mathbf{v},\mathbf{m})$ are not independent of one another and therefore it is not clear whether (\ref{TrialitySchemat}) can be interpreted as (consistent) power series expansions in a set of independent parameters that can be identified with the gauge coupling parameters of a gauge theory. For this to be the case, we need to show that there exists a co-dimension $n$ region in the K\"ahler moduli space of $X_{N,M}$ in which either all $\mathbf{h}$ or $\mathbf{v}$ or $\mathbf{m}$ become infinite, while all the remaining parameters are finite. From the gauge theory perspective, this region corresponds to the weak-coupling regime and in a finite neighbourhood of it either $Z^{(N,M)}_{\text{hor}}$ or $Z^{(N,M)}_{\text{vert}}$ or $Z^{(N,M)}_{\text{diag}}$ in (\ref{TrialitySchemat}) respectively are well defined power series expansions of $\pf{N}{M}$ that can be identified as an instanton expansion.  

%%%%%%%%%%%%%%%%%%%%%%%%
\subsubsection{K\"ahler Cone}\label{Sect:KaehlerCone}

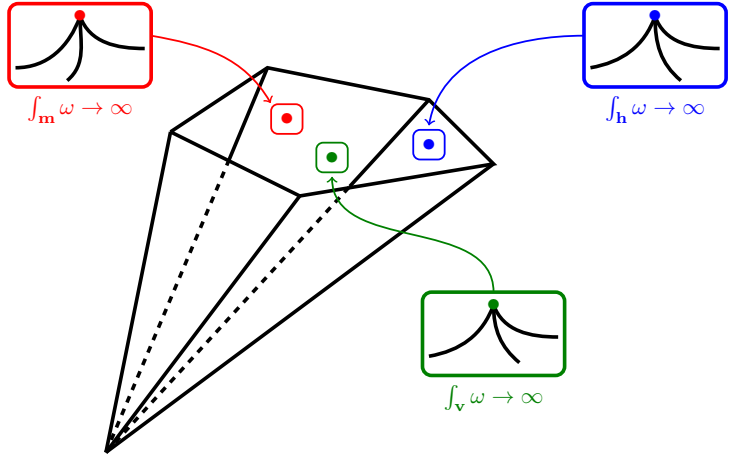
\begin{figure}
\begin{center}
\scalebox{0.85}{\parbox{11.2cm}{\begin{tikzpicture}[scale = 1]
\draw[ultra thick] (0,0) -- (1,5);
\draw[ultra thick] (0,0) -- (3,4);
\draw[ultra thick] (0,0) -- (6,4.5);
\draw[ultra thick,dashed] (0,0) -- (3.775,4.15);
\draw[ultra thick,dashed] (0,0) -- (1.9,4.55);
\draw[ultra thick] (1,5) -- (3,4) -- (6,4.5) -- (5,5.5) -- (2.5,6) -- (1,5);
\draw[ultra thick] (2.5,6) -- (1.9,4.55);
\draw[ultra thick] (5,5.5) -- (3.775,4.15);
%%%%%%%
\node[red, thick, rounded corners=3pt, draw] at (2.8,5.2) {$\bullet$};
\node[blue, thick, rounded corners=3pt, draw] at (5,4.8) {$\bullet$};
\node[green!50!black, thick, rounded corners=3pt, draw] at (3.5,4.6) {$\bullet$};
%%%%%%%
%\draw[thick, red,<-] (2.5,5.4) to [out=120,in=350] (1,6.5) to [out =180,in=300] (0,7);
\draw[thick, red,<-] (2.55,5.45) to [out=120,in=350] (0.7,6.5);
\begin{scope}
\draw[ultra thick,rounded corners,red] (-1.5,5.7) -- (0.7,5.7) -- (0.7,7) -- (-1.5,7) -- cycle;
\draw[ultra thick] (-1.4,6) to [out =0,in=250] (-0.4,6.8) to [out=290,in=180] (0.6,6.3); 
\draw[ultra thick] (-0.6,5.8) to [out=40,in=270] (-0.4,6.8);
\node[red] at (-0.4,6.8) {$\bullet$};
\node[red] at (-0.4,5.35) {{\footnotesize $\int_{\mathbf{m}}\omega\to \infty$}};
\end{scope}
%%%%%%%%%%%%%%%%%%%%
\draw[thick,blue,<-] (5,5.1) to [out=80,in=180] (7.4,6.5);
\begin{scope}[xshift=9cm]
\draw[ultra thick,rounded corners,blue] (-1.6,5.7) -- (0.6,5.7) -- (0.6,7) -- (-1.6,7) -- cycle;
\draw[ultra thick] (-1.5,6) to [out =10,in=250] (-0.5,6.8) to [out=290,in=180] (0.5,6.3); 
\draw[ultra thick] (-0.1,5.8) to [out=140,in=270] (-0.5,6.8);
\node[blue] at (-0.5,6.8) {$\bullet$};
\node[blue] at (-0.5,5.35) {{\footnotesize $\int_{\mathbf{h}}\omega\to \infty$}};
\end{scope} 
%%%%%%%%%%%%%%%%%%%%
%%%%%%%%%%%%%%%%%%%%
\draw[thick,green!50!black,<-] (3.5,4.3) to [out=270,in=90] (6,2.5);
\begin{scope}[xshift=6.5cm,yshift=-4.5cm]
\draw[ultra thick,rounded corners,green!50!black] (-1.6,5.7) -- (0.6,5.7) -- (0.6,7) -- (-1.6,7) -- cycle;
\draw[ultra thick] (-1.5,6) to [out =10,in=250] (-0.5,6.8) to [out=290,in=180] (0.5,6.3); 
\draw[ultra thick] (-0.1,5.9) to [out=140,in=270] (-0.5,6.8);
\node[green!50!black] at (-0.5,6.8) {$\bullet$};
\node[green!50!black] at (-0.5,5.35) {{\footnotesize $\int_{\mathbf{v}}\omega\to \infty$}};
\end{scope} 
\end{tikzpicture}}}
\caption{\sl Different regions in the K\"ahler cone of $X_{N,M}$.}
\label{Fig:RegionKaehlerCone}
\end{center}
\end{figure}

To give a more geometric picture of the situation, we first recall that the K\"ahler moduli space of $X_{N,M}$ takes the form of a cone, as shown in~\figref{Fig:RegionKaehlerCone}. Denoting the K\"ahler form of $X_{N,M}$ by $\omega$, the interior of this cone is parametrised by
\begin{align}
&\int_{X_{N,M}}\omega\wedge \omega\wedge\omega>0\,,&&\text{and} &&\int_{P_a}\omega\wedge\omega>0\,,&&\text{and}&&\int_{\Sigma_i}\omega>0\,.
\end{align} 
where $P_a$ are two-complex-dimensional submanifolds and $\Sigma_i\in\{\mathbf{h},\mathbf{v},\mathbf{m}\}$ are holomorphic curves in $X_{N,M}$.\footnote{By abuse of notation, we use the same name for the curve as well as its area.}
The walls of the cone are given by
\begin{align}
&\int_{X_{N,M}}\omega\wedge \omega\wedge\omega=0\,,&&\text{and} &&\int_{P_a}\omega\wedge\omega=0\,,&&\text{and}&&\int_{\Sigma_i}\omega=0\,.\label{LociKaehlerCone}
\end{align} 
and in particular include loci in which any of the parameters $(\mathbf{h},\mathbf{v},\mathbf{m})$ vanish. In these loci, the Calabi-Yau $X_{N,M}$ develops a singularity. 

The weak coupling regions (from the point of view of the gauge theories engineered by $X_{N,M}$) proposed above are shown as the three coloured areas in \figref{Fig:RegionKaehlerCone}. They  are characterised by (for $i=1,\ldots, MN$)
\begin{align}
&\bullet\text{ red:} &&\int_{m_i}\omega\longrightarrow \infty\,,&&\text{and} &&\int_{h_i,v_i}\omega=\text{finite}\,,\nonumber\\
&\bullet\text{ blue:} &&\int_{h_i}\omega\longrightarrow \infty\,,&&\text{and} &&\int_{m_i,v_i}\omega=\text{finite}\,,\nonumber\\
&\bullet\text{ green:} &&\int_{v_i}\omega\longrightarrow \infty\,,&&\text{and} &&\int_{v_i,m_i}\omega=\text{finite}\,.\label{WeakCouplingRegionsXNM}
\end{align}
respectively. Notice that the co-dimension of these regions are in general not the same. In the following section we will give evidence that such three regions indeed exist in the interior of the K\"ahler cone of $X_{N,M}$ for generic $(N,M)$.

%%%%%%%%%%%%%%%%%%%%%%%%%%%%%
\subsubsection{Type IIB 5-Brane Web Interpretation}\label{Sect:BraneInterpretation}
The triality of gauge theories alluded to in the previous subsubsection can also be argued for from the point of view of the $(p,q)$-brane web, which is dual to $X_{N,M}$. Indeed, the web contains D5-branes, NS5-branes and $(1,1)$ branes, associated with the horizontal, vertical and diagonal line segments in \figref{Fig:WebToric} respectively. The $SL(2,\mathbb{Z})$ symmetry of type IIB string theory acts in the form of rotations on the web and allows us to exchange the roles of these branes respectively. This not only gives rise to a duality between the horizontal and vertical description in the notation above (which has been discussed in the literature at various places~\cite{Haghighat:2013gba,Haghighat:2013tka,Hohenegger:2015btj}), the S-duality also allows us to convert $(1,1)$ branes to D5-branes, thus providing another non-locally dual description (\emph{i.e.} the diagonal expansion in (\ref{TrialitySchemat})). 

With each brane web we can associate the gauge theory that lives on the world-volume of the D5-branes. Since different line segments (from the perspective of the original $(p,q)$-brane web) play the role of the D5-branes in the various S-duality frames, we therefore also expect that the above $SL(2,\mathbb{Z})$ symmetry allows us to relate different gauge theories. As already mentioned above (see section~\ref{Sect:TheoryData}), while specific details of the theories (\emph{e.g.} the precise form of the gauge group) are different, they share certain quantities in common (\emph{e.g.} the rank of the gauge group). 

Specifically, in the $(p,q)$ 5-brane web shown in \figref{Fig:WebToric} with $M$ horizontal D5-branes and $N$ NS5-branes intersecting them vertically, the theory on the former is a $G_{\text{hor}}=[U(M)]^N$ quiver gauge theory of $A_{N-1}$ type, with bi-fundamental matter arising at the intersections. Furthermore, for non-zero separation between the branes, the gauge group is broken to $U(1)^{NM}$. Using the S-duality of the Type IIB string theory, we can map the above $(p,q)$ 5-brane web to the S-dual web in which the identification of D5-branes and NS5-branes get exchanged so that we have $N$ D5-branes in the background of $M$ NS5-branes. This configuration then gives rise to another five-dimensional $G_{\text{vert}}=[U(N)]^M$ quiver gauge theory of $A_{M-1}$ type. As we shall discuss below, a similar duality frame allows us to associate a gauge theory with gauge group $G_{\text{diag}}=[U(NM/k)]^k$ (for $k=\text{gcd}(N,M)$) with the diagonal line segments.

%%%%%%%%%%%%%%%%%%%%%%%%%%%%%%%%%%%%%%%%%%
\section{Weak Coupling Limit}\label{Sect:WeakCouplingLimit}
In the previous section we have seen that a crucial step in associating a quiver gauge theory with a given Calabi-Yau manifold is to identify a region in the K\"ahler moduli space of the latter that can be identified with the weak coupling regime. In this section we will analyse this weak coupling limit in more detail to provide evidence for the triality of gauge theories suggested by (\ref{TrialitySchemat}): we will first consider two examples in detail (namely $(N,M)=(2,2)$ and $(3,2)$) and generalise to generic $(N,M)$ later on.
%%%%%%%
\subsection{Configuration $(N,M)=(2,2)$}\label{Sect:Basis22}
As remarked above, the key aspect of interpreting $Z^{(N,M)}_{\text{hor}}$, $Z^{(N,M)}_{\text{vert}}$ or $Z^{(N,M)}_{\text{diag}}$ in (\ref{TrialitySchemat}) as instanton gauge theory partition functions, is to find a region in the K\"ahler moduli space of $X_{N,M}$ in which either all $\mathbf{h}$ or all $\mathbf{v}$ or all $\mathbf{m}$ become infinitely large, while the remaining parameters remain finite. In order to find such a region in the moduli space, we require a particular basis for the K\"ahler parameters, which provides a solution for the consistency conditions discussed in section~\ref{Sect:KaehlerParameters}. While such a basis is very involved for generic $(N,M)$ (see section~\ref{App:BasisofKaehler} for a proposal in the most general case), we first consider as a simple example the configuration $(N,M)=(2,2)$ (with $k=\text{gcd}(N,M)=2$), to illustrate the point. In this case, three different parametrisations (suitable for the horizontal, vertical and diagonal gauge theory) along with a graphical interpretation of the weak coupling limit are shown in table~\ref{Tab:ChoiceBasis22}.
\begin{table}[htbp]
\begin{center}
\begin{tabular}{c|c|c}
{\bf horizontal} & {\bf vertical} & {\bf diagonal}\\\hline
\scalebox{0.5}{\parbox{10.8cm}{\begin{tikzpicture}[scale = 1.5]
\draw[ultra thick] (-6,0) -- (-5,0);
\draw[ultra thick] (-5,-1) -- (-5,0);
\draw[ultra thick] (-5,0) -- (-4,1);
\draw[ultra thick] (-4,1) -- (-3,1);
\draw[ultra thick] (-4,1) -- (-4,2);
\draw[ultra thick] (-3,1) -- (-3,0);
\draw[ultra thick] (-3,1) -- (-2,2);
\draw[ultra thick] (-4,2) -- (-3,3);
\draw[ultra thick] (-5,2) -- (-4,2);
\draw[ultra thick] (-3,3) -- (-3,4);
\draw[ultra thick] (-3,3) -- (-2,3);
\draw[ultra thick] (-2,2) -- (-2,3);
\draw[ultra thick] (-2,2) -- (-1,2);
\draw[ultra thick] (-2,3) -- (-1,4);
\draw[ultra thick] (-1,4) -- (0,4);
%
%\draw[ultra thick] (0,4) -- (1,5);
\draw[ultra thick] (-1,4) -- (-1,5);
\node at (-6.2,0) {{\small \bf $a$}};
\node at (-5.2,2) {{\small \bf $b$}};
\node at (-0.8,2) {{\small \bf $a$}};
\node at (0.2,4) {{\small \bf $b$}};
\node at (-5,-1.3) {{\small \bf $1$}};
\node at (-3,-0.3) {{\small \bf $2$}};
\node at (-3,4.3) {{\small \bf $1$}};
\node at (-1,5.3) {{\small \bf $2$}};
\node at (-4.2,0.4) {{\small $m_1$}};
\node at (-2.2,1.45) {{\small $m_2$}};
\node at (-3.7,2.6) {{\small $m_3$}};
\node at (-1.2,3.4) {{\small $m_4$}};
\node at (-3.5,1.2) {{\small $h_1$}};
\node at (-1.5,2.2) {{\small $h_2$}};
\node at (-5.5,0.2) {{\small $h_2$}};
\node at (-2.5,3.2) {{\small $h_3$}};
\node at (-0.5,4.2) {{\small $h_4$}};
\node at (-4.5,2.2) {{\small $h_4$}};
\node at (-5.2,-0.8) {{\small $v_1$}};
\node at (-3.2,3.8) {{\small $v_1$}};
\node at (-3.2,0.2) {{\small $v_2$}};
\node at (-1.2,4.8) {{\small $v_2$}};
\node at (-4.2,1.5) {{\small $v_3$}};
\node at (-2.2,2.5) {{\small $v_4$}};
%
%%%%%%%%%%%%%%%%%%%%%%%%%%%%%%%
%basis
\draw[thick,red,<->] (-6.5,0) -- (-6.5,4);
\node[red] at (-6.7,2) {{\small $\tau$}};
\draw[thick,red,<->] (-5,-1.6) -- (-1,-1.6);
\node[red] at (-3,-1.8) {{\small $\rho$}};
\draw[thick,red,<->] (-4.9,0) -- (-3.1,0);
\node[red] at (-4,-0.2) {{\small $\widehat{b}_1$}};
\draw[thick,red,<->] (-5,0.1) -- (-5,1.9);
\node[red] at (-5.2,1) {{\small $\widehat{c}_1$}};
\draw[thick,red,<->] (-3,1.1) -- (-3,2.9);
\node[red] at (-3.2,2) {{\small $\widehat{c}_2$}};
\draw[thick,red,<->] (-0.5,2) -- (-0.5,0);
\node[red] at (-0.3,1) {{\small $E$}};
\end{tikzpicture}}}
%%%%%%%%%%%%%
&
%%%%%%%%%%%%%
%
\scalebox{0.5}{\parbox{10.8cm}{\begin{tikzpicture}[scale = 1.5]
\draw[ultra thick] (-6,0) -- (-5,0);
\draw[ultra thick] (-5,-1) -- (-5,0);
\draw[ultra thick] (-5,0) -- (-4,1);
\draw[ultra thick] (-4,1) -- (-3,1);
\draw[ultra thick] (-4,1) -- (-4,2);
\draw[ultra thick] (-3,1) -- (-3,0);
\draw[ultra thick] (-3,1) -- (-2,2);
\draw[ultra thick] (-4,2) -- (-3,3);
\draw[ultra thick] (-5,2) -- (-4,2);
\draw[ultra thick] (-3,3) -- (-3,4);
\draw[ultra thick] (-3,3) -- (-2,3);
\draw[ultra thick] (-2,2) -- (-2,3);
\draw[ultra thick] (-2,2) -- (-1,2);
\draw[ultra thick] (-2,3) -- (-1,4);
\draw[ultra thick] (-1,4) -- (0,4);
%
%\draw[ultra thick] (0,4) -- (1,5);
\draw[ultra thick] (-1,4) -- (-1,5);
\node at (-6.2,0) {{\small \bf $a$}};
\node at (-5.2,2) {{\small \bf $b$}};
\node at (-0.8,2) {{\small \bf $a$}};
\node at (0.2,4) {{\small \bf $b$}};
\node at (-5,-1.3) {{\small \bf $1$}};
\node at (-3,-0.3) {{\small \bf $2$}};
\node at (-3,4.3) {{\small \bf $1$}};
\node at (-1,5.3) {{\small \bf $2$}};
\node at (-4.2,0.4) {{\small $m_1$}};
\node at (-2.2,1.45) {{\small $m_2$}};
\node at (-3.1,2.45) {{\small $m_3$}};
\node at (-1.2,3.4) {{\small $m_4$}};
\node at (-3.5,1.2) {{\small $h_1$}};
\node at (-1.5,2.2) {{\small $h_2$}};
\node at (-5.5,0.2) {{\small $h_2$}};
\node at (-2.5,3.2) {{\small $h_3$}};
\node at (-0.5,4.2) {{\small $h_4$}};
\node at (-4.5,2.2) {{\small $h_4$}};
\node at (-5.2,-0.8) {{\small $v_1$}};
\node at (-3.2,3.8) {{\small $v_1$}};
\node at (-3.2,0.2) {{\small $v_2$}};
\node at (-1.2,4.8) {{\small $v_2$}};
\node at (-4.2,1.5) {{\small $v_3$}};
\node at (-2.2,2.5) {{\small $v_4$}};
%
%%%%%%%%%%%%%%%%%%%%%%%%%%%%%%%
%basis
\draw[thick,red,<->] (-6.5,0) -- (-6.5,4);
\node[red] at (-6.7,2) {{\small $\tau$}};
\draw[thick,red,<->] (-5,-1.6) -- (-1,-1.6);
\node[red] at (-3,-1.8) {{\small $\rho$}};
\draw[thick,red,<->] (-4.9,0) -- (-3.1,0);
\node[red] at (-4,-0.2) {{\small $\widehat{b}_1$}};
\draw[thick,red,<->] (-3.9,2) -- (-2.1,2);
\node[red] at (-3,1.8) {{\small $\widehat{b}_2$}};
\draw[thick,red,<->] (-5,0.1) -- (-5,1.9);
\node[red] at (-5.2,1) {{\small $\widehat{c}_1$}};
\draw[thick,red,<->] (-5,4.5) -- (-3,4.5);
\node[red] at (-4,4.7) {{\small $D$}};
\end{tikzpicture}}}
%%%%%%%%%
&
%%%%%%%%%
%
\scalebox{0.5}{\parbox{10.1cm}{\begin{tikzpicture}[scale = 1.5]
\draw[ultra thick] (-6,0) -- (-5,0);
\draw[ultra thick] (-5,-1) -- (-5,0);
\draw[ultra thick] (-5,0) -- (-4,1);
\draw[ultra thick] (-4,1) -- (-3,1);
\draw[ultra thick] (-4,1) -- (-4,2);
\draw[ultra thick] (-3,1) -- (-3,0);
\draw[ultra thick] (-3,1) -- (-2,2);
\draw[ultra thick] (-4,2) -- (-3,3);
\draw[ultra thick] (-5,2) -- (-4,2);
\draw[ultra thick] (-3,3) -- (-3,4);
\draw[ultra thick] (-3,3) -- (-2,3);
\draw[ultra thick] (-2,2) -- (-2,3);
\draw[ultra thick] (-2,2) -- (-1,2);
\draw[ultra thick] (-2,3) -- (-1,4);
\draw[ultra thick] (-1,4) -- (0,4);
%
%\draw[ultra thick] (0,4) -- (1,5);
\draw[ultra thick] (-1,4) -- (-1,5);
\node at (-6.2,0) {{\small \bf $a$}};
\node at (-5.2,2) {{\small \bf $b$}};
\node at (-0.8,2) {{\small \bf $a$}};
\node at (0.2,4) {{\small \bf $b$}};
\node at (-5,-1.3) {{\small \bf $1$}};
\node at (-3,-0.3) {{\small \bf $2$}};
\node at (-3,4.3) {{\small \bf $1$}};
\node at (-1,5.3) {{\small \bf $2$}};
\node at (-4.45,0.2) {{\small $m_1$}};
\node at (-2.2,1.45) {{\small $m_2$}};
\node at (-3.7,2.7) {{\small $m_3$}};
\node at (-1.2,3.4) {{\small $m_4$}};
\node at (-3.5,1.2) {{\small $h_1$}};
\node at (-1.5,2.2) {{\small $h_2$}};
\node at (-5.5,0.2) {{\small $h_2$}};
\node at (-2.5,3.2) {{\small $h_3$}};
\node at (-0.5,4.2) {{\small $h_4$}};
\node at (-4.5,2.2) {{\small $h_4$}};
\node at (-5.2,-0.4) {{\small $v_1$}};
\node at (-2.8,3.8) {{\small $v_1$}};
\node at (-3.2,0.2) {{\small $v_2$}};
\node at (-1.2,4.8) {{\small $v_2$}};
\node at (-4.2,1.5) {{\small $v_3$}};
\node at (-2.2,2.5) {{\small $v_4$}};
%
%%%%%%%%%%%%%%%%%%%%%%%%
%basis
\draw[thick,<->,red] (-3.3,-0.2) -- (-4.2,0.7);
\node[red] at (-3.7,0) {{\small $\widehat{a}_{2}$}};
\draw[thick,<->,red] (-2.3,1.8) -- (-3.2,2.7);
\node[red] at (-2.9,2.2) {{\small $\widehat{a}_{1}$}};
\draw[dashed] (-3,4) -- (-5.5,1.5);
\draw[red,<->] (-4.45,0.65) -- (-5.35,1.55);
\node[red] at (-5.6,1) {{\small $F=v_1+v_3$}};
\draw[dashed] (-6,0) -- (-6.5,-0.5);
\draw[dashed] (-3,1) -- (-5.5,-1.5);
\draw[red,<->] (-6.45,-0.55) -- (-5.55,-1.45);
\node[red] at (-4.6,-1) {{\small $M=h_2+v_1+h_3+v_4$}};
%%%%%%%%%%%%%%%%%%%
\node[red] at (-1,0) {{\small $\begin{array}{l}V_1=m_1+h_1+h_2\,,\\V_2=m_3+h_3+h_4\,.\end{array}$}};
\end{tikzpicture}}}\\
%%%%%%%%%%%%%%%%%%%%%%%%%%%%%%
%%%%%%%%%%%%%%%%%%%%%%%%%%%%%%
&&\\
%%%%%%%%%%%%%%%%%%%%%%%%%%%%%%
%%%%%%%%%%%%%%%%%%%%%%%%%%%%%%
\parbox{3.8cm}{\begin{align}
&h_1=h_3=\widehat{b}_1-\tfrac{\widehat{c}_1-\widehat{c}_2+E}{2}\,,\nonumber\\
&h_2=h_4=\rho-\widehat{b}_1+\tfrac{\widehat{c}_1-\widehat{c}_2-E}{2},\nonumber\\
&v_1=v_2=\tau-\tfrac{\widehat{c}_1+\widehat{c}_2+E}{2}\,,\nonumber\\
&v_3=v_4=\tfrac{\widehat{c}_1+\widehat{c}_2-E}{2}\,,\nonumber\\
&m_1=m_4=\tfrac{\widehat{c}_1-\widehat{c}_2+E}{2}\,,\nonumber\\
&m_2=m_3=-\tfrac{\widehat{c}_1-\widehat{c}_2-E}{2}\,,\nonumber
\end{align}}
&
\parbox{3.8cm}{\begin{align}
&h_1=h_3=\tfrac{\widehat{b}_1+\widehat{b}_2-D}{2}\,,\nonumber\\
&h_2=h_4=\rho-\tfrac{\widehat{b}_1+\widehat{b}_2+D}{2}\,,\nonumber\\
&v_1=v_2=\tau-\widehat{c}_1+\tfrac{\widehat{b}_1-\widehat{b}_2-D}{2},\nonumber\\
&v_3=v_4=\widehat{c}_1-\tfrac{\widehat{b}_1-\widehat{b}_2+D}{2}\,,\nonumber\\
&m_1=m_4=\tfrac{\widehat{b}_1-\widehat{b}_2+D}{2}\,,\nonumber\\
&m_2=m_3=-\tfrac{\widehat{b}_1-\widehat{b}_2-D}{2}\,,\nonumber
\end{align}}
&
\parbox{4cm}{\begin{align}
&h_1=h_3=\tfrac{\widehat{a}_1+\widehat{a}_2-F}{2}\,,\nonumber\\
&h_2=h_4=M-\tfrac{\widehat{a}_1+\widehat{a}_2+F}{2}\,,\nonumber\\
&v_1=v_2=-\tfrac{\widehat{a}_1-\widehat{a}_2-F}{2}\,,\nonumber\\
&v_3=v_4=\tfrac{\widehat{a}_1-\widehat{a}_2+F}{2}\,,\nonumber\\
&m_1=m_4=V_1+F-M\,,\nonumber\\
&m_2=m_3=V_2+F-M\,,\nonumber
\end{align}}
\\
%%%%%%%%%%%%%%%%%%%%%%%%%%%%%%
%%%%%%%%%%%%%%%%%%%%%%%%%%%%%%
&&\\
%%%%%%%%%%%%%%%%%%%%%%%%%%%%%%
%%%%%%%%%%%%%%%%%%%%%%%%%%%%%%
\parbox{2.6cm}{\begin{tikzpicture}
\draw[thick,->] (0,0) -- (0,-2);
\node at (1.3,-1) {$\begin{array}{r}\rho-\widehat{b}_1\longrightarrow \infty \\ \widehat{b}_1\longrightarrow \infty\end{array}$};
\end{tikzpicture} }
&
\parbox{2.6cm}{\begin{tikzpicture}
\draw[thick,->] (0,0) -- (0,-2);
\node at (1.3,-1) {$\begin{array}{r}\tau-\widehat{c}_1\longrightarrow \infty \\ \widehat{c}_1\longrightarrow \infty\end{array}$};
\end{tikzpicture}}
&
\parbox{2.5cm}{\begin{tikzpicture}
\draw[thick,->] (0,0) -- (0,-2);
\node at (1.3,-1) {$V_{1,2}\longrightarrow \infty$};
\end{tikzpicture}}\\
%%%%%%%%%%%%%%%%%%%%%%%%%%%%%%
%%%%%%%%%%%%%%%%%%%%%%%%%%%%%%
&&\\
%%%%%%%%%%%%%%%%%%%%%%%%%%%%%%
%%%%%%%%%%%%%%%%%%%%%%%%%%%%%%
\scalebox{0.51}{\parbox{11cm}{\begin{tikzpicture}[scale = 1.5]
\draw[ultra thick,dashed] (-5.5,0) -- (-5,0);
\draw[ultra thick] (-5,-1) -- (-5,0);
\draw[ultra thick] (-5,0) -- (-4,1);
\draw[ultra thick,dashed] (-4,1) -- (-3.5,1);
\draw[ultra thick] (-4,1) -- (-4,2);
\draw[ultra thick] (-4,2) -- (-3,3);
\draw[ultra thick,dashed] (-4.5,2) -- (-4,2);
\draw[ultra thick] (-3,3) -- (-3,4);
\draw[ultra thick,dashed] (-3,3) -- (-2.5,3);
%%%%
\draw[ultra thick,xshift=1.5cm,yshift=-1cm] (-3,1) -- (-3,0);
\draw[ultra thick, dashed] (-2,0) -- (-1.5,0);
\draw[ultra thick,xshift=1.5cm,yshift=-1cm] (-3,1) -- (-2,2);
\draw[ultra thick,xshift=1.5cm,yshift=-1cm] (-2,2) -- (-2,3);
\draw[ultra thick,dashed,xshift=1.5cm,yshift=-1cm] (-2,2) -- (-1.5,2);
\draw[ultra thick,xshift=1.5cm,yshift=-1cm] (-2,3) -- (-1,4);
\draw[ultra thick, dashed] (-1,2) -- (-0.5,2);
\draw[ultra thick,dashed,xshift=1.5cm,yshift=-1cm] (-1,4) -- (-0.5,4);
\draw[ultra thick,xshift=1.5cm,yshift=-1cm] (-1,4) -- (-1,5);
\node at (-5.7,0) {{\small \bf $a$}};
\node at (-4.7,2) {{\small \bf $b$}};
\node at (-3.3,1) {{\small \bf $c$}};
\node at (-2.3,3) {{\small \bf $d$}};
\node at (0.2,1) {{\small \bf $a$}};
\node at (1.2,3) {{\small \bf $b$}};
\node at (-2.2,0) {{\small \bf $c$}};
\node at (-1.2,2) {{\small \bf $d$}};
\node at (-5,-1.3) {{\small \bf $1$}};
\node at (-1.5,-1.3) {{\small \bf $2$}};
\node at (-3,4.3) {{\small \bf $1$}};
\node at (0.5,4.3) {{\small \bf $2$}};
\node at (-4.2,0.4) {{\small $m_1$}};
\node[xshift=2.2cm,yshift=-1.5cm] at (-2.2,1.45) {{\small $m_2$}};
\node at (-3.1,2.45) {{\small $m_2$}};
\node[xshift=2.2cm,yshift=-1.5cm] at (-1.2,3.4) {{\small $m_1$}};
%
%\node at (-3.5,1.2) {{\small $h_1$}};
%\node at (-1.5,2.2) {{\small $h_2$}};
%\node at (-5.5,0.2) {{\small $h_2$}};
%\node at (-2.5,3.2) {{\small $h_1$}};
%\node at (-0.5,4.2) {{\small $h_2$}};
%\node at (-4.5,2.2) {{\small $h_2$}};
%
\node at (-5.2,-0.8) {{\small $v_1$}};
\node at (-3.2,3.8) {{\small $v_1$}};
\node[xshift=2.2cm,yshift=-1.5cm] at (-3.2,0.2) {{\small $v_1$}};
\node[xshift=2.2cm,yshift=-1.5cm] at (-1.2,4.8) {{\small $v_1$}};
\node at (-4.2,1.5) {{\small $v_3$}};
\node[xshift=2.2cm,yshift=-1.5cm] at (-2.2,2.5) {{\small $v_3$}};
\end{tikzpicture}}}
%%%%%%%%%%%%%
&
%%%%%%%%%%%%%
%
\scalebox{0.51}{\parbox{8.5cm}{\begin{tikzpicture}[scale = 1.5]
\draw[ultra thick] (-6,0) -- (-5,0);
\draw[ultra thick,dashed] (-5,-0.5) -- (-5,0);
\draw[ultra thick] (-5,0) -- (-4,1);
\draw[ultra thick] (-4,1) -- (-3,1);
\draw[ultra thick,dashed] (-4,1) -- (-4,1.5);
\draw[ultra thick,dashed] (-2,2) -- (-2,2.5);
\draw[ultra thick,dashed] (-3,1) -- (-3,0.5);
\draw[ultra thick] (-3,1) -- (-2,2);
\draw[ultra thick] (-2,2) -- (-1,2);
\draw[ultra thick,xshift=-1cm,yshift=1cm] (-5,2) -- (-4,2);
\draw[ultra thick,xshift=-1cm,yshift=1cm] (-3,3) -- (-2,3);
\draw[ultra thick,xshift=-1cm,yshift=1cm] (-2,3) -- (-1,4);
\draw[ultra thick,xshift=-1cm,yshift=1cm] (-1,4) -- (0,4);
\draw[ultra thick,dashed,xshift=-1cm,yshift=1cm] (-1,4) -- (-1,4.5);
\draw[ultra thick,dashed,xshift=-1cm,yshift=1cm] (-3,3) -- (-3,3.5);
\draw[ultra thick,dashed,xshift=-1cm,yshift=1cm] (-2,2.5) -- (-2,3);
\draw[ultra thick,dashed,xshift=-1cm,yshift=1cm] (-4,1.5) -- (-4,2);
\draw[ultra thick,xshift=-1cm,yshift=1cm] (-4,2) -- (-3,3);
\node at (-6.2,0) {{\small \bf $a$}};
\node at (-6.2,3) {{\small \bf $b$}};
\node at (-0.8,2) {{\small \bf $a$}};
\node at (-0.8,5) {{\small \bf $b$}};
\node at (-5,-0.8) {{\small \bf $1$}};
\node at (-3,0.2) {{\small \bf $2$}};
\node at (-4,1.7) {{\small \bf $3$}};
\node at (-2,2.7) {{\small \bf $4$}};
\node at (-4,4.7) {{\small \bf $1$}};
\node at (-2,5.7) {{\small \bf $2$}};
\node at (-5,2.2) {{\small \bf $3$}};
\node at (-3,3.2) {{\small \bf $4$}};
\node at (-4.2,0.4) {{\small $m_1$}};
\node at (-2.2,1.45) {{\small $m_2$}};
\node[xshift=-1.8cm,yshift=1.2cm] at (-3.1,2.45) {{\small $m_2$}};
\node[xshift=-1.8cm,yshift=1.2cm] at (-1.2,3.4) {{\small $m_1$}};
\node at (-3.5,1.2) {{\small $h_1$}};
\node at (-1.5,2.2) {{\small $h_2$}};
\node at (-5.5,0.2) {{\small $h_2$}};
\node[xshift=-1.5cm,yshift=0.9cm] at (-2.5,3.2) {{\small $h_1$}};
\node[xshift=-1.5cm,yshift=0.9cm] at (-0.5,4.2) {{\small $h_2$}};
\node[xshift=-1.5cm,yshift=0.9cm] at (-4.5,2.2) {{\small $h_2$}};
\end{tikzpicture}}}
%%%%%%%%%%%%%%%%%
&
%%%%%%%%%%%%%%%%%
\scalebox{0.51}{\parbox{10cm}{\begin{tikzpicture}[scale = 1.5]
\draw[ultra thick] (-6,0) -- (-5,0) -- (-5,-1) -- (-4,-1) -- (-4,-2) -- (-3,-2);
\draw[ultra thick, dashed] (-5,0) -- (-4.5,0.5);
\draw[ultra thick, dashed] (-4,-1) -- (-3.5,-0.5);
\draw[ultra thick, dashed] (-5,-1) -- (-5.5,-1.5);
\draw[ultra thick, dashed] (-4,-2) -- (-4.5,-2.5);
\draw[ultra thick,xshift=1cm,yshift=1cm] (-4,2) -- (-3,2) -- (-3,1) -- (-2,1) -- (-2,0) -- (-1,0);
\draw[ultra thick, dashed,xshift=1cm,yshift=1cm] (-3,2) -- (-2.5,2.5);
\draw[ultra thick, dashed,xshift=1cm,yshift=1cm] (-2,1) -- (-1.5,1.5);
\draw[ultra thick, dashed,xshift=1cm,yshift=1cm] (-3,1) -- (-3.5,0.5);
\draw[ultra thick, dashed,xshift=1cm,yshift=1cm] (-2,0) -- (-2.5,-0.5);
\node at (-5.6,-1.6) {{\footnotesize \bf I}};
\node at (-4.6,-2.6) {{\footnotesize \bf II}};
\node at (-4.4,0.6) {{\footnotesize \bf III}};
\node at (-3.4,-0.4) {{\footnotesize \bf IV}};
\node at (-1.4,3.6) {{\footnotesize \bf I}};
\node at (-0.4,2.6) {{\footnotesize \bf II}};
\node at (-2.6,1.4) {{\footnotesize \bf III}};
\node at (-1.6,0.4) {{\footnotesize \bf IV}};
\node at (-6.2,0) {{\small \bf $a$}};
\node at (-3.2,3) {{\small \bf $b$}};
\node at (-2.8,-2) {{\small \bf $a$}};
\node at (0.2,1) {{\small \bf $b$}};
\node at (-5.5,0.2) {{\small $h_2$}};
\node at (-4.5,-0.8) {{\small $h_1$}};
\node at (-3.5,-1.8) {{\small $h_2$}};
\node at (-5.2,-0.5) {{\small $v_1$}};
\node at (-4.2,-1.5) {{\small $v_3$}};
\node at (-2.5,3.2) {{\small $h_2$}};
\node at (-1.5,2.2) {{\small $h_1$}};
\node at (-0.5,1.2) {{\small $h_2$}};
\node at (-2.2,2.5) {{\small $v_3$}};
\node at (-1.2,1.5) {{\small $v_1$}};

\end{tikzpicture}}}
\end{tabular}
\end{center}
\caption{\sl Three different choices of maximal sets of independent K\"ahler parameters for the configuration $(N,M)=(2,2)$. In each case, the 12 lines $(h_{1,\ldots,4},v_{1,\ldots,4},m_{1,\ldots 4})$ are parametrised by 6 independent variables. The last row shows the weak coupling limit, which is obtained by sending two of the parameters (related to the coupling constants of the respective gauge theories) to infinity.}
\label{Tab:ChoiceBasis22}
\end{table}
The three different expansions (and in particular the weak coupling limit) can be interpreted as follows:
\begin{itemize}
\item \underline{horizontal expansion in the basis $(\rho,\widehat{b}_1;\widehat{c}_1,\widehat{c}_2, \tau; E)$}\\
Upon taking the limit
\begin{align}
&\rho-\widehat{b}_1\longrightarrow \infty\,,&& \text{and} &&\widehat{b}_1\longrightarrow \infty\,,
\end{align}
all horizontal lines of the toric diagram are effectively cut, since $h_{1,\ldots,4}\longrightarrow \infty$, while $v_{1,\ldots,4}$ and $m_{1,\ldots,4}$ remain finite.\footnote{Notice in particular that there exists a region in the parameter space of $(\tau,\widehat{c}_1,\widehat{c}_2,E)$ in which $(v_1,v_2,v_3,v_4,m_1,m_2,m_3,m_4)$ are positive, which is important for their interpretation from the point of view of gauge theory as Coulomb branch parameters and hypermultiplet masses respectively.} The remaining diagram takes the form of two vertical strips\footnote{While we call them strips, we point out that they are still defined on a cylinder, \emph{i.e.} their ends are being identified (see also \figref{Fig:StripVertical}).}, thus implying that the gauge group associated with the horizontal expansion is
\begin{align}
G_{\text{hor}}=U(2)\times U(2)\,.\label{Horizontal22GaugeGroup}
\end{align}
Indeed, $\rho-\widehat{b}_1$ and $\widehat{b}_1$ are related to the gauge couplings associated with each of the two $U(2)$ factors, while the parameters $\widehat{c}_{1,2}$ can be interpreted as the (simple, positive) roots of each of two $\mathfrak{a}_1$ related to the two $U(2)$ factors. Furthermore, $\tau$ can be interpreted as an additional root, that extends each of these algebras to affine $\widehat{\mathfrak{a}}_1$, while $E$ is a parameter associated with the compactification of the toric web on a torus.
\item \underline{vertical expansion in the basis $(\tau,\widehat{c}_1;\widehat{b}_1,\widehat{b}_2;\rho,D)$}\\
In the limit
\begin{align}
&\tau-\widehat{c}_1\longrightarrow \infty\,,&& \text{and} &&\widehat{c}_1\longrightarrow \infty\,,
\end{align}
all vertical lines of the toric diagram are cut, since $v_{1,\ldots,4}\longrightarrow \infty$, while  $h_{1,\ldots,4}$ and $m_{1,\ldots,4}$ remain finite (and positive for certain values of $(\rho,\widehat{b}_1,\widehat{b}_2,D)$). In this way, the diagram decomposes into two horizontal strips, which implies that the gauge group associated with the vertical expansion is in fact
\begin{align}
G_{\text{vert}}=U(2)\times U(2)\,.
\end{align}
This group is in fact the electro-magnetic S-dual (\emph{i.e.} the Langlands dual) of the horizontal gauge group (\ref{Horizontal22GaugeGroup}). Indeed, $\tau-\widehat{c}_1$ and $\widehat{c}_1$ can be related to the gauge couplings associated with each of the two $U(2)$ factors, and the parameters $\widehat{b}_{1,2}$ can be interpreted as the (simple, positive) roots of each of two $\mathfrak{a}_1$ related to the two $U(2)$ factors. The parameter, $\tau$ can be interpreted as an additional root, that extends each of these algebras to affine $\widehat{\mathfrak{a}}_1$, while $D$ is a parameter associated with the compactification of the toric web on a torus.

\item \underline{diagonal expansion in the basis $(V_1,V_2;\widehat{a}_1,\widehat{a}_2;F,L)$}\\
In the limit $V_{1,2}\longrightarrow\infty$, all diagonal lines (along direction $(1,1)$) of the toric diagram are cut, since $m_{1,\ldots,4}\longrightarrow \infty$, while  $h_{1,\ldots,4}$ and $v_{1,\ldots,4}$ remain finite (and positive for certain values ). In this way, the diagram decomposes into two diagonal strips (which were called 'staircase strips' in \cite{Bastian:2017ing}), which implies that the gauge group associated with the vertical expansion is as well
\begin{align}
G_{\text{diag}}=U(2)\times U(2)\,.
\end{align}
Indeed, $V_{1,2}$ can be related to the gauge couplings associated with each of the two $U(2)$ factors, and the parameters $\widehat{a}_{1,2}$ can be interpreted as the (simple, positive) roots of each of two $\mathfrak{a}_1$ related to the two $U(2)$ factors. The parameter, $L$ can be interpreted as an additional root, that extends each of these algebras to affine $\widehat{\mathfrak{a}}_1$, while $F$ is a parameter associated with the compactification of the toric web on a torus.
\end{itemize}

%%%%%%%%%%%%%%%%%%%
\noindent
It is important to notice that in all three cases, the connection to a certain gauge theory relies on the fact that in the weak coupling limit the web diagram decomposes into a number of \emph{parallel} strips (either horizontally, vertically or diagonally): physically, the latter can be interpreted as parallel $NS5$ branes with semi-infinite D5-branes ending on either side in equal numbers \cite{Witten:1997sc}. When the strips are glued together the world-volume theory on these D5-branes is the corresponding gauge theory, as explained above in section~\ref{Sect:BraneInterpretation}.

In the current case, since the orientation of the strips can be changed through an $SL(2,\mathbb{Z})$ transformation, the diagrams in the last line of table~\ref{Tab:ChoiceBasis22} are identical up to a relabelling of the parameters. This indicates that the gauge theories engineered from the three expansions in (\ref{TrialitySchemat}) have the same gauge group, \emph{i.e.}
\begin{align}
G_{\text{hor}}=G_{\text{vert}}=G_{\text{diag}}=U(2)\times U(2)\,.
\end{align}
This is a peculiarity of the configuration $(N,M)=(2,2)$ as in general the three gauge groups are different (albeit that their rank is the same as argued above), as we shall see from the next example $(N,M)=(3,2)$.
%%%%%%%%%%%%%%%%%%%%%
\subsection{Configuration $(N,M)=(3,2)$}\label{Sect:Basis32}
The next non-trivial configuration corresponds to $(N,M)=(3,2)$ with $k=\text{gcd}(3,2)=1$. The corresponding web diagram contains $18$ lines which are the K\"ahler parameters of various rational curves in the Calabi-Yau threefold $X_{3,2}$:
\begin{align}
&{\bf h} = (h_1, \cdots, h_6) \,, &&{\bf v} = (v_1,\cdots, v_6)\,,
&&{\bf m} = (m_1, \cdots, m_6 )\, .
\end{align}
As discussed before, these parameters are not linearly independent but they can be parametrised by choosing $8$ independent variables (for more details see also \cite{Bastian:2017ing}). Three different such parametrisations are shown in table~\ref{Tab:Bases32}, leading to the following expansions:
\begin{table}[htbp]
\begin{center}
\begin{tabular}{c|c|c}
\multicolumn{3}{c}{
%%%%%%%%%%%%%%%%%%%%
\scalebox{0.7}{\parbox{15.5cm}{\begin{tikzpicture}[scale = 1.50]
\draw[ultra thick] (-6,0) -- (-5,0);
\draw[ultra thick] (-5,-1) -- (-5,0);
\draw[ultra thick] (-5,0) -- (-4,1);
\draw[ultra thick] (-4,1) -- (-3,1);
\draw[ultra thick] (-4,1) -- (-4,2);
\draw[ultra thick] (-3,1) -- (-3,0);
\draw[ultra thick] (-3,1) -- (-2,2);
\draw[ultra thick] (-4,2) -- (-3,3);
\draw[ultra thick] (-5,2) -- (-4,2);
\draw[ultra thick] (-3,3) -- (-3,4);
\draw[ultra thick] (-3,3) -- (-2,3);
\draw[ultra thick] (-2,2) -- (-2,3);
\draw[ultra thick] (-2,2) -- (-1,2);
\draw[ultra thick] (-2,3) -- (-1,4);
\draw[ultra thick] (-1,2) -- (0,3);
\draw[ultra thick] (-1,2) -- (-1,1);
\draw[ultra thick] (-1,4) -- (0,4);
\draw[ultra thick] (0,3) -- (1,3);
\draw[ultra thick] (0,3) -- (0,4);
\draw[ultra thick] (0,4) -- (1,5);
\draw[ultra thick] (1,5) -- (2,5);
\draw[ultra thick] (1,5) -- (1,6);
\draw[ultra thick] (-1,4) -- (-1,5);
\node[rotate=90] at (-5.5,0) {$=$};
\node at (-5.4,-0.15) {{\tiny$1$}};
\node[rotate=90] at (-4.5,2) {$=$};
\node at (-4.4,1.85) {{\tiny$2$}};
\node[rotate=90] at (0.6,3) {$=$};
\node at (0.7,2.8) {{\tiny$1$}};
\node[rotate=90] at (1.6,5) {$=$};
\node at (1.7,4.8) {{\tiny$2$}};
\node at (-5,-0.5) {$-$};
\node at (-4.85,-0.55) {{\tiny $1$}};
\node at (-3,0.5) {$-$};
\node at (-2.85,0.45) {{\tiny $2$}};
\node at (-1,1.5) {$-$};
\node at (-0.85,1.45) {{\tiny $3$}};
\node at (-3,3.7) {$-$};
\node at (-2.85,3.6) {{\tiny $1$}};
\node at (-1,4.7) {$-$};
\node at (-0.85,4.6) {{\tiny $2$}};
\node at (1,5.7) {$-$};
\node at (1.15,5.6) {{\tiny $3$}};
\node at (-4.3,0.3) {{\small $m_1$}};
\node at (-2.4,1.2) {{\small $m_2$}};
\node at (-0.3,2.3) {{\small $m_3$}};
\node at (-3.4,2.2) {{\small $m_4$}};
\node at (-1.4,3.2) {{\small $m_5$}};
\node at (0.7,4.3) {{\small $m_6$}};
\node at (-3.5,1.2) {{\small $h_1$}};
\node at (-1.5,2.2) {{\small $h_2$}};
\node at (0.5,3.3) {{\small $h_3$}};
\node at (-2.5,3.2) {{\small $h_4$}};
\node at (-0.5,4.2) {{\small $h_5$}};
\node at (1.5,5.3) {{\small $h_6$}};
\node at (-5.3,-0.5) {{\small $v_1$}};
\node at (-3.3,0.5) {{\small $v_2$}};
\node at (-0.65,1.6) {{\small $v_3$}};
\node at (-3.8,1.5) {{\small $v_4$}};
\node at (-1.8,2.5) {{\small $v_5$}};
\node at (0.2,3.5) {{\small $v_6$}};
%roots diagonal
\draw[thick,<->,red] (-3.3,2.8) -- (-4.2,3.7);
\node[red] at (-3.6,2.9) {{\small $\widehat{a}_1$}};
\draw[thick,<->,red] (-2.3,1.8) -- (-3.2,2.7);
\node[red] at (-2.6,2.4) {{\small $\widehat{a}_2$}};
\draw[thick,<->,red] (-1.3,0.8) -- (-2.2,1.7);
\node[red] at (-1.8,1.6) {{\small $\widehat{a}_3$}};
\draw[thick,<->,red] (1.7,3.8) -- (0.8,4.7);
\node[red] at (1.4,4.4) {{\small $\widehat{a}_4$}};
\draw[thick,<->,red] (-3.3,-0.2) -- (-4.2,0.7);
\node[red] at (-3.6,0.4) {{\small $\widehat{a}_5$}};
%roots vertical
\draw[green!50!black,<->] (-5,0.1) -- (-5,1.9);
\node[green!50!black,<->] at (-4.8,1.6) {$\widehat{c}_1$};
\draw[green!50!black,<->] (-3,1.1) -- (-3,2.9);
\node[green!50!black,<->] at (-3.2,1.8) {$\widehat{c}_2$};
\draw[green!50!black,<->] (-1,2.1) -- (-1,3.9);
\node[green!50!black,<->] at (-1.2,2.7) {$\widehat{c}_3$};
%roots horizontal
\draw[blue,<->] (-4.9,0) -- (-3.1,0);
\node[blue,<->] at (-4,-0.25) {$\widehat{b}_1$};
\draw[blue,<->] (-2.9,1) -- (-1.1,1);
\node[blue,<->] at (-2,0.75) {$\widehat{b}_2$};
\draw[blue,<->] (-3.9,2) -- (-2.1,2);
\node[blue,<->] at (-2.65,1.75) {$\widehat{b}_3$};
\draw[blue,<->] (-1.9,3) -- (-0.1,3);
\node[blue,<->] at (-0.65,3.25) {$\widehat{b}_4$};
\draw[dashed] (0,3) -- (-5,-2);
\draw[dashed] (-6,0) -- (-6.5,-0.5);
\draw[thick,<->] (-6.45,-0.55) -- (-5.05,-1.95);
\node at (-5.9,-1.5) {{\small $M$}};
\draw[thick,<->] (-4.9,6.25) -- (0.9,6.25);
\node at (-2,6.05) {{\small $\rho$}};
\draw[dashed] (-5,6.25) -- (-5,2);
\draw[dashed] (1,6.25) -- (1,6);
\draw[thick,<->] (-4.8,0.3) -- (-5.7,1.2);
\draw[dashed] (-3,4) -- (-5.75,1.25);
\node at (-4.8,1.1) {{\small $F=v_1+v_4$}};
\draw[thick,<->] (-5,-1.4) -- (-3,-1.4);
\node at (-4,-1.6) {{\small $D=m_1+m_4$}};
\draw[thick,<->] (1.5,3) -- (1.5,0);
\node at (2.7,1.5) {{\small $E=m_1+m_2+m_3$}};
\node at (0.7,-1) {{\small $V=m_1+(3-1)(h_1+h_3)+(3-2)(v_2+h_5+v_6+h_5)$}};
\draw[thick,<->] (-6.5,0.1) -- (-6.5,3.9);
\node at (-6.3,2) {{\small $\tau$}};
\draw[dashed] (-6.5,0) -- (-6,0);
\draw[dashed] (-6.5,4) -- (-3,4);
%%%%%%%%%%%%%%%%%%%%%%%%
 \draw [line width=1mm,->] (-4,-2.5) -- (-6,-4.5);
 \node at (-6,-5) {$\phantom{x}$};
  \draw [line width=1mm,->] (-1,-2.5) -- (-1,-4.5);
 \node at (1,-5) {$\phantom{x}$};
  \draw [line width=1mm,->] (2,-2.5) -- (4,-4.5);
 \node at (4,-5) {$\phantom{x}$};
\end{tikzpicture}}}
}\\
%%%%%%%%%%%%%%%%%%%%
%%%%%%%%%%%%%%%%%%%%%%%%%%%%%%%%%%%%%%%%%%%%%
%%%%%%%%%%%%%%%%%%%%%%%%%%%%%%%%%%%%%%%%%%%%%
{\bf horizontal }&{\bf vertical}&{\bf diagonal}\\
&&\\
$(\rho,\widehat{b}_1,\widehat{b}_2;\widehat{c}_1,\widehat{c}_2,\widehat{c}_3;\tau,E)$ & $(\tau,\widehat{c}_1;\widehat{b}_1,\widehat{b}_2,\widehat{b}_3,\widehat{b}_4;\rho,D)$& $(V;\widehat{a}_1,\widehat{a}_2,\widehat{a}_3,\widehat{a}_4,\widehat{a}_5;M,F)$\\
&&\\
\parbox{2.6cm}{\begin{tikzpicture}
\draw[line width=0.4mm,->] (0,0) -- (0,-2);
\node at (1.75,-1) {{\small $\begin{array}{r}\rho-\widehat{b}_1-\widehat{b}_2\longrightarrow \infty \\ \widehat{b}_1\longrightarrow \infty\\ \widehat{b}_2\longrightarrow \infty\end{array}$}};
\end{tikzpicture} }
&
\parbox{2.6cm}{\begin{tikzpicture}
\draw[line width=0.4mm,->] (0,0) -- (0,-2);
\node at (1.3,-1) {{\small $\begin{array}{r}\tau-\widehat{c}_1\longrightarrow \infty \\ \widehat{c}_1\longrightarrow \infty\end{array}$}};
\end{tikzpicture}}
&
\parbox{2.5cm}{\begin{tikzpicture}
\draw[line width=0.4mm,->] (0,0) -- (0,-2);
\node at (0.9,-1) {{\small $V\longrightarrow \infty$}};
\end{tikzpicture}}\\
&&\\
%%%%%%%%%%%%%%%%%%%%%%%%%%%%%%%%%%%%%%%%%%%%%
%%%%%%%%%%%%%%%%%%%%%%%%%%%%%%%%%%%%%%%%%%%%%
%%%%%%%%%%%%%%%%%%%%%%%%%%%%%%
\scalebox{0.43}{\parbox{14.5cm}{\begin{tikzpicture}[scale = 1.5]
\draw[ultra thick,dashed] (-5.5,0) -- (-5,0);
\draw[ultra thick] (-5,-1) -- (-5,0);
\draw[ultra thick] (-5,0) -- (-4,1);
\draw[ultra thick,dashed] (-4,1) -- (-3.5,1);
\draw[ultra thick] (-4,1) -- (-4,2);
\draw[ultra thick] (-4,2) -- (-3,3);
\draw[ultra thick,dashed] (-4.5,2) -- (-4,2);
\draw[ultra thick] (-3,3) -- (-3,4);
\draw[ultra thick,dashed] (-3,3) -- (-2.5,3);
\node at (-5.7,0) {{\small \bf $a$}};
\node at (-4.7,2) {{\small \bf $b$}};
\node at (-3.3,1) {{\small \bf $c$}};
\node at (-2.3,3) {{\small \bf $d$}};
\node at (-5,-1.3) {{\small \bf $1$}};
\node at (-3,4.3) {{\small \bf $1$}};
\node at (-4.2,0.4) {{\small $m_1$}};
\node at (-3.1,2.45) {{\small $m_4$}};
\node at (-5.2,-0.8) {{\small $v_1$}};
\node at (-3.2,3.8) {{\small $v_1$}};
\node at (-4.2,1.5) {{\small $v_4$}};
%%%%
\begin{scope}[xshift=-0.5cm]
\draw[ultra thick,xshift=1.5cm,yshift=-1cm] (-3,1) -- (-3,0);
\draw[ultra thick, dashed] (-2,0) -- (-1.5,0);
\draw[ultra thick,xshift=1.5cm,yshift=-1cm] (-3,1) -- (-2,2);
\draw[ultra thick,xshift=1.5cm,yshift=-1cm] (-2,2) -- (-2,3);
\draw[ultra thick,dashed,xshift=1.5cm,yshift=-1cm] (-2,2) -- (-1.5,2);
\draw[ultra thick,xshift=1.5cm,yshift=-1cm] (-2,3) -- (-1,4);
\draw[ultra thick, dashed] (-1,2) -- (-0.5,2);
\draw[ultra thick,dashed,xshift=1.5cm,yshift=-1cm] (-1,4) -- (-0.5,4);
\draw[ultra thick,xshift=1.5cm,yshift=-1cm] (-1,4) -- (-1,5);
\node at (0.2,1) {{\small \bf $e$}};
\node at (1.2,3) {{\small \bf $f$}};
\node at (-2.2,0) {{\small \bf $c$}};
\node at (-1.2,2) {{\small \bf $d$}};
\node at (-1.5,-1.3) {{\small \bf $2$}};
\node at (0.5,4.3) {{\small \bf $2$}};
\node[xshift=2.2cm,yshift=-1.5cm] at (-2.2,1.45) {{\small $m_2$}};
\node[xshift=2.2cm,yshift=-1.5cm] at (-1.2,3.4) {{\small $m_5$}};
%
%\node at (-3.5,1.2) {{\small $h_1$}};
%\node at (-1.5,2.2) {{\small $h_2$}};
%\node at (-5.5,0.2) {{\small $h_2$}};
%\node at (-2.5,3.2) {{\small $h_1$}};
%\node at (-0.5,4.2) {{\small $h_2$}};
%\node at (-4.5,2.2) {{\small $h_2$}};
%

\node[xshift=2.2cm,yshift=-1.5cm] at (-3.2,0.2) {{\small $v_2$}};
\node[xshift=2.2cm,yshift=-1.5cm] at (-1.2,4.8) {{\small $v_2$}};
\node[xshift=2.2cm,yshift=-1.5cm] at (-2.2,2.5) {{\small $v_5$}};
\end{scope}
\begin{scope}[xshift=6cm]
\draw[ultra thick,dashed] (-5.5,0) -- (-5,0);
\draw[ultra thick] (-5,-1) -- (-5,0);
\draw[ultra thick] (-5,0) -- (-4,1);
\draw[ultra thick,dashed] (-4,1) -- (-3.5,1);
\draw[ultra thick] (-4,1) -- (-4,2);
\draw[ultra thick] (-4,2) -- (-3,3);
\draw[ultra thick,dashed] (-4.5,2) -- (-4,2);
\draw[ultra thick] (-3,3) -- (-3,4);
\draw[ultra thick,dashed] (-3,3) -- (-2.5,3);
\node at (-5,-1.3) {{\small \bf $3$}};
\node at (-3,4.3) {{\small \bf $3$}};
\node at (-5.2,-0.8) {{\small $v_3$}};
\node at (-3.2,3.8) {{\small $v_3$}};
\node at (-4.2,1.5) {{\small $v_6$}};
\node at (-4.2,0.4) {{\small $m_3$}};
\node at (-3.1,2.45) {{\small $m_6$}};
\node at (-5.7,0) {{\small \bf $e$}};
\node at (-4.7,2) {{\small \bf $f$}};
\node at (-3.3,1) {{\small \bf $a$}};
\node at (-2.3,3) {{\small \bf $b$}};
\end{scope}
\end{tikzpicture}}}
%%%%%%%%%%%%%
&
%%%%%%%%%%%%%
%
\scalebox{0.43}{\parbox{11.5cm}{\begin{tikzpicture}[scale = 1.5]
\draw[ultra thick] (-6,0) -- (-5,0);
\draw[ultra thick,dashed] (-5,-0.5) -- (-5,0);
\draw[ultra thick] (-5,0) -- (-4,1);
\draw[ultra thick] (-4,1) -- (-3,1);
\draw[ultra thick,dashed] (-4,1) -- (-4,1.5);
\draw[ultra thick,dashed] (-2,2) -- (-2,2.5);
\draw[ultra thick,dashed] (0,3) -- (0,3.5);
\draw[ultra thick,dashed] (-3,1) -- (-3,0.5);
\draw[ultra thick,dashed] (-1,2) -- (-1,1.5);
\draw[ultra thick] (-3,1) -- (-2,2);
\draw[ultra thick] (-2,2) -- (-1,2);
\draw[ultra thick] (-1,2) -- (0,3);
\draw[ultra thick] (0,3) -- (1,3);
\draw[ultra thick,xshift=-1cm,yshift=1cm] (-5,2) -- (-4,2);
\draw[ultra thick,xshift=-1cm,yshift=1cm] (-3,3) -- (-2,3);
\draw[ultra thick,xshift=-1cm,yshift=1cm] (-2,3) -- (-1,4);
\draw[ultra thick,xshift=-1cm,yshift=1cm] (-1,4) -- (0,4);
\draw[ultra thick,dashed,xshift=-1cm,yshift=1cm] (-1,4) -- (-1,4.5);
\draw[ultra thick,dashed,xshift=-1cm,yshift=1cm] (-3,3) -- (-3,3.5);
\draw[ultra thick,dashed,xshift=-1cm,yshift=1cm] (-2,2.5) -- (-2,3);
\draw[ultra thick,dashed,xshift=-1cm,yshift=1cm] (-4,1.5) -- (-4,2);
\draw[ultra thick,xshift=-1cm,yshift=1cm] (-4,2) -- (-3,3);
\draw[ultra thick,yshift=3cm] (-1,2) -- (0,3);
\draw[ultra thick,yshift=3cm] (0,3) -- (1,3);
\draw[ultra thick,dashed,yshift=3cm] (0,3) -- (0,3.5);
\draw[ultra thick,dashed,yshift=3cm] (-1,2) -- (-1,1.5);
\node at (-6.2,0) {{\small \bf $a$}};
\node at (-6.2,3) {{\small \bf $b$}};
\node at (1.2,3) {{\small \bf $a$}};
\node at (1.2,6) {{\small \bf $b$}};
\node at (-5,-0.8) {{\small \bf $1$}};
\node at (-3,0.2) {{\small \bf $2$}};
\node at (-1,1.2) {{\small \bf $3$}};
\node at (-4,1.7) {{\small \bf $4$}};
\node at (-2,2.7) {{\small \bf $5$}};
\node at (0,3.7) {{\small \bf $6$}};
\node at (-4,4.7) {{\small \bf $1$}};
\node at (-2,5.7) {{\small \bf $2$}};
\node at (0,6.7) {{\small \bf $3$}};
\node at (-5,2.2) {{\small \bf $4$}};
\node at (-3,3.2) {{\small \bf $5$}};
\node at (-1,4.2) {{\small \bf $6$}};
\node at (-4.2,0.4) {{\small $m_1$}};
\node at (-2.2,1.45) {{\small $m_2$}};
\node at (-0.2,2.45) {{\small $m_3$}};
\node[xshift=-1.8cm,yshift=1.2cm] at (-3.1,2.45) {{\small $m_4$}};
\node[xshift=-1.8cm,yshift=1.2cm] at (-1.2,3.4) {{\small $m_5$}};
\node[xshift=-1.8cm,yshift=1.2cm] at (0.8,4.4) {{\small $m_6$}};
\node at (-3.5,1.2) {{\small $h_1$}};
\node at (-1.5,2.2) {{\small $h_2$}};
\node at (0.5,3.2) {{\small $h_3$}};
\node at (-5.5,0.2) {{\small $h_3$}};
\node[xshift=-1.5cm,yshift=0.9cm] at (-2.5,3.2) {{\small $h_4$}};
\node[xshift=-1.5cm,yshift=0.9cm] at (-0.5,4.2) {{\small $h_5$}};
\node[xshift=-1.5cm,yshift=0.9cm] at (1.5,5.2) {{\small $h_6$}};
\node[xshift=-1.5cm,yshift=0.9cm] at (-4.5,2.2) {{\small $h_6$}};
\end{tikzpicture}}}
%%%%%%%%%%%%%%%%%
&
%%%%%%%%%%%%%%%%%
\scalebox{0.43}{\parbox{11.5cm}{\begin{tikzpicture}[scale = 1.50]
\draw[ultra thick] (0,0) -- (1,0) -- (1,-1) -- (2,-1) -- (2,-2) -- (3,-2) -- (3,-3) -- (4,-3) -- (4,-4) -- (5,-4) -- (5,-5) -- (6,-5) -- (6,-6) -- (7,-6);
%diagonals up
\draw[ultra thick,dashed] (1,0) -- (1.5,0.5);
\draw[ultra thick,dashed] (2,-1) -- (2.5,-0.5);
\draw[ultra thick,dashed] (3,-2) -- (3.5,-1.5);
\draw[ultra thick,dashed] (4,-3) -- (4.5,-2.5);
\draw[ultra thick,dashed] (5,-4) -- (5.5,-3.5);
\draw[ultra thick,dashed] (6,-5) -- (6.5,-4.5);
%diagonals down
\draw[ultra thick,dashed] (1,-1) -- (0.5,-1.5);
\draw[ultra thick,dashed] (2,-2) -- (1.5,-2.5);
\draw[ultra thick,dashed] (3,-3) -- (2.5,-3.5);
\draw[ultra thick,dashed] (4,-4) -- (3.5,-4.5);
\draw[ultra thick,dashed] (5,-5) -- (4.5,-5.5);
\draw[ultra thick,dashed] (6,-6) -- (5.5,-6.5);
%labels up
\node at (1.65,0.65) {{\small \bf IV}};
\node at (2.6,-0.4) {{\small \bf V}};
\node at (3.6,-1.4) {{\small \bf VI}};
\node at (4.6,-2.4) {{\small \bf I}};
\node at (5.6,-3.4) {{\small \bf II}};
\node at (6.65,-4.35) {{\small \bf III}};
%labels down
\node at (0.4,-1.6) {{\small \bf I}};
\node at (1.4,-2.6) {{\small \bf II}};
\node at (2.4,-3.6) {{\small \bf III}};
\node at (3.4,-4.6) {{\small \bf IV}};
\node at (4.4,-5.6) {{\small \bf V}};
\node at (5.35,-6.65) {{\small \bf VI}};
%labels horizontal
\node at (-0.2,0) {{\small \bf $a$}};
\node at (7.2,-6) {{\small \bf $a$}};
%labels horizontal
\node at (0.5,0.2) {{\small $h_3$}};
\node at (1.5,-0.8) {{\small $h_4$}};
\node at (2.5,-1.8) {{\small $h_2$}};
\node at (3.5,-2.8) {{\small $h_6$}};
\node at (4.5,-3.8) {{\small $h_1$}};
\node at (5.5,-4.8) {{\small $h_5$}};
\node at (6.5,-5.8) {{\small $h_3$}};
%labels vertical
\node at (0.8,-0.5) {{\small $v_1$}};
\node at (1.8,-1.5) {{\small $v_5$}};
\node at (2.8,-2.5) {{\small $v_3$}};
\node at (3.8,-3.5) {{\small $v_4$}};
\node at (4.8,-4.5) {{\small $v_2$}};
\node at (5.8,-5.5) {{\small $v_6$}};
\end{tikzpicture}}}
\end{tabular}
\end{center}
\caption{{\sl Three different parametrisations of the web diagram $(N,M)=(3,2)$. The last line shows the decomposition of the diagram in the weak coupling limit in the horizontal, vertical and diagonal description respectively.}}
\label{Tab:Bases32}
\end{table}
%%%%%%%%%%%%%%%%%%%%%%%%%%%%%%%%%%%%%%%%%%%%
%%%%%%%%%%%%%%%%%%%%%%%%%%%%%%%%%%%%%%%%%%%%
\begin{itemize}
\item \underline{horizontal expansion in the basis $(\rho,\widehat{b}_1,\widehat{b}_2;\widehat{c}_1,\widehat{c}_2,\widehat{c}_3;\tau,E)$}\\
In the limit
\begin{align}
&\rho-\widehat{b}_1-\widehat{b}_2\longrightarrow \infty\,,&&\text{and} &&\widehat{b}_1\longrightarrow \infty\,,&&\widehat{b}_2\longrightarrow \infty\,,\label{WeakHorizon32}
\end{align}
we find $h_{1,\ldots,6}\longrightarrow\infty$, while $v_{1,\ldots,6}$ and $m_{1,\ldots,6}$ remain finite. Therefore, as indicated in table~\ref{Tab:Bases32}, in the limit (\ref{WeakHorizon32}) the toric web diagram decomposes into 3 vertical strips, implying that the horizontal expansion gives rise to a gauge theory with gauge group
\begin{align}
G_{\text{hor}}=U(2)\times U(2)\times U(2)\,,\label{GaugeHor32}
\end{align}
More specifically, the parameters $(\rho-\widehat{b}_1-\widehat{b}_2,\widehat{b}_1,\widehat{b}_2)$ are related to the gauge coupling constants, while $\widehat{c}_1$, $\widehat{c}_2$ and $\widehat{c}_3$ can be interpreted as the (simple positive) roots of $\mathfrak{a}_1$ algebras associated with each of the $U(2)$ factors. Each of these algebras is further extended to affine $\widehat{a}_1$ through the parameter $\tau$.
\item \underline{vertical expansion in the basis $(\tau,\widehat{c}_1;\widehat{b}_1,\widehat{b}_2,\widehat{b}_3,\widehat{b}_4;\rho,D)$}\\
In the limit
\begin{align}
&\tau-\widehat{c}_1\longrightarrow \infty\,,&&\widehat{c}_2\longrightarrow\infty\,,
\end{align}
we find $v_{1,\ldots,6}\longrightarrow\infty$, while $h_{1,\ldots,6}$ and $m_{1,\ldots,6}$ remain finite. Therefore, the $(3,2)$ web diagram decomposes into two horizontal strips, indicating that the vertical expansion is associated with a gauge theory with gauge group
\begin{align}
G_{\text{vert}}=U(3)\times U(3)\,,\label{GaugeVert32}
\end{align}
In this manner, $(\tau-\widehat{c}_1,\widehat{c}_1)$ are related to the coupling constants and $(\widehat{b}_1,\widehat{b_2})$ and $(\widehat{b}_3,\widehat{b}_4)$ correspond to the (simple positive) roots of two copies of $\mathfrak{a}_2$, associated with the to the $U(3)$ factors in (\ref{GaugeVert32}). These algebras are extended to affine $\widehat{\mathfrak{a}}_2$ by the parameter $\rho$.
\item \underline{diagonal expansion in the basis $(V;\widehat{a}_1,\widehat{a}_2,\widehat{a}_3,\widehat{a}_4,\widehat{a}_5;M,F)$}\\
In the limit $V\longrightarrow\infty$ we find $m_{1,\ldots,6}\longrightarrow\infty$, while $h_{1,\ldots,6}$ and $v_{1,\ldots,6}$ remain finite, such that the web diagram decomposes into a single diagonal strip. This indicates that the diagonal expansion is associated with a gauge theory with gauge group
\begin{align}
G_{\text{diag}}=U(6)\,.
\end{align}
Here $V$ is related to the coupling constant, while $(\widehat{a}_1,\ldots,\widehat{a}_5)$ play the role of the (simple positive) roots of $\mathfrak{a}_5$ associated with $G_{\text{diag}}$.

The fact that such a gauge theory exists also outside of the weak coupling limit $V\to \infty$ can be inferred from the duality of the web diagram shown in table~\ref{Tab:Bases32} with the toric web of $X_{6,1}$ through a series of flop- and symmetry transformations. The latter was first discussed in \cite{Hohenegger:2016yuv} and is schematically shown in \figref{Fig:FlopSymTrafo}.
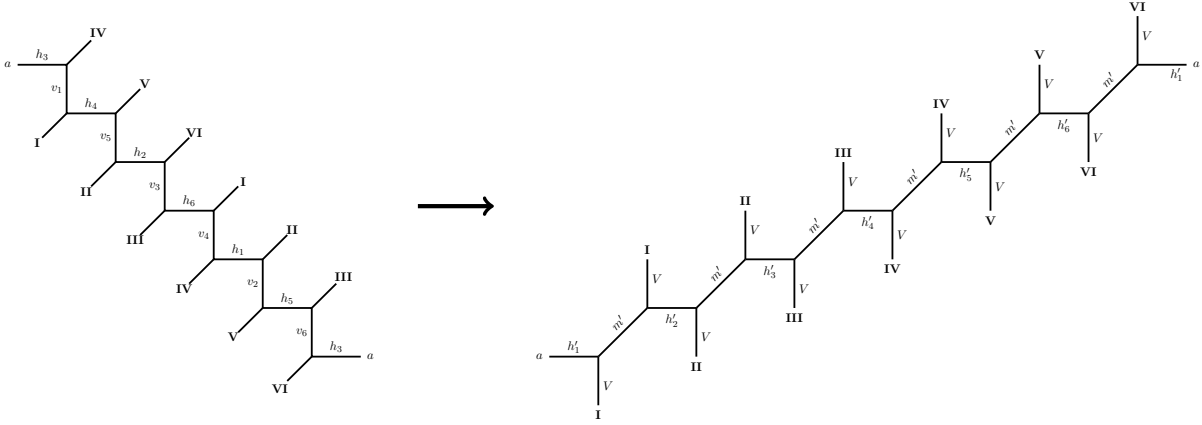
\begin{figure}
\begin{center}
\scalebox{0.43}{\parbox{11.5cm}{\begin{tikzpicture}[scale = 1.50]
\draw[ultra thick] (0,0) -- (1,0) -- (1,-1) -- (2,-1) -- (2,-2) -- (3,-2) -- (3,-3) -- (4,-3) -- (4,-4) -- (5,-4) -- (5,-5) -- (6,-5) -- (6,-6) -- (7,-6);
%diagonals up
\draw[ultra thick] (1,0) -- (1.5,0.5);
\draw[ultra thick] (2,-1) -- (2.5,-0.5);
\draw[ultra thick] (3,-2) -- (3.5,-1.5);
\draw[ultra thick] (4,-3) -- (4.5,-2.5);
\draw[ultra thick] (5,-4) -- (5.5,-3.5);
\draw[ultra thick] (6,-5) -- (6.5,-4.5);
%diagonals down
\draw[ultra thick] (1,-1) -- (0.5,-1.5);
\draw[ultra thick] (2,-2) -- (1.5,-2.5);
\draw[ultra thick] (3,-3) -- (2.5,-3.5);
\draw[ultra thick] (4,-4) -- (3.5,-4.5);
\draw[ultra thick] (5,-5) -- (4.5,-5.5);
\draw[ultra thick] (6,-6) -- (5.5,-6.5);
%labels up
\node at (1.65,0.65) {{\small \bf IV}};
\node at (2.6,-0.4) {{\small \bf V}};
\node at (3.6,-1.4) {{\small \bf VI}};
\node at (4.6,-2.4) {{\small \bf I}};
\node at (5.6,-3.4) {{\small \bf II}};
\node at (6.65,-4.35) {{\small \bf III}};
%labels down
\node at (0.4,-1.6) {{\small \bf I}};
\node at (1.4,-2.6) {{\small \bf II}};
\node at (2.4,-3.6) {{\small \bf III}};
\node at (3.4,-4.6) {{\small \bf IV}};
\node at (4.4,-5.6) {{\small \bf V}};
\node at (5.35,-6.65) {{\small \bf VI}};
%labels horizontal
\node at (-0.2,0) {{\small \bf $a$}};
\node at (7.2,-6) {{\small \bf $a$}};
%labels horizontal
\node at (0.5,0.2) {{\small $h_3$}};
\node at (1.5,-0.8) {{\small $h_4$}};
\node at (2.5,-1.8) {{\small $h_2$}};
\node at (3.5,-2.8) {{\small $h_6$}};
\node at (4.5,-3.8) {{\small $h_1$}};
\node at (5.5,-4.8) {{\small $h_5$}};
\node at (6.5,-5.8) {{\small $h_3$}};
%labels vertical
\node at (0.8,-0.5) {{\small $v_1$}};
\node at (1.8,-1.5) {{\small $v_5$}};
\node at (2.8,-2.5) {{\small $v_3$}};
\node at (3.8,-3.5) {{\small $v_4$}};
\node at (4.8,-4.5) {{\small $v_2$}};
\node at (5.8,-5.5) {{\small $v_6$}};
\end{tikzpicture}}}
\hspace{0.25cm}
\parbox{1cm}{\begin{tikzpicture}
\draw[ultra thick,->] (0,0) -- (1,0); 
\end{tikzpicture}}
\hspace{0.25cm}
\scalebox{0.43}{\parbox{22.7cm}{\begin{tikzpicture}[scale = 1.5]
%vertical
\draw[ultra thick] (5,2) -- (5,1);
\draw[ultra thick] (6,3) -- (6,4);
\draw[ultra thick] (7,3) -- (7,2);
\draw[ultra thick] (8,4) -- (8,5);
\draw[ultra thick] (9,4) -- (9,3);
\draw[ultra thick] (10,5) -- (10,6);
\draw[ultra thick] (11,5) -- (11,4);
\draw[ultra thick] (12,6) -- (12,7);
\draw[ultra thick] (13,6) -- (13,5);
\draw[ultra thick] (14,7) -- (14,8);
\draw[ultra thick] (15,7) -- (15,6);
\draw[ultra thick] (16,8) -- (16,9);
%horizontal
\draw[ultra thick] (4,2) -- (5,2);
\draw[ultra thick] (6,3) -- (7,3);
\draw[ultra thick] (8,4) -- (9,4);
\draw[ultra thick] (10,5) -- (11,5);
\draw[ultra thick] (12,6) -- (13,6);
\draw[ultra thick] (14,7) -- (15,7);
\draw[ultra thick] (16,8) -- (17,8);
%diagonals
\draw[ultra thick] (5,2) -- (6,3);
\draw[ultra thick] (7,3) -- (8,4);
\draw[ultra thick] (9,4) -- (10,5);
\draw[ultra thick] (11,5) -- (12,6);
\draw[ultra thick] (13,6) -- (14,7);
\draw[ultra thick] (15,7) -- (16,8);
\node[rotate=45] at (5.4,2.7) {{\small $m'$}};
\node[rotate=45] at (7.4,3.7) {{\small $m'$}};
\node[rotate=45] at (9.4,4.7) {{\small $m'$}};
\node[rotate=45] at (11.4,5.7) {{\small $m'$}};
\node[rotate=45] at (13.4,6.7) {{\small $m'$}};
\node[rotate=45] at (15.4,7.7) {{\small $m'$}};
\node at (4.5,2.2) {{\small $h'_1$}};
\node at (6.5,2.75) {{\small $h'_2$}};
\node at (8.5,3.75) {{\small $h'_3$}};
\node at (10.5,4.75) {{\small $h'_4$}};
\node at (12.5,5.75) {{\small $h'_5$}};
\node at (14.5,6.75) {{\small $h'_6$}};
\node at (16.8,7.8) {{\small $h'_1$}};
\node at (5.2,1.4) {{\small $V$}};
\node at (7.2,2.4) {{\small $V$}};
\node at (9.2,3.4) {{\small $V$}};
\node at (11.2,4.5) {{\small $V$}};
\node at (13.2,5.5) {{\small $V$}};
\node at (15.2,6.5) {{\small $V$}};
\node at (6.2,3.6) {{\small $V$}};
\node at (8.2,4.6) {{\small $V$}};
\node at (10.2,5.6) {{\small $V$}};
\node at (12.2,6.6) {{\small $V$}};
\node at (14.2,7.6) {{\small $V$}};
\node at (16.2,8.6) {{\small $V$}};
\node at (3.8,2) {{\small \bf $a$}};
\node at (17.2,8) {{\small \bf $a$}};
\node at (5,0.8) {{\small \bf I}};
\node at (7,1.8) {{\small \bf II}};
\node at (9,2.8) {{\small \bf III}};
\node at (11,3.8) {{\small \bf IV}};
\node at (13,4.8) {{\small \bf V}};
\node at (15,5.8) {{\small \bf VI}};
\node at (6,4.2) {{\small \bf I}};
\node at (8,5.2) {{\small \bf II}};
\node at (10,6.2) {{\small \bf III}};
\node at (12,7.2) {{\small \bf IV}};
\node at (14,8.2) {{\small \bf V}};
\node at (16,9.2) {{\small \bf VI}};
\end{tikzpicture}}
}
\end{center}
\caption{\sl Flop- and symmetry transformations relating the diagonal presentation (rightmost column in table~\ref{Tab:Bases32}) to the toric web of $X_{6,1}$.}
\label{Fig:FlopSymTrafo}
\end{figure}
In this duality transformation, the diagonal lines in the left part of the \figref{Fig:FlopSymTrafo} do not undergo flop transitions, such that the K\"ahler parameters $h'_{1,2,3,4,5,6}$ and $m'$ of $X_{6,1}$ are independent of $V$. Therefore, the vertical expansion of the partition function $\pf{6}{1}$ is a power series in $Q_V=e^{-V}$, which can be interpreted as the instanton partition function $Z_{\text{vert}}^{(6,1)}$ of a gauge theory with gauge group $U(6)$. Moreover, through the duality map implied by \figref{Fig:FlopSymTrafo}, $Z_{\text{vert}}^{(6,1)}$ can also be related to $Z_{\text{diag}}^{(3,2)}$
\begin{align}
Z_{\text{vert}}^{(6,1)}(V,h'_{1,\ldots,6},m)=Z_{\text{diag}}^{(3,2)}(V,\widehat{a}_{1,\ldots,5},M,F)\,,
\end{align}
which was proven explicitly in \cite{Bastian:2017ing}. This shows that $Z_{\text{diag}}^{(3,2)}$ (as a power series expansion in $Q_V$) can be read as the instanton partition function of a gauge theory with gauge group $U(6)$ also outside of the region $V\to \infty$.
\end{itemize}
In the case of $(N,M)=(3,2)$, the gauge groups $G_{\text{hor}}$, $G_{\text{vert}}$ and $G_{\text{diag}}$ are different\footnote{We remark, however, that at a generic point in moduli space, the gauge groups are broken down to $U(1)^6$ respectively.}, however, as discussed in the previous section, their rank is identical.

Furthermore, we stress that in all three cases the specific form of the parametrisation is not unique: Different choices of parameters leading to the same decomposition of the toric web diagram as in table~\ref{Tab:Bases32} are possible. Indeed, in \cite{Bastian:2017ing}  we have chosen a slightly different parametrisation suitable for the explicit computations of the general building block of the partition function.

%%%%%%%%%%%%%%%%%%%%%%%%%%%%%%%%%%%%%%%%%%%%%%
%%%%%%%%%%%%%%%%%%%%%%%%%%%%%%%%%%%%%%%%%%%%%%
%%%%%%%%%%%%%%%%%%%%%%%%%%%%%%%%%%%%%%%%%%%%%%

\subsection{General Web Configuration}\label{App:BasisofKaehler}
The discussion of the previous examples $(2,2)$ and $(3,2)$ can be generalised to a web diagram with generic $(N,M)$. Indeed, in the following we make a proposal for three different parametrisations of the K\"ahler moduli space of $X_{N,M}$, facilitating the three expansions of $\pf{N}{M}$ that were schematically written in (\ref{TrialitySchemat}). In the following, we present sets -- in general not unique -- of $NM+2$ independent parameters (which we shall refer to as a basis in the following) suitable for the description of the horizontal, vertical and diagonal theory.\footnote{This choice of bases is motivated by studying numerous examples with small values of $N$ or $M$ such as those in sections~\ref{Sect:Basis22} and \ref{Sect:Basis32}. A proof of the linear independence for generic $(N,M)$ is currently still missing.} 

\begin{figure}[htbp]
\begin{center}
\rotatebox{90}{\scalebox{0.58}{\parbox{36cm}{\begin{tikzpicture}[scale = 1.50]
%%%%%%%%%%%%%%%%left lines
\draw[ultra thick] (-1,0) -- (0,0) -- (1,1) -- (2,1) -- (3,2) -- (4,2);
%verticals
\draw[ultra thick] (0,-1) -- (0,0);
\draw[ultra thick] (2,0) -- (2,1);
\draw[ultra thick] (1,1) -- (1,2);
\draw[ultra thick] (3,2) -- (3,3);
%dots
\node[rotate=90] at (1,2.5) {{\Huge$\ldots$}};
\node[rotate=90] at (3,3.5) {{\Huge$\ldots$}};
\draw[ultra thick] (0,4) -- (1,4) -- (2,5) -- (3,5) -- (4,6) -- (5,6);
\draw[ultra thick] (1,6) -- (2,6) -- (3,7) -- (4,7) -- (5,8) -- (6,8);
\draw[ultra thick] (1,3) -- (1,4);
\draw[ultra thick] (3,4) -- (3,5);
\draw[ultra thick] (2,5) -- (2,6);
\draw[ultra thick] (4,6) -- (4,7);
\draw[ultra thick] (3,7) -- (3,8);
\draw[ultra thick] (5,8) -- (5,9);
%dots
\node[rotate=90] at (3,8.5) {{\Huge$\ldots$}};
\node[rotate=90] at (5,9.5) {{\Huge$\ldots$}};
\draw[ultra thick] (3,9) -- (3,10);
\draw[ultra thick] (5,10) -- (5,11);
\draw[ultra thick] (2,10) -- (3,10) -- (4,11) -- (5,11) -- (6,12) -- (7,12);
\draw[ultra thick] (4,11) -- (4,12);
\draw[ultra thick] (6,12) -- (6,13);
%%%%%%%%%%%%%%%%%horizontal dotes
%dots
\node at (4.5,2) {{\Huge$\ldots$}};
\node at (5.5,6) {{\Huge$\ldots$}};
\node at (6.5,8) {{\Huge$\ldots$}};
\node at (7.5,12) {{\Huge$\ldots$}};
%%%%%%%%%%%%centerlines
\draw[ultra thick] (5,2) -- (6,2) -- (7,3) -- (8,3) -- (9,4) -- (10,4);
\draw[ultra thick] (6,1) -- (6,2);
\draw[ultra thick] (8,3) -- (8,2);
\draw[ultra thick] (7,3) -- (7,4);
\draw[ultra thick] (9,4) -- (9,5);
%dots
\node[rotate=90] at (7,4.5) {{\Huge$\ldots$}};
\node[rotate=90] at (9,5.5) {{\Huge$\ldots$}};
\draw[ultra thick] (7,5) -- (7,6);
\draw[ultra thick] (9,6) -- (9,7);
\draw[ultra thick] (6,6) -- (7,6) -- (8,7) -- (9,7) -- (10,8) -- (11,8);
\draw[ultra thick] (7,8) -- (8,8) -- (9,9) -- (10,9) -- (11,10) -- (12,10);
\draw[ultra thick] (8,7) -- (8,8);
\draw[ultra thick] (10,8) -- (10,9);
\draw[ultra thick] (9,9) -- (9,10);
\draw[ultra thick] (11,10) -- (11,11);
%dots
\node[rotate=90] at (9,10.5) {{\Huge$\ldots$}};
\node[rotate=90] at (11,11.5) {{\Huge$\ldots$}};
\draw[ultra thick] (9,11) -- (9,12);
\draw[ultra thick] (11,12) -- (11,13);
\draw[ultra thick] (8,12) -- (9,12) -- (10,13) -- (11,13) -- (12,14) -- (13,14);
\draw[ultra thick] (10,13) -- (10,14);
\draw[ultra thick] (12,14) -- (12,15);
%%%%%%
\node at (10.5,4) {{\Huge$\ldots$}};
\node at (11.5,8) {{\Huge$\ldots$}};
\node at (12.5,10) {{\Huge$\ldots$}};
\node at (13.5,14) {{\Huge$\ldots$}};
%%%%%%%%%%%%right lines
\draw[ultra thick] (12,4) -- (12,3);
\draw[ultra thick] (14,5) -- (14,4);
\draw[ultra thick] (11,4) -- (12,4) -- (13,5) -- (14,5) -- (15,6) -- (16,6);
\draw[ultra thick] (13,5) -- (13,6);
\draw[ultra thick] (15,7) -- (15,6);
\node[rotate=90] at (13,6.5) {{\Huge$\ldots$}};
\node[rotate=90] at (15,7.5) {{\Huge$\ldots$}};
\draw[ultra thick] (13,7) -- (13,8);
\draw[ultra thick] (15,9) -- (15,8);
\draw[ultra thick] (12,8) -- (13,8) -- (14,9) -- (15,9) -- (16,10) -- (17,10);
\draw[ultra thick] (13,10) -- (14,10) -- (15,11) -- (16,11) -- (17,12) -- (18,12);
\draw[ultra thick] (14,9) -- (14,10);
\draw[ultra thick] (16,11) -- (16,10);
\draw[ultra thick] (15,11) -- (15,12);
\draw[ultra thick] (17,13) -- (17,12);
\node[rotate=90] at (15,12.5) {{\Huge$\ldots$}};
\node[rotate=90] at (17,13.5) {{\Huge$\ldots$}};
\draw[ultra thick] (15,13) -- (15,14);
\draw[ultra thick] (17,15) -- (17,14);
\draw[ultra thick] (14,14) -- (15,14) -- (16,15) -- (17,15) -- (18,16) -- (19,16);
\draw[ultra thick] (16,15) -- (16,16);
\draw[ultra thick] (18,16) -- (18,17);
%%%%%%%%%%%%%%%%%%%%%%%%%%%%%%%%%%%%%%%%%%%
%%%%%%%%%%%%%%%%%%%%%%%%%%%%%%%%%%%%%%%%%%
%ID labels bottom
\node at (0,-1.2) {{\small \bf $1$}};
\node at (2,-0.2) {{\small \bf $2$}};
\node at (6,0.8) {{\small \bf $s$}};
\node at (8,1.8) {{\small \bf $s+1$}};
\node at (12,2.8) {{\small \bf $N-1$}};
\node at (14,3.8) {{\small \bf $N$}};
%ID labels top
\node at (4,12.2) {{\small \bf $1$}};
\node at (6,13.2) {{\small \bf $2$}};
\node at (10,14.2) {{\small \bf $s$}};
\node at (12,15.2) {{\small \bf $s+1$}};
\node at (16,16.2) {{\small \bf $N-1$}};
\node at (18,17.2) {{\small \bf $N$}};
%ID labels left
\node at (-1.25,0) {{\small \bf $a_1$}};
\node at (-0.25,4) {{\small \bf $a_{r+1}$}};
\node at (0.65,6) {{\small \bf $a_{r+2}$}};
\node at (1.75,10) {{\small \bf $a_{M}$}};
%ID labels right
\node at (16.25,6) {{\small \bf $a_1$}};
\node at (17.35,10) {{\small \bf $a_{r+1}$}};
\node at (18.35,12) {{\small \bf $a_{r+2}$}};
\node at (19.25,16) {{\small \bf $a_{M}$}};
%%%%%%%%%%%%%%%%%%%%%%%%%%%%%%%%%%%%%%%%%%%
%%%%%%%%%%%%%%%%%%%%%%%%%%%%%%%%%%diagonal m-lables
%diagonal-labels bottom layer
\node[rotate=45] at (0.25,0.45) {{\small \bf $m_{1}$}};
\node[rotate=45] at (2.3,1.5) {{\small \bf $m_{2}$}};
\node[rotate=45] at (6.3,2.5) {{\small \bf $m_{s}$}};
\node[rotate=45] at (8.3,3.5) {{\small \bf $m_{s+1}$}};
\node[rotate=45] at (12.3,4.5) {{\small \bf $m_{N-1}$}};
\node[rotate=45] at (14.3,5.5) {{\small \bf $m_{N}$}};
%diagonal-labels second layer
\node[rotate=45] at (1.4,4.6) {{\small \bf $m_{rN+1}$}};
\node[rotate=45] at (3.4,5.6) {{\small \bf $m_{rN+2}$}};
\node[rotate=45] at (7.35,6.55) {{\small \bf $m_{rN+s}$}};
\node[rotate=45] at (9.45,7.65) {{\small \bf $m_{rN+s+1}$}};
\node[rotate=45] at (13.5,8.7) {{\small \bf $m_{(r+1)N-1}$}};
\node[rotate=45] at (15.6,9.3) {{\small \bf $m_{(r+1)N}$}};
%diagonal-labels third layer
\node[rotate=45] at (2.5,6.7) {{\small \bf $m_{(r+1)N+1}$}};
\node[rotate=45] at (4.5,7.7) {{\small \bf $m_{(r+1)N+2}$}};
\node[rotate=45] at (8.5,8.7) {{\small \bf $m_{(r+1)N+s}$}};
\node[rotate=45] at (10.4,9.6) {{\small \bf $m_{(r+1)N+s+1}$}};
\node[rotate=45] at (14.5,10.7) {{\small \bf $m_{(r+2)N-1}$}};
\node[rotate=45] at (16.4,11.6) {{\small \bf $m_{(r+2)N}$}};
%diagonal-labels top layer
\node[rotate=45] at (3.35,10.55) {{\small \bf $m_{(M-1)N+1}$}};
\node[rotate=45] at (5.35,11.55) {{\small \bf $m_{(M-1)N+2}$}};
\node[rotate=45] at (9.35,12.55) {{\small \bf $m_{(M-1)N+s}$}};
\node[rotate=45] at (11.4,13.6) {{\small \bf $m_{(M-1)N+s+1}$}};
\node[rotate=45] at (15.35,14.55) {{\small \bf $m_{MN-1}$}};
\node[rotate=45] at (17.35,15.55) {{\small \bf $m_{MN}$}};
%%%%%%%%%%%%%%%%%%%%%%%%%%%%%vertical v-labels
%vertical-labels bottom layer
\node at (-0.2,-0.5) {{\small \bf $v_1$}};
\node at (1.8,0.5) {{\small \bf $v_2$}};
\node at (5.8,1.5) {{\small \bf $v_s$}};
\node at (7.7,2.5) {{\small \bf $v_{s+1}$}};
\node at (11.65,3.5) {{\small \bf $v_{N-1}$}};
\node at (13.8,4.5) {{\small \bf $v_{N}$}};
%vertical-labels first layer
\node at (0.6,1.5) {{\small \bf $v_{N+1}$}};
\node at (2.6,2.5) {{\small \bf $v_{N+2}$}};
\node at (6.6,3.5) {{\small \bf $v_{N+s}$}};
\node at (8.5,4.5) {{\small \bf $v_{N+s+1}$}};
\node at (12.6,5.5) {{\small \bf $v_{2N-1}$}};
\node at (14.7,6.5) {{\small \bf $v_{2N}$}};
%vertical-labels second layer
\node at (0.6,3.5) {{\small \bf $v_{rN+1}$}};
\node at (2.6,4.5) {{\small \bf $v_{rN+2}$}};
\node at (6.6,5.5) {{\small \bf $v_{rN+s}$}};
\node at (9.55,6.4) {{\small \bf $v_{rN+s+1}$}};
\node at (12.4,7.5) {{\small \bf $v_{(r+1)N-1}$}};
\node at (14.5,8.5) {{\small \bf $v_{(r+1)N}$}};
%vertical-labels third layer
\node[rotate=45] at (2.55,5.8) {{\small \bf $v_{(r+1)N+1}$}};
\node at (4.65,6.7) {{\small \bf $v_{(r+1)N+2}$}};
\node at (7.4,7.5) {{\small \bf $v_{(r+1)N+s}$}};
\node at (10.8,8.5) {{\small \bf $v_{(r+1)N+s+1}$}};
\node at (13.4,9.5) {{\small \bf $v_{(r+2)N-1}$}};
\node at (15.5,10.5) {{\small \bf $v_{(r+2)N}$}};
%vertical-labels fourth layer
\node[rotate=45] at (2.5,7.5) {{\small \bf $v_{(r+2)N+1}$}};
\node at (4.4,8.5) {{\small \bf $v_{(r+2)N+2}$}};
\node at (8.4,9.6) {{\small \bf $v_{(r+2)N+s}$}};
\node at (10.3,10.5) {{\small \bf $v_{(r+2)N+s+1}$}};
\node at (14.4,11.5) {{\small \bf $v_{(r+3)N-1}$}};
\node at (16.5,12.5) {{\small \bf $v_{(r+3)N}$}};
%vertical-labels fifth layer
\node[rotate=90] at (2.8,9.3) {{\small \bf $v_{(M-1)N+1}$}};
\node[rotate=90] at (4.8,10.3) {{\small \bf $v_{(M-1)N+2}$}};
\node at (8.3,11.5) {{\small \bf $v_{(M-1)N+s}$}};
\node at (11.85,12.5) {{\small \bf $v_{(M-1)N+s+1}$}};
\node at (14.55,13.5) {{\small \bf $v_{MN-1}$}};
\node at (17.4,14.5) {{\small \bf $v_{MN}$}};
%vertical-labels top layer
\node at (3.8,11.5) {{\small \bf $v_{1}$}};
\node at (5.8,12.5) {{\small \bf $v_{2}$}};
\node at (9.8,13.5) {{\small \bf $v_{s}$}};
\node at (11.7,14.5) {{\small \bf $v_{s+1}$}};
\node at (15.65,15.5) {{\small \bf $v_{N-1}$}};
\node at (17.75,16.5) {{\small \bf $v_{N}$}};
%%%%%%%%%%%%%%%%%%%%%%%%%%%%%horizontal h-lables
%horizontal-labels first layer
\node at (-0.5,0.2) {{\small \bf $h_{N}$}};
\node at (1.5,1.2) {{\small \bf $h_{1}$}};
\node at (3.5,2.2) {{\small \bf $h_{2}$}};
\node at (5.5,2.2) {{\small \bf $h_{s-1}$}};
\node at (7.5,3.2) {{\small \bf $h_{s}$}};
\node at (9.5,4.2) {{\small \bf $h_{s+1}$}};
\node at (11.5,4.2) {{\small \bf $h_{N-2}$}};
\node at (13.5,5.2) {{\small \bf $h_{N-1}$}};
\node at (15.5,6.2) {{\small \bf $h_{N}$}};
%horizontal-labels second layer
\node at (0.5,4.2) {{\small \bf $h_{(r+1)N}$}};
\node at (2.5,5.2) {{\small \bf $h_{rN+1}$}};
\node at (4.5,6.2) {{\small \bf $h_{rN+2}$}};
\node at (6.5,6.2) {{\small \bf $h_{rN+s-1}$}};
\node at (8.5,7.2) {{\small \bf $h_{rN+s}$}};
\node[rotate=90] at (10.8,7.45) {{\small \bf $h_{rN+s+1}$}};
\node at (12.45,8.2) {{\small \bf $h_{(r+1)N-2}$}};
\node[rotate=15] at (14.7,9.3) {{\small \bf $h_{(r+1)N-1}$}};
\node at (16.5,10.2) {{\small \bf $h_{(r+1)N}$}};
%horizontal-labels second layer
\node at (1.5,6.2) {{\small \bf $h_{(r+2)N}$}};
\node[rotate=90] at (3.5,7.65) {{\small \bf $h_{(r+1)N+1}$}};
\node at (5.65,8.2) {{\small \bf $h_{(r+1)N+2}$}};
\node at (7.35,8.2) {{\small \bf $h_{(r+1)N+s-1}$}};
\node[rotate=30] at (9.35,8.6) {{\small \bf $h_{(r+1)N+s}$}};
\node at (11.8,10.2) {{\small \bf $h_{(r+1)N+s+1}$}};
\node at (13.4,10.2) {{\small \bf $h_{(r+2)N-2}$}};
\node[rotate=90] at (15.4,11.65) {{\small \bf $h_{(r+2)N-1}$}};
\node at (17.5,12.2) {{\small \bf $h_{(r+2)N}$}};
%horizontal-labels second layer
\node at (2.4,10.2) {{\small \bf $h_{MN}$}};
\node[rotate=90] at (4.4,11.65) {{\small \bf $h_{(M-1)N+1}$}};
\node at (6.7,12.2) {{\small \bf $h_{(M-1)N+2}$}};
\node[rotate=90] at (8.4,12.8) {{\small \bf $h_{(M-1)N+s-1}$}};
\node at (6.7,12.2) {{\small \bf $h_{(M-1)N+2}$}};
\node[rotate=90] at (10.4,13.65) {{\small \bf $h_{(M-1)N+s}$}};
\node[rotate=90] at (12.4,14.8) {{\small \bf $h_{(M-1)N+s+1}$}};
\node at (14.5,14.2) {{\small \bf $h_{MN-2}$}};
\node at (16.5,15.2) {{\small \bf $h_{MN-1}$}};
\node at (18.5,16.2) {{\small \bf $h_{MN}$}};
%%%%%%%%%%%%%%%%%%%%%%%%%%%%%%%%%%%%%%%%%%%%
%vertical basis
%%%%%%%%%%%%%%%%%%%%%%%%%%%%%%%%%%%%%%%%%%%%
%roots first layer
\draw[thick,blue,<->] (0.1,0) -- (1.9,0);
\node[blue] at (0.8,-0.2) {{\small $\widehat{b}_{M-1,1}$}};
\draw[thick,blue,<->] (2.1,1) -- (3.9,1);
\node[blue] at (3,1.2) {{\small $\widehat{b}_{M-1,2}$}};
\draw[thick,blue,<->] (6.1,2) -- (7.9,2);
\node[blue] at (7,2.2) {{\small $\widehat{b}_{M-1,s}$}};
\draw[thick,blue,<->] (8.1,3) -- (9.9,3);
\node[blue] at (9,3.2) {{\small $\widehat{b}_{M-1,s+1}$}};
\draw[thick,blue,<->] (12.2,4) -- (13.9,4);
\node[blue] at (13,4.2) {{\small $\widehat{b}_{M-1,N-1}$}};
%roots second layer
\draw[thick,blue,<->] (1.1,2) -- (2.9,2);
\node[blue] at (1.5,1.8) {{\small $\widehat{b}_{0,1}$}};
\draw[thick,blue,<->] (7.1,4) -- (8.9,4);
\node[blue] at (8,4.2) {{\small $\widehat{b}_{0,s}$}};
%\draw[thick,blue,<->] (9.1,5) -- (10.9,5);
%\node[blue] at (10,5.2) {{\small $\widehat{b}_{N+s}$}};
\draw[thick,blue,<->] (13.2,6) -- (14.9,6);
\node[blue] at (14,6.2) {{\small $\widehat{b}_{0,N-1}$}};
%%%%%%%%%%%%%%%%%%%%%%%%%%%%%%%%%%%%%%%%%%%%
%horizontal basis
%%%%%%%%%%%%%%%%%%%%%%%%%%%%%%%%%%%%%%%%%%%%
%roots first column
\draw[thick,green!50!black,<->] (0,1.9) -- (0,0.1);
\node[green!50!black] at (-0.3,1) {{\small $\widehat{c}_{0,N}$}};
\draw[thick,green!50!black,<->] (1,5.9) -- (1,4.1);
\node[green!50!black] at (0.7,4.8) {{\small $\widehat{c}_{r,N}$}};
\draw[thick,green!50!black,<->] (2,9.9) -- (2,8.1);
\node[green!50!black] at (1.5,9) {{\small $\widehat{c}_{M-2,N}$}};
%
%roots second column
\draw[thick,green!50!black,<->] (2,2.9) -- (2,1.1);
\node[green!50!black] at (1.8,2.3) {{\small $\widehat{c}_{0,1}$}};
\draw[thick,green!50!black,<->] (3,6.9) -- (3,5.1);
\node[green!50!black] at (3.25,6.6) {{\small $\widehat{c}_{r,1}$}};
\draw[thick,green!50!black,<->] (4,10.9) -- (4,9.1);
\node[green!50!black,rotate=270] at (4.2,10.4) {{\small $\widehat{c}_{M-2,1}$}};
%%%%%%%%%%%%%%%%%%%%%%%%%%%%%%%%%%%%%%%%%%%
%diagonal basis
%%%%%%%%%%%%%%%%%%%%%%%%%%%%%%%%%%%%%%%%%%%
%roots bottom layer
%\draw[thick,<->,red] (-0.5,-1.4) -- (-1.4,-0.5);
%\node[red] at (-1.05,-1.15) {{\small $\widehat{a}_{1}$}};
\draw[thick,<->,red] (3.5,10.6) -- (2.6,11.5);
\node[red] at (2.75,10.85) {{\small $\widehat{a}_{M-1,N}$}};
\draw[thick,<->,red] (15.5,4.6) -- (14.6,5.5);
\node[red] at (14.75,4.85) {{\small $\widehat{a}_{M-1,N}$}};
\draw[thick,<->,red] (4.5,9.6) -- (3.6,10.5);
\node[red] at (3.55,9.95) {{\small $\widehat{a}_{M-2,1}$}};
\draw[thick,<->,red] (0.5,0.6) -- (-0.4,1.5);
\node[red] at (0.45,1.1) {{\small $\widehat{a}_{0,N}$}};
\draw[thick,<->,red] (1.5,-0.4) -- (0.6,0.5);
\node[red] at (1.3,0.35) {{\small $\widehat{a}_{M-1,1}$}};
\draw[thick,<->,red] (5.5,11.6) -- (4.6,12.5);
\node[red] at (5.25,12.35) {{\small $\widehat{a}_{M-1,1}$}};
\draw[thick,<->,red] (6.5,10.6) -- (5.6,11.5);
\node[red] at (6.5,11.2) {{\small $\widehat{a}_{M-2,2}$}};
\draw[thick,<->,red] (1.5,4.6) -- (0.6,5.5);
\node[red,rotate=45] at (1.5,5.35) {{\small $\widehat{a}_{r+1,N}$}};
\draw[thick,<->,red] (2.5,3.6) -- (1.6,4.5);
\node[red] at (1.75,3.8) {{\small $\widehat{a}_{r-1,1}$}};
\draw[thick,<->,red] (2.5,6.6) -- (1.6,7.5);
\node[red] at (1.65,6.9) {{\small $\widehat{a}_{r+2,N}$}};
\draw[thick,<->,red] (3.5,5.6) -- (2.6,6.5);
\node[red,rotate=-45] at (3.2,6.2) {{\fontsize{10}{50}{ $\widehat{a}_{r,1}$}}};
%%%%%%%%%%%%%%%%%%%%%%%%%%%%%%%%%%%%%%%%%%%
%%%%%%%%%%%%%%%%%%%%%%%%%%%%%%%%%%%%%%%%%%%
%D-line
\draw[thick,<->] (6,15) -- (10,15);
\node at (8,15.2) {{\small $D$}};
%D-line
\draw[thick,<->] (16.5,0) -- (16.5,6);
\node at (16.7,3) {{\small $E$}};
%rho-line
\draw[thick,<->] (0,-1.5) -- (16,-1.5);
\node at (8,-1.7) {{\small $\rho$}};
%tau-line
\draw[thick,<->] (19.7,6) -- (19.7,16);
\node at (19.9,11) {{\small $\tau$}};
%F-line
\draw[dashed] (17,15) -- (-0.25,-2.25);
\draw[dashed] (3,10) -- (-4.75,2.25);
\draw[thick,<->] (-0.3,-2.2) -- (-4.7,2.2);
\node at (-2.25,0.25) {{\small $F$}};
%L-line
\draw[dashed] (4,12) -- (-4,4);
\draw[thick,<->] (-3.95,3.95) -- (-0.05,0.05);
\node at (-1.8,2.2) {{\small $L$}};
%%%%%%%%%%%%%%%%%%%%%%%%%%%%%%%%%%%%%%%%%%%
%%%%%%%%%%%%%%%%%%%%%%%%%%%%%%%%%%%%%%%%%%%
%%%%%%%%%%%%%%%%%%%%%%%%%%%%%%%%%%%%%%%%%%%
%highlight
\draw[line width=1.25mm,orange,rounded corners=10mm] (7.6,6.6) rectangle (10.4,9.4);
\draw[line width=1.25mm,orange] (7.785,9.215) to [out=115,in=305] (5.2,11.5) to [out=120,in=330] (3.75,12.7);
\draw[line width=1.25mm,orange,rounded corners=10mm] (4,12.5) rectangle (-1,17.5);
\draw[line width=0.9mm] (-0.775,12.725) -- (0,13.5) -- (1.5,13.5) -- (3,15) -- (4,15);
\draw[line width=0.9mm] (-1,15) -- (0,15) -- (1.5,16.5) -- (3,16.5) -- (3.775,17.275);
\draw[line width=0.9mm] (1.5,12.5) -- (1.5,13.5);
\draw[line width=0.9mm] (1.5,16.5) -- (1.5,17.5);
\draw[line width=0.9mm] (0,13.5) -- (0,15);
\draw[line width=0.9mm] (3,15) -- (3,16.5);
\node at (0.75,13.25) {{\fontsize{15}{50}{\bf $h_{rN+s}$}}};
\node[rotate=90] at (-0.25,14.15) {{\fontsize{15}{50}{\bf $v_{(r+1)N+s}$}}};
\node[rotate=45] at (0.6,16) {{\fontsize{15}{50}{\bf $m_{(r+1)N+s}$}}};
\node at (2.3,16.75) {{\fontsize{15}{50}{\bf $h_{(r+1)N+s}$}}};
\node[rotate=45] at (2.5,14.1) {{\fontsize{15}{50}{\bf $m_{rN+s+1}$}}};
\node[rotate=70] at (3.5,16) {{\fontsize{15}{50}{\bf $v_{(r+1)N+s+1}$}}};
\draw[green!50!black,line width=0.8mm,<->] (1.5,13.65) -- (1.5,16.35);
\node[green!50!black] at (1.85,14.5) {{\fontsize{15}{50}{\bf $\widehat{c}_{r,s}$}}};
\draw[blue,line width=0.8mm,<->] (0.15,15) -- (2.85,15);
\node[blue] at (0.85,14.65) {{\fontsize{15}{50}{\bf $\widehat{b}_{r,s}$}}};
\draw[red,line width=0.8mm,<->] (1.05,15.9) -- (2.45,14.6);
\node[red] at (1.95,15.5) {{\fontsize{15}{50}{\bf $\widehat{a}_{r,s}$}}};
%%%%%%%%%%%%%%%%%%%%%%%%%%%%%%%%%%%%%%%%%%
%%%%%%%%%%%%%%%%%%%%%%%%%%%%%%%%%%%%%%%%%%
\end{tikzpicture}}}}
\caption{\sl Three different maximally independent sets of K\"ahler parameters for a generic toric web $(N,M)$. For concreteness we assume $N\geq M$. Furthermore, for the sets $\widehat{a}$, $\widehat{b}$ and $\widehat{c}$ (which will constitute the roots in the three different gauge theory descriptions), we have only shown the first few explicitly in the diagram, along with an assignment for a generic hexagon in the web. The latter is labelled by two integers $(r,s)$ whose range is specified in eq.~(\ref{rsParameters}).}
\label{Fig:WebDiagGenericBasis}
\end{center}
\end{figure}
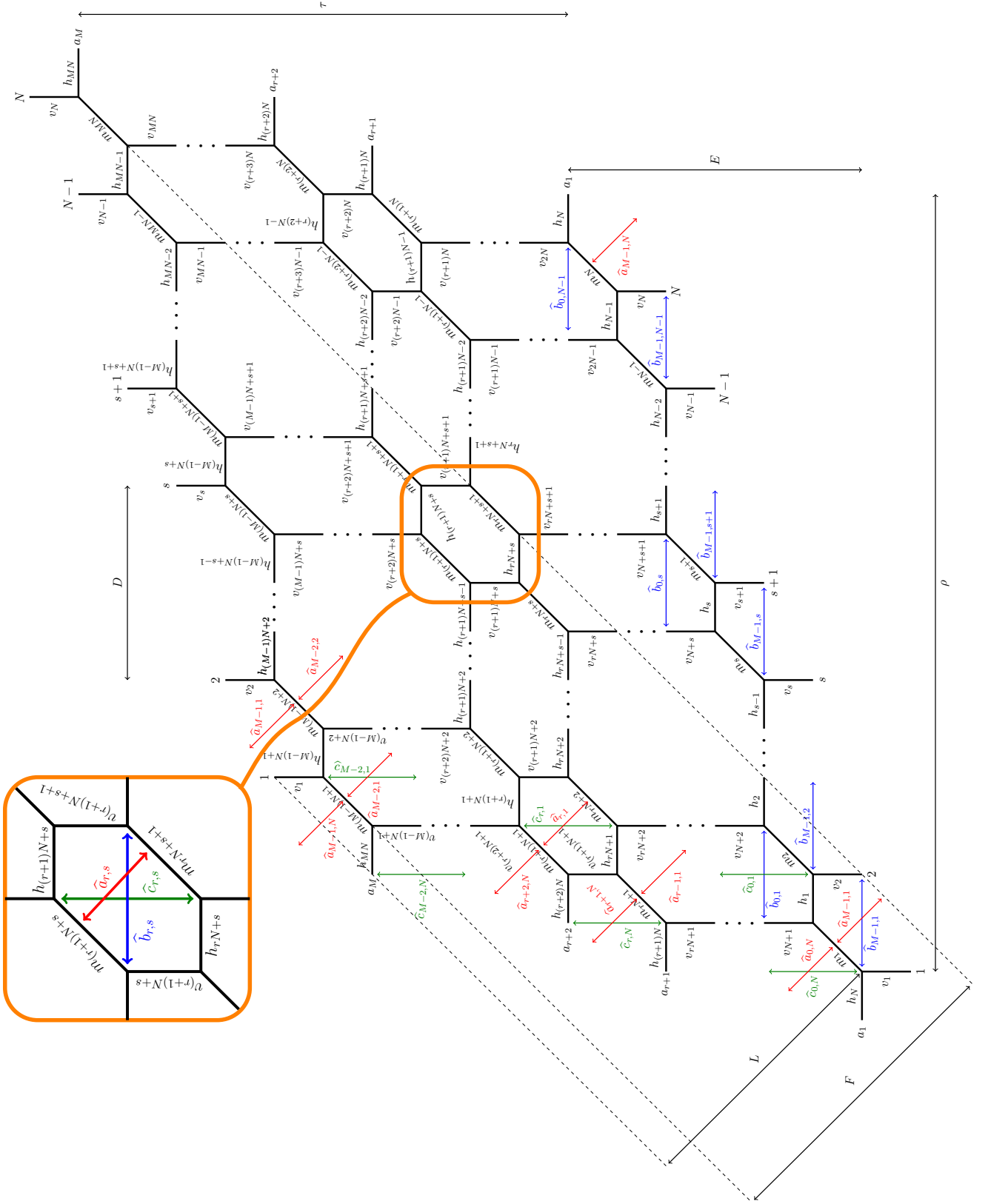

The geometric interpretation of (some of) the parameters in the bases is shown in \figref{Fig:WebDiagGenericBasis}.
The orange box in \figref{Fig:WebDiagGenericBasis}) highlights a generic hexagon in the $(N,M)$ web-diagram, which can be labelled by two integers 
\begin{align}
&r\in \{0,1\ldots, M-1\}\text{ mod }M\,,&&\text{and} &&s\in\{1,\ldots,N\}\text{ mod }N\,.\label{rsParameters}
\end{align}
With the parameters shown in \figref{Fig:WebDiagGenericBasis}, we propose the following three (inequivalent) bases
\begin{itemize}
\item \underline{horizontal basis}\\
We propose as a basis suitable for the description of the horizontal expansion $Z_{\text{hor}}^{(N,M)}$ in (\ref{TrialitySchemat}) the following
\begin{align}
&\mathfrak{B}_{\text{hor}}=\left\{\widehat{b}_{M-1,s=1,\ldots,N-1}\,,\rho\,,\{\widehat{c}\}_{u=1,\ldots,N}\,,\tau\,,E\right\}\,,
\end{align}
with
\begin{align}
&\{\widehat{c}\}_u=\{\widehat{c}_{r,u}|r\in \{0,1,\ldots,M-2\}\}\,,&&\forall u=1,\ldots,N\,.\label{RootsHorizontalNonAffine}
\end{align}
This basis indeed suggests that $Z_{\text{hor}}^{(N,M)}$ is the instanton partition function of gauge theory with gauge group $G_{\text{hor}}=[U(M)]^N$: indeed $\{c\}_{u=1,\ldots,N}$ furnish $N$ sets of (simple positive) roots for the $N$ factors of $U(M)$, while the $N$ decoupling parameters,
\begin{align}
\left\{\widehat{b}_{M-1,1},\widehat{b}_{M-1,2},\ldots,\widehat{b}_{M-1,N-1}\,,\rho-\sum_{i=1}^{N-1}\widehat{b}_{M-1i}\right\}\label{HorizontalGaugeCoupling}
\end{align}
are related with the gauge coupling constants as mentioned in Eq. (\ref{DecouplingDef}) (one associated with every factor of $U(M)$ in $G_{\text{hor}}$), in the sense\footnote{Notice, depending on the explicit realisation of the gauge theory, the coupling constants can be shifted by some of the remaining parameters as long as they lead to the same weak coupling theory. We will be more specific on this point later on, when we discuss explicit examples.} that in the limit
\begin{align}
&\rho-\sum_{i=1}^{N-1}\widehat{b}_{M-1,i}\longrightarrow \infty\,,&&\text{and} &&\widehat{b}_{M-1,i}\longrightarrow\infty&&\forall i=1,\ldots,N-1\,,
\end{align}
we have $h_{1,\ldots,NM}\longrightarrow \infty$, while $\{v_{1,\ldots,NM},m_{1,\ldots,NM}\}$ in the diagram in \figref{Fig:WebDiagGenericBasis} remain finite. Graphically, the diagram therefore composes into $N$ vertical strips of length $M$, each of which associated with the theory corresponding to a single $U(M)$. The expansion of the partition function $Z_{\text{hor}}^{(N,M)}$ (schematically written in (\ref{TrialitySchemat})) can therefore be more concisely be written as an instanton expansion in (\ref{HorizontalGaugeCoupling}).

Finally, the parameter $\tau$ extends each of the algebras $\mathfrak{a}_{M-1}$ (whose roots are given in (\ref{RootsHorizontalNonAffine})) to affine $\widehat{\mathfrak{a}}_{M-1}$.
%%%%%%%%%%%%%%%%%%%%%%%%%%%%%%%%%%%%
%%%%%%%%%%%%%%%%%%%%%%%%%%%%%%%%%%%%
\item \underline{vertical basis}\\
A basis suitable for describing the vertical expansion $Z_{\text{vert}}^{(N,M)}$ in (\ref{TrialitySchemat})  can be found through a judicious exchange of vertical and horizontal parameters of the horizontal basis. Indeed, we propose the vertical basis to be
\begin{align}
\mathfrak{B}_{\text{vert}}=\left\{\widehat{c}_{r=0,\ldots, M-2,N}\,,\tau\,,\{\widehat{b}\}_{u=0,\ldots,M-1}\,,\rho\,,D\right\}\,,
\end{align}
with
\begin{align}
&\{\widehat{b}\}_u=\left\{\widehat{b}_{u,s}|s\in\{1,\ldots,N-1\}\right\}\,,&&\forall\, u=0,\ldots,M-1\,,\label{RootsVerticalNonAffine}
\end{align}
which suggests that $Z_{\text{vert}}^{(N,M)}$ can be interpreted as the instanton partition function of a gauge theory with gauge group $G_{\text{vert}}=[U(N)]^M$. Specifically $\{b\}_{u=0,\ldots,M-1}$ furnish $M$ sets of (simple positive) roots for the $M$ different factors of $U(N)$. Moreover, the $M$ parameters
\begin{align}
\left\{\widehat{c}_{0,N},\ldots,\widehat{c}_{M-2,N}\,,\tau-\sum_{i=0}^{M-2}\widehat{c}_{i,N}\right\}\label{VerticalGaugeCoupling}
\end{align}
are the decoupling parameters (Eq. (\ref{DecouplingDef})) which are associated with the gauge coupling constants (one associated with each of the $M$ factors $U(N)$) in the sense that in the limit
\begin{align}
&\tau-\sum_{i=0}^{M-2}\widehat{c}_{i,N}\longrightarrow \infty\,,&&\text{and} &&\widehat{c}_{i,N}\longrightarrow\infty&&\forall i=0,\ldots,M-2\,,
\end{align}
we have $v_{1,\ldots,NM}$ while $\{h_{1,\ldots,NM},m_{1,\ldots,NM}\}$ remain finite. Thus, the diagram in \figref{Fig:WebDiagGenericBasis} therefore decomposes into $M$ horizontal strips of length $N$ each of which begin associated with the theory corresponding to a single $U(N)$. The series expansion $Z_{\text{vert}}^{(N,M)}$ (which is schematically given in (\ref{TrialitySchemat})) can therefore be more concisely be written as an instanton expansion in (\ref{VerticalGaugeCoupling}).

Finally, the parameter $\rho$ extends each of algebras $\mathfrak{a}_{N-1}$ (whose roots are given in (\ref{RootsVerticalNonAffine})) to affine $\widehat{\mathfrak{a}}_{N-1}$.
%%%%%%%%%%%%%%%%%%%%%%%%%%%%%%%%%%%%%%%%
\item \underline{diagonal basis}\\
The diagonal expansion is somewhat more involved to describe. Indeed, we propose the following $NM+2$ parameters as a basis (with $k=\text{gcd}(N,M)$) for the diagonal expansion
\begin{align}
&\mathfrak{B}_{\text{diag}}=\left\{V_1,\ldots,V_k\,,\{\widehat{a}\}_{u=0,\ldots,k-1}\,,L\,,F\right\}\,,\label{BasisDiagonal}
\end{align}
with
\begin{align}
&\{\widehat{a}\}_u=\left\{\widehat{a}_{M-1-a-u,N+a}|a\in\{0,\ldots,\tfrac{MN}{k}-2\}\right\}\,,\label{RootsDiagonalNonAffine}
\end{align}
which suggest that $Z_{\text{diag}}^{(N,M)}$ in (\ref{TrialitySchemat}) is the instanton partition function of a gauge theory with gauge group $G_{\text{diag}}=\left[U(NM/k)\right]^k$.
In (\ref{BasisDiagonal}), the parameters $V_{1,\ldots,k}$ are difficult to directly identify in the web-diagram in
\figref{Fig:WebDiagGenericBasis}. They can, however, be written as a linear combination of $(\mathbf{h},\mathbf{v},\mathbf{m})$. To this end we introduce a similar notation as in \cite{Hohenegger:2016yuv}: for any diagonal line of area $m_a$ (with $a=1,\ldots,NM$) stretched between two vertices $A$ and $B$
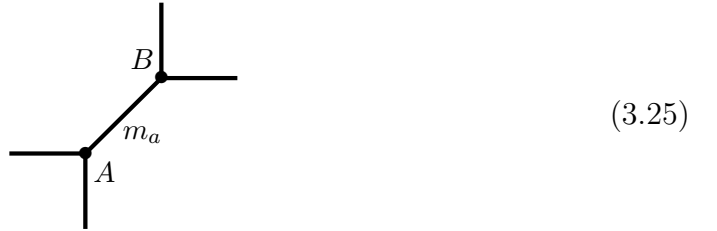
\begin{align}
\parbox{2cm}{\begin{tikzpicture}
\draw[ultra thick] (-1,0) -- (0,0) -- (1,1) -- (2,1);
\draw[ultra thick] (0,-1) -- (0,0);
\draw[ultra thick] (1,1) -- (1,2);
\node at (0,0) {$\bullet$};
\node at (0.25,-0.25) {{\small $A$}};
\node at (1,1) {$\bullet$};
\node at (0.75,1.25) {{\small $B$}};
\node at (0.75,0.25) {{\small $m_a$}};
\end{tikzpicture}}
\end{align}
we define $\mathcal{P}_{L}(m_a)$ as the path starting at $A$ and following $N$ distinct horizontal and $N-1$ distinct vertical lines (going to the left), as well as $\mathcal{P}_{R}(m_a)$ the path starting at $B$ and following $N$ distinct horizontal and $N-1$ distinct vertical lines (going to the right). Furthermore, we denote $(\mathfrak{p}_{L}(m_a))_{i=1,\ldots,2N-1}$ and $(\mathfrak{p}_{R}(m_a))_{i=1,\ldots,2N-1}$ as the components of $\mathcal{P}_{L}(m_a)$ and $\mathcal{P}_{R}(m_a)$ respectively.\footnote{We refer the reader to section 5 of \cite{Hohenegger:2016yuv}, pointing out, however, that in the latter work $N<M$ had been assumed such that the roles of the horizontal and vertical lines have been exchanged.} With this notation, we define\footnote{While the definition (\ref{DefGaugeParaDiag}) is very abstract, it is inspired by the definition of the gauge coupling constants of the vertical expansion associated with the dual Calabi-Yau $X_{\frac{NM}{k},k}$ as explained in~\cite{Hohenegger:2016yuv}.} the decoupling parameters (Eq. (\ref{DecouplingDef})) ($a=0,\ldots,k$)
\begin{align}
V_{a+1}=m_{1+aN}&+\left(\frac{N}{k}-1\right)\left((\mathfrak{p}_{L}(m_{1+aN}))_1+(\mathfrak{p}_{R}(m_{1+aN}))_1\right)\nonumber\\
&+\sum_{i=1}^{\frac{N}{k}-2}\left(\frac{N}{k}-1-i\right)\left[(\mathfrak{p}_{L}(m_{1+aN}))_{2i}+(\mathfrak{p}_{R}(m_{1+aN}))_{2i}\right]\nonumber\\
&+\sum_{i=1}^{\frac{N}{k}-2}\left(\frac{N}{k}-1-i\right)\left[(\mathfrak{p}_{L}(m_{1+aN}))_{2i+1}+(\mathfrak{p}_{R}(m_{1+aN}))_{2i+1}\right]\,.\label{DefGaugeParaDiag}
\end{align}
Indeed, for $V_{1,\ldots,k}\longrightarrow \infty$ we have $m_{1,\ldots,NM}\longrightarrow\infty$, while $(h_{1,\ldots,MN},v_{1,\ldots,MN})$ remain finite. In this way, the $(N,M)$ web-diagram decomposes into $k$ diagonal strips of length $\tfrac{NM}{k}$, which can be interpreted as the weak coupling limit of a quiver gauge theory whose gauge group is $G_{\text{diag}}=[U(\tfrac{NM}{k})]^k$.\footnote{Each strip can be associated with an individual $U(\tfrac{NM}{k})\subset G_{\text{diag}}$.}  The existence of this theory outside of the weak coupling limit can be argued by the fact that $X_{N,M}$ is dual to $X_{NM/k,k}$ through a combination of flop- and symmetry transformations proposed in~\cite{Hohenegger:2016yuv}. Throughout this series of transformations, the diagonal lines (labelled by $m_{1,\ldots NM}$) do not undergo flop transitions, such that the $V_{1,\ldots, k}$ are related to the coupling constants of the $[U(\tfrac{NM}{k})]^k$ quiver gauge theory furnished by the vertical expansion of $\pf{NM/k}{k}$. Moreover, due to the fact that the partition function is expected to be invariant under the duality proposed in \cite{Hohenegger:2016yuv} (this was explicitly proven for $k=1$ in \cite{Bastian:2017ing}), we propose that the expansion of $\pf{N}{M}$ in powers of $Q_{V_a}=e^{-V_a}$ (for $a=1,\ldots,k$) can also be interpreted as the instanton partition function of a quiver gauge theory with gauge group $[U(\tfrac{NM}{k})]^k$. From this perspective, the $\{\widehat{a}\}_{u=0,\ldots,k-1}$ in (\ref{RootsDiagonalNonAffine}) furnish $k$ sets of (simple positive) roots, each associated with a factor $U(\tfrac{NM}{k})\subset G_{\text{diag}}$. 

Finally, the parameter $L$ extends each of algebras $\mathfrak{a}_{NM/k-1}$ (whose roots are given in (\ref{RootsDiagonalNonAffine})) to affine $\widehat{\mathfrak{a}}_{NM/k-1}$.
\end{itemize}

\noindent
To summarise, based on the proposed bases $\mathfrak{B}_{\text{hor}}$, $\mathfrak{B}_{\text{vert}}$ and $\mathfrak{B}_{\text{diag}}$ (as well as the examples discussed above) we conjecture that for given $(N,M)$ we can engineer three different gauge theories
\begin{itemize}
\item horizontal gauge theory with gauge group $G_{\text{hor}}=[U(M)]^N$
\item vertical gauge theory with gauge group $G_{\text{vert}}=[U(N)]^M$
\item diagonal gauge theory with gauge group $G_{\text{diag}}=[U(NM/k)]^k$ with $k=\text{gcd}(N,M)$
\end{itemize}
whose gauge groups have the same rank. Moreover, since the partition functions of these three theories are identical (indeed, by construction they are simply different expansions of $\pf{N}{M}$, namely $\mathcal{Z}^{(N,M)}_{\text{hor}}$, $\mathcal{Z}^{(N,M)}_{\text{vert}}$ and $\mathcal{Z}^{(N,M)}_{\text{diag}}$ respectively) they are mutually dual to each other leading to the triality
\begin{align}
&G_{\text{hor}} = [U(M)]^N && \longleftrightarrow && G_{\text{vert}} = [U(N)]^M &&\longleftrightarrow &&
G_{\text{diag}} = [U(MN/k)]^k\,.\label{TrialityRel}
\end{align}
Notice that this duality is not limited to the weak coupling limit: by decomposing the web diagram in parallel strips (whose explicit form was recently computed in \cite{Bastian:2017ing} in full generality), the triality can be extended to the full non-perturbative partition function.
%%%%%%%%%%%%%%%%%%%%%%%%%%%%%%%%%%%%%%%%%%%%%%%%%%%%

\section{Triality and W-algebras}
In this section we briefly discuss the implication of the triality of gauge theories, that we discussed in the previous sections, for the W-algebras. Recall that the 
six dimensional $(2,0)$ theory can be compactified on a Riemann surface $\Sigma$  with punctures to obtain the four dimensional class ${\cal S}$ theories \cite{Gaiotto}. Since the $A_{N-1}$ type six dimensional $(2,0)$ theory can be realized on the world-volume of the $N$ coincident M5-branes therefore we can take the world-volume to be,
\bea
\mathbb{R}_{\epsilon_{1},\epsilon_2}^{4}\times \Sigma\,,
\eea
to obtain an $\Omega$ deformed four dimensional gauge theory whose details such as the gauge group and the matter content depend on the details of $\Sigma$ (its punctures and its pair of pants decomposition). In \cite{AGT} it was conjectured (the AGT conjecture) that the compactification of the $A_{1}$ $(2,0)$ theory on the Riemann surface $\Sigma$ gives rise to Liouville theory such that  the instanton partition function of the four dimensional theory can be obtained from the conformal blocks of the two dimensional Liouville theory. The conjecture was extended to the case of $A_{N-1}$ $(2,0)$ six dimensional theory in which case the two dimensional theory on the Riemann surface is the $A_{N-1}$ Toda theory \cite{AGT2}. It was shown in \cite{Dijkgraaf:2009pc} that the AGT conjecture and its extension to the $A_{N-1}$ case can be understood using B-model topological strings.

The two dimensional theory on the Riemann surface has the Virasoro symmetry for the $N=2$ case which generalizes to $W_{N}$ for the general $A_{N-1}$. In lifting the four dimensional instanton counting to five dimensions, with ${\cal N}=1$ theory on $\mathbb{R}^{4}\times S^1$, a q-deformation is introduced which corresponds to considering the K-theory of instanton moduli spaces rather than the cohomology. It was shown in \cite{Awata:2009ur,Schiappa:2009cc, Awata:2010yy} that the q-deformed instanton counting or the K-theoretic instanton partition functions \cite{Nekrasov:2002qd,Nakajima:2005fg} satisfy the q-deformed $W_{N}$ constraints. More recently it was shown in \cite{Kimura:2016ebq} that for a quiver gauge theory with quiver $\Gamma$ the partition function can be written as a correlation function in the free field representation of the $W(\Gamma)$ algebra. Lifting it further to six dimensions gives an elliptic deformation of the instanton counting which can be related to the elliptic deformation of the Virasoro and the $W_{N}$ algebras \cite{Tan1, Iqbal:2015fvd,Nieri:2015dts,Tan2, Kimura:2016dys}.

There was a parallel development on the mathematics side in the study of geometry of instanton moduli spaces which was greatly helped by the AGT conjecture. Maulik-Okounkov \cite{math1} and Schiffmann-Vesserot \cite{math2} proved that for the instanton moduli space of type $G$, $M_{G}$, one can construct an action of the W-algebra of type $G$ on the cohomology of $M_{G}$ such that the unit cohomology class is related to the Gaiotto state and its pairing with itself gives the Nekrasov instanton partition function. It was shown in \cite{Braverman:2014xca,Aganagic:2017smx} that the construction of Maulik-Okounkov and Schiffmann-Vesserot can be realized for the K-theory of the instanton moduli spaces giving the analog of the AGT conjecture for the five dimensional ${\cal N}=1$ theories.

In \cite{Nekrasov:2012xe} ${\cal N}=2$ the Seiberg-Witten geometry of the four dimensional gauge theories with gauge group given by a quiver were studied. As was discussed in \cite{Katz:1997eq} some of these
theories, depending on the quiver and the gauge group factors associated with the nodes of the quiver, can be realized geometrically
in terms of Calabi-Yau threefolds and are dual to other gauge theories via the fiber-base duality. It was argued in \cite{Bao:2011rc,Aganagic:2013tta,Aganagic:2015cta} that
because of the fiber base duality two different W-algebras can be associated with the gauge theory. If we denote the gauge group of the two dual theories by $G_1$ and $G_2$ respectively then both theories realize $W_{G_1}$ and $W_{G_2}$. For theories coming from six dimensional theories on $T^2$ the W-algebra is deformed to the elliptic W-algebras \cite{Kimura:2016dys}. The triality we discussed in the previous sections extends the class of (elliptic) W-algebra associated with a theory by identifying new dual theories. From the results of the previous sections it follows that if we consider the $(N,M)$ web then associated to it are several dual gauge theories and hence several  elliptic W-algebras. Since these theories have the same partition function ${\cal Z}_{N,M}$, which is the topological string partition function of $X_{N,M}$, with gauge theory parameters associated with K\"ahler parameters of $X_{N,M}$ in different ways therefore the identification of generators of different elliptic W-algebras is in terms of distinct parameters \cite{paper3}.

\begin{figure}
\begin{center}
\scalebox{0.58}{\parbox{20.5cm}{\begin{tikzpicture}[scale = 1.5]
\%second row
\draw[ultra thick] (14,4) -- (16,5);
\draw[ultra thick] (12,2) -- (12,0.5);
\draw[ultra thick] (10,4) -- (8,5);
%\draw[ultra thick] (11,2) -- (10,1);

%dots for the cutting
\draw[ultra thick, red] (8.5,3.8) -- (15.5,3.8) -- (15.5,2.2) -- (8.5,2.2) -- (8.5,3.8); 
\node[rotate=0] at (12,3) {\parbox{10cm}{\Huge Toric Calabi-Yau threefold $X_{N,M}$ with $(N,M)$ web}};
\node[rotate=0] at (16.7,5.5) {\Huge ($U(N)^M$, $W^{\text{ell}}_{A_{N-1}}$)};
\node[rotate=0] at (7.5,5.5) {\Huge ($U(M)^N$, $W^{\text{ell}}_{A_{M-1}}$)};
\node[rotate=0] at (12,-0.5) {\Huge ($U(\frac{MN}{k})^k$, $W^{\text{ell}}_{A_{\frac{NM}{k}-1}}$)};
%\node[rotate=0] at (14.7,0.3) {\Huge ($U(k)^{\frac{NM}{k}}$, $W^{ell}_{A_{k-1}}$)};
\end{tikzpicture}}}
\end{center} 
\caption{\sl The gauge groups and W-algebras associated to the gauge theories coming from the $(N,M)$ web. There are potentially more branches of this tree depending on the product $NM$ and the $\mbox{gcd}(N,M)$ as discussed in previous sections.}
\label{Fig:Walgebra}
\end{figure}
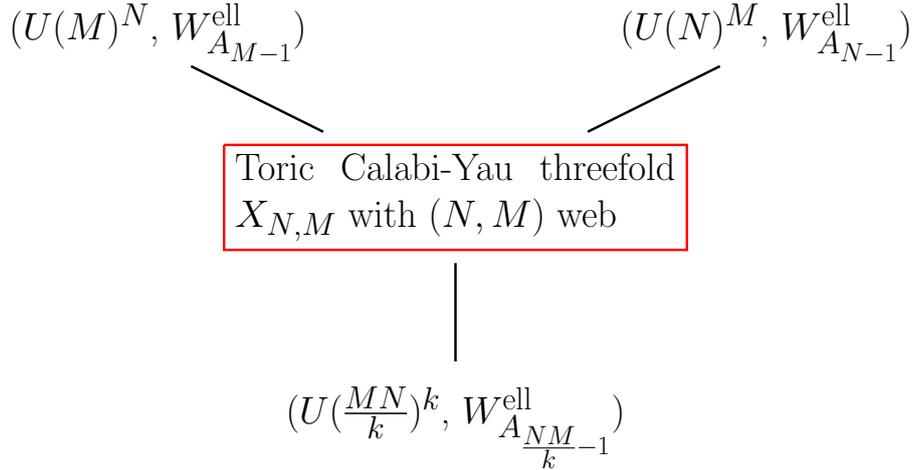

\section{Conclusions}
In this paper, we have analysed dualities of a class of little string orbifold theories with 8 supercharges as well as the supersymmetric gauge theories given by their low-energy limit below the string tension scale. Among other approaches, the (non-perturbative) gauge theory partition functions are captured by the refined topological string partition function $\pf{N}{M}(\mathbf{h},\mathbf{v},\mathbf{m},\epsilon_{1,2})$ of a class of toric Calabi-Yau threefolds $X_{N,M}$, whose generic web diagram is shown in \figref{Fig:WebToric}. The latter consists of three sets of parallel lines whose associated K\"ahler parameters are denoted collectively by $\{\mathbf{h}\}$, $\{\mathbf{v}\}$ and $\{\mathbf{m}\}$ respectively.\footnote{Of these $3NM$ parameters, however, only $NM+2$ are independent of each other.} We have argued that for a given such toric web there exist three (generically distinct) dual $(p,q)$ 5-brane web descriptions in type II string theory in which each of these three sets of parallel lines is identified with the D5-branes respectively. In each of these three brane configurations, the world-volume theory on the latter can be identified with a quiver gauge theory. Therefore, we find a \emph{triality} of gauge theories associated with a given toric Calabi-Yau manifold $X_{N,M}$, whose gauge groups are 
\begin{align}
&G_{\text{hor}} = [U(M)]^{N}\,,&&\text{and} && G_{\text{vert}} = [U(N)]^{M}\,, &&\text{and} && G_{\text{diag}} = [U(MN/k)]^k\,.\label{GaugeGroupsConclusions}
\end{align}
The duality between the horizontal and vertical theory has been discussed previously in the literature \cite{Hohenegger:2013ala,Haghighat:2013tka,Hohenegger:2015btj} and is a consequence of an $SL(2,\mathbb{Z})$ symmetry of type IIB string theory. However, to give evidence for the full triality (including in particular the diagonal theory) we have argued\footnote{We have supplemented our reasoning by a number of examples in which the existence of these regions was shown explicitly.} for the existence of three different regions in the K\"ahler cone of $X_{N,M}$ in which the toric web diagram decomposes into a number of horizontal, vertical or diagonal strips respectively. From the point of view of the $(p,q)$ five-brane theories, a strip correspond to the NS5-brane with equal number of D5-branes ending on it from either side. Since the distance between the NS5-branes correspond to the inverse gauge coupling the configuration of disconnected strips is therefore identified with the weak coupling limit of the associated world-volume theory. More precisely, for each of these three descriptions, we have identified the K\"ahler parameters of the toric web of $X_{N,M}$ (see (\ref{HorizontalGaugeCoupling}), (\ref{VerticalGaugeCoupling}) and (\ref{DefGaugeParaDiag}) respectively) that govern the distances of the NS5-branes in the brane web picture and which are related to the inverse coupling constants of the respective gauge theories. In turn, expanding $\pf{N}{M}(\mathbf{h},\mathbf{v},\mathbf{m},\epsilon_{1,2})$ as a power series in (the exponential of) these parameters respectively can be interpreted as the Nekrasov instanton partition function of the three gauge theories (\ref{GaugeGroupsConclusions}), as explained in (\ref{TrialitySchemat}). From the point of view of the (refined) topological vertex (which is used to compute $\pf{N}{M}(\mathbf{h},\mathbf{v},\mathbf{m},\epsilon_{1,2})$) these three different expansions correspond to choosing a preferred direction in the web diagram (which can either be horizontally, vertically or diagonally). The diagonal expansion has recently been discussed in \cite{Bastian:2017ing}, where moreover a generic building block has been computed that allows to determine these three expansions of $\pf{N}{M}(\mathbf{h},\mathbf{v},\mathbf{m},\epsilon_{1,2})$ explicitly for generic $(N,M)$.

For the sake of clarity, we provide an overview over the three different gauge theories (along with an overview of which K\"ahler parameters of $X_{N,M}$ govern the coupling constants and constitute the Coulomb branches and hypermultiplet masses respectively) in the following table
%%%%%%%
\begin{center}
\begin{tabular}{c|c|c|c|c|c}
{\bf theory} & {\bf partition fct.} & {\bf gauge gr.} & {\bf decoup. para.} & {\bf Coul. br.} & {\bf hyperm. masses} \\\hline
&&&&&\\[-10pt]
horizontal & $Z_{\text{hor}}^{(N,M)}$ & $[U(M)]^N$ & eq.~(\ref{HorizontalGaugeCoupling}) & $\{\mathbf{v}\}$ & $\{\mathbf{m}\}$\\[10pt]
%%%
&&&&&\\[-10pt]
vertical & $Z_{\text{vert}}^{(N,M)}$ & $[U(N)]^M$ & eq.~(\ref{VerticalGaugeCoupling}) & $\{\mathbf{h}\}$ & $\{\mathbf{m}\}$\\[10pt]
%%%
&&&&&\\[-10pt]
diagonal & $Z_{\text{diag}}^{(N,M)}$ & $\left[U\left(\tfrac{NM}{k}\right)\right]^k$ & eq.~(\ref{DefGaugeParaDiag}) & $\{\mathbf{h},\mathbf{v}\}$ & $\{\mathbf{h},\mathbf{v}\}$\\[10pt]
\end{tabular}
\end{center}

\noindent
From the point of view of the gauge theories, however, the above table in general does not exhaust all possible dual theories. As will be discussed in \cite{paper3} a larger set of dual theories can be found by studying the \emph{extended moduli space} associated with $X_{N,M}$.

It would be interesting to extend the web of dualities we have found to other classes of little string theories (or supersymmetric gauge theories). The theories we have studied in this paper can be geometrically realised as resolutions of a $\mathbb{Z}_N\times \mathbb{Z}_M$ orbifold of $X_{1,1}$, which is a toric Calabi-Yau manifold that resembles the conifold in a certain decompactification limit. The dualities we have studied here, are based on the fact that different superconformal gauge theories can be associated with different resolutions of these A-type singularities, which, however, all lie in a common (extended) moduli space. They can therefore be related through various symmetry transformations or geometric transitions. It is an interesting question, whether similar relations can be obtained for different types of orbifolds, where $\mathbb{Z}_N\times \mathbb{Z}_M$ is replaced by $G_1\times G_2$, with $G_i$ some other discrete subgroup of $SU(2)$.

It would also be interesting to compactify the theories studied in this work to four-dimensional ${\cal N}=2$ supersymmetric gauge theories. As long as the compactification keeps the Coulomb branch compact, these theories would also exhibit an extensive web of dualities: the duality between the theories with $[U(M)]^N$ and $[U(N)]^M$ gauge group is known to correspond to exchange of quiver group and gauge group. The extended duality that encompasses $[U(MN/k)]^k$ is to our knowledge new.

%%%%%%%%%%%%%%%%%%%%%%%%%%%%%%%%%%%%%

\section*{Acknowledgement}
We are grateful to Dongsu Bak, Kang-Sin Choi, Taro Kimura, Jeong-Hyuck Park, Washington Taylor and the participants of the Focus Program ``Liouville, Integrability and Branes (13)"  of the Asia-Pacific Center for Theoretical Physics for useful discussions. A.I. would like to acknowledge the ``2017 Simons Summer Workshop on Mathematics and Physics" for hospitality during this work. A.I. was supported in part by the Higher Education Commission grant HEC-20-2518.

%%%%%%%%%%%%%%%%%%%%%%%%%%%%%%
\appendix
\section{Special Cases: Self-Triality and SCFT Limit}
In sections 3 and 4, we have discussed different ways of associating gauge theories to a generic web-diagram characterised by the parameters $(N,M)$. In general the three gauge theories with gauge groups given in (\ref{TrialityRel}):
\begin{align}
&G_{\text{hor}} = [U(M)]^{N}\,,&&\text{and} && G_{\text{vert}} = [U(N)]^{M}\,, &&\text{and} && G_{\text{diag}} = [U(\tfrac{MN}{k})]^k
\end{align}
are distinct, as can be seen by the fact the gauge groups are typically different. However, for particular web configurations the gauge groups of two or even all three of these gauge theories may become identical and  the latter can be related to one another  through a mere exchange of  the parameters involved. In this sense some of these theories may be self-dual. In the following, we shall list examples of configurations in which this indeed happens.

\subsection{Web-diagrams $(N,N)$}
A class of examples in which horizontal, vertical and diagonal expansion yield gauge theories with the same gauge group (albeit with a different choice of parameters) are the configurations with $M=N$, giving rise to gauge theories with gauge group $[U(N)]^N$. This is obvious for the vertical and horizontal points of view. However it is less obvious for the diagonal point of view. Below, we show that the diagonal description again yields $[U(N)]^N$ quiver gauge theory, but the parameters are different from either the vertical or horizontal description. 

Starting from the $(N,N)$ web diagram, the idea is that there always exists a way to cut $N$ diagonal lines simultaneously, as shown in the figure below
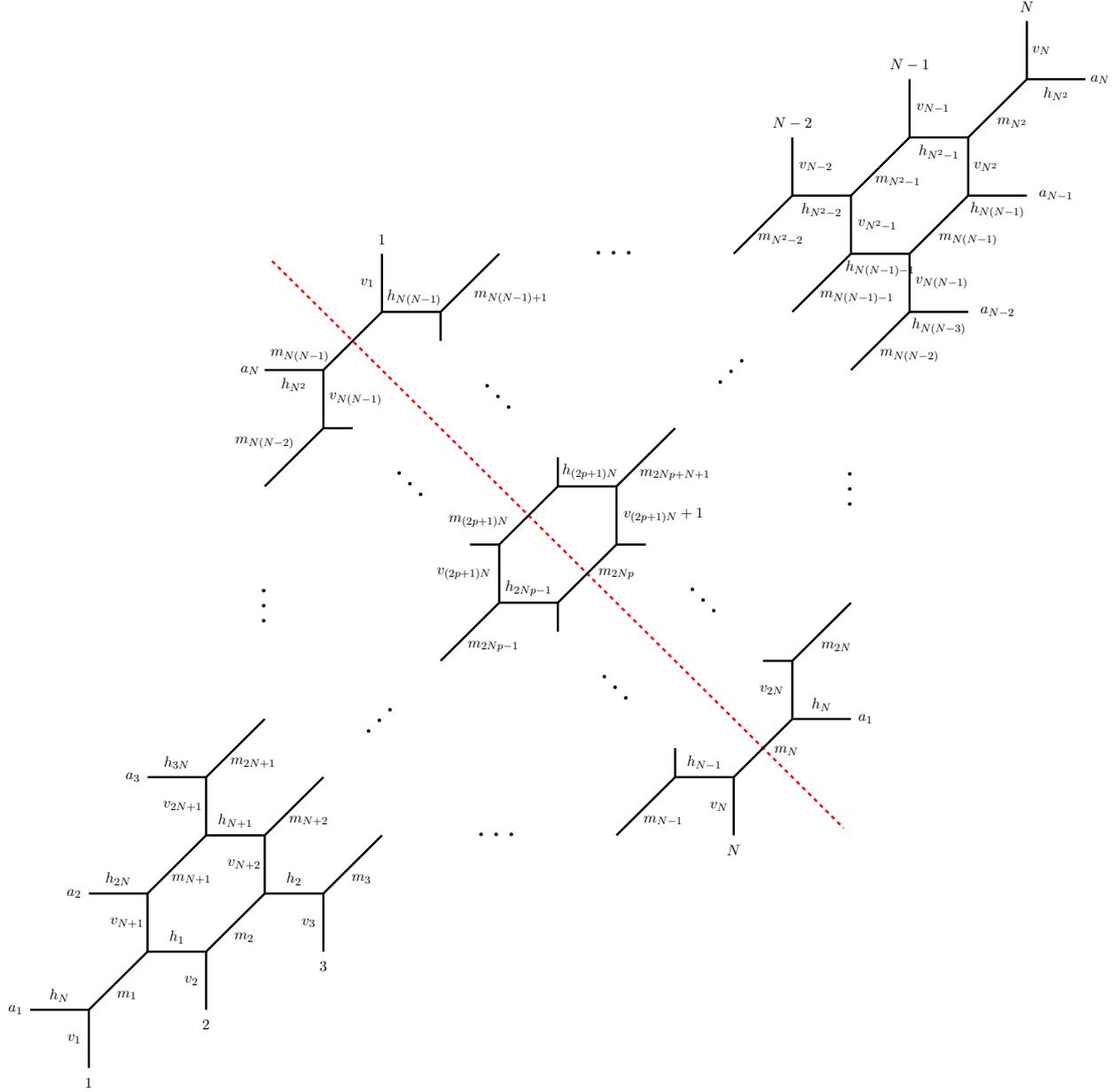
\begin{center}
\scalebox{0.58}{\parbox{28.25cm}{\begin{tikzpicture}[scale = 1.5]
\draw[dashed,ultra thick,red] (3.125,12.875) -- (12.875,3.125);
\draw[ultra thick] (-1,0) -- (0,0) -- (1,1) -- (2,1) -- (3,2) -- (4,2) -- (5,3);
\draw[ultra thick] (0,2) -- (1,2) -- (2,3) -- (3,3) -- (4,4);
\draw[ultra thick] (1,4) -- (2,4) -- (3,5);
%bottom vertical
\draw[ultra thick] (0,-1) -- (0,0);
\draw[ultra thick] (2,0) -- (2,1);
\draw[ultra thick] (4,1) -- (4,2);
%first row
\draw[ultra thick] (1,1) -- (1,2);
\draw[ultra thick] (3,2) -- (3,3);
%second row
\draw[ultra thick] (2,3) -- (2,4);
%dots for the cutting
\node[rotate=90] at (3,7) {\Huge $\cdots$};
\node[rotate=45] at (5,5) {\Huge $\cdots$};
\node[rotate=0] at (7,3) {\Huge $\cdots$};
%Cells to be cut
\draw[ultra thick] (7,8) -- (8,9) -- (9,9) -- (10,10);
\draw[ultra thick] (6,6) -- (7,7) -- (8,7) -- (9,8) -- (9.5,8);
\draw[ultra thick] (8,9) -- (8,9.5);
\draw[ultra thick] (8,6.5) -- (8,7);
\draw[ultra thick] (7,8) -- (6.5,8);
%\draw[ultra thick] (10,7.5) -- (10,8);
\draw[ultra thick] (7,7) -- (7,8);
\draw[ultra thick] (9,8) -- (9,9);
%Cells bottom to be cut
\begin{scope}[xshift=1cm,yshift=-1cm]
\draw[ultra thick] (8,4) -- (9,5) -- (10,5) -- (11,6) -- (12,6);
\draw[ultra thick] (10,5) -- (10,4);
\draw[ultra thick] (11,6) -- (11,7) -- (12,8);
\draw[ultra thick] (11,7) -- (10.5,7);
\draw[ultra thick] (9,5) -- (9,5.5);
\node[rotate=-45] at (8,6.5) {\Huge $\cdots$};
\node[rotate=-45] at (9.5,8) {\Huge $\cdots$};
\end{scope}
%Cells top to be cut
\begin{scope}[xshift=-1cm,yshift=1cm]
\draw[ultra thick] (5.5,9) -- (5,9) -- (5,10) -- (4,10);
\draw[ultra thick] (4,8) -- (5,9);
\draw[ultra thick] (5,10) -- (6,11) -- (7,11) -- (8,12);
\draw[ultra thick] (6,11) -- (6,12);
\draw[ultra thick] (7,11) -- (7,10.5);
\node[rotate=-45] at (8,9.5) {\Huge $\cdots$};
\node[rotate=-45] at (6.5,8) {\Huge $\cdots$};
\end{scope}
\begin{scope}[rotate=180,xshift=-16cm,yshift=-16cm]
\draw[ultra thick] (-1,0) -- (0,0) -- (1,1) -- (2,1) -- (3,2) -- (4,2) -- (5,3);
\draw[ultra thick] (0,2) -- (1,2) -- (2,3) -- (3,3) -- (4,4);
\draw[ultra thick] (1,4) -- (2,4) -- (3,5);
%bottom vertical
\draw[ultra thick] (0,-1) -- (0,0);
\draw[ultra thick] (2,0) -- (2,1);
\draw[ultra thick] (4,1) -- (4,2);
%first row
\draw[ultra thick] (1,1) -- (1,2);
\draw[ultra thick] (3,2) -- (3,3);
%second row
\draw[ultra thick] (2,3) -- (2,4);
\node[rotate=90] at (3,7) {\Huge $\cdots$};
\node[rotate=45] at (5,5) {\Huge $\cdots$};
\node[rotate=0] at (7,3) {\Huge $\cdots$};
\end{scope}
\node at (0,-1.25) {{\bf \small $1$}};
\node at (2,-0.25) {{\bf \small $2$}};
\node at (4,0.75) {{\bf \small $3$}};
\node at (11,2.75) {{\bf \small $N$}};
\node at (5,13.25) {{\bf \small $1$}};
\node at (12,15.25) {{\bf \small $N-2$}};
\node at (14,16.25) {{\bf \small $N-1$}};
\node at (16,17.25) {{\bf \small $N$}};
\node at (-1.25,0) {{\bf \small $a_1$}};
\node at (-0.25,2) {{\bf \small $a_2$}};
\node at (0.75,4) {{\bf \small $a_3$}};
\node at (2.75,11) {{\bf \small $a_N$}};
\node at (13.25,5) {{\bf \small $a_1$}};
\node at (15.5,12) {{\bf \small $a_{N-2}$}};
\node at (16.5,14) {{\bf \small $a_{N-1}$}};
\node at (17.25,16) {{\bf \small $a_{N}$}};
%diagonal bottom
\node at (0.65,0.25) {{\bf \small $m_1$}};
\node at (2.65,1.25) {{\bf \small $m_2$}};
\node at (4.65,2.25) {{\bf \small $m_3$}};
\node at (9.8,3.25) {{\bf \small $m_{N-1}$}};
\node at (11.9,4.45) {{\bf \small $m_{N}$}};
%diagonal second row
\node at (1.75,2.25) {{\bf \small $m_{N+1}$}};
\node at (2.8,4.25) {{\bf \small $m_{2N+1}$}};
\node at (3.75,3.25) {{\bf \small $m_{N+2}$}};
\node at (12.75,6.25) {{\bf \small $m_{2N}$}};
\node at (6.9,6.25) {{\bf \small $m_{2Np-1}$}};
\node at (9,7.5) {{\bf \small $m_{2Np}$}};
\node at (6.65,8.4) {{\bf \small $m_{(2p+1)N}$}};
\node at (10,9.2) {{\bf \small $m_{2Np+N+1}$}};
%
%vertical nodes bottom
\node at (-0.25,-0.5) {{\bf \small $v_1$}};
\node at (1.75,0.5) {{\bf \small $v_2$}};
\node at (3.75,1.5) {{\bf \small $v_3$}};
\node at (10.75,3.5) {{\bf \small $v_N$}};
\node at (0.65,1.5) {{\bf \small $v_{N+1}$}};
\node at (2.65,2.5) {{\bf \small $v_{N+2}$}};
\node at (11.65,5.5) {{\bf \small $v_{2N}$}};
\node at (1.6,3.5) {{\bf \small $v_{2N+1}$}};
\node at (6.4,7.5) {{\bf \small $v_{(2p+1)N}$}};
\node at (9.8,8.5) {{\bf \small $v_{(2p+1)N}+1$}};
%
%horizontal nodes bottom
\node at (-0.5,0.25) {{\bf \small $h_N$}};
\node at (1.5,1.25) {{\bf \small $h_1$}};
\node at (3.5,2.25) {{\bf \small $h_2$}};
\node at (0.5,2.25) {{\bf \small $h_{2N}$}};
\node at (2.5,3.25) {{\bf \small $h_{N+1}$}};
\node at (1.5,4.25) {{\bf \small $h_{3N}$}};
\node at (10.5,4.25) {{\bf \small $h_{N-1}$}};
\node at (12.5,5.25) {{\bf \small $h_{N}$}};
\node at (7.5,7.25) {{\bf \small $h_{2Np-1}$}};
\node at (8.55,9.25) {{\bf \small $h_{(2p+1)N}$}};
%diagonal lines top corner
\node at (15.75,15.25) {{\bf \small $m_{N^2}$}};
\node at (13.8,14.25) {{\bf \small $m_{N^2-1}$}};
\node at (15,13.25) {{\bf \small $m_{N(N-1)}$}};
\node at (13.1,12.25) {{\bf \small $m_{N(N-1)-1}$}};
\node at (11.8,13.25) {{\bf \small $m_{N^2-2}$}};
\node at (14,11.25) {{\bf \small $m_{N(N-2)}$}};
%vertical lines top corner
\node at (16.25,16.5) {{\bf \small $v_{N}$}};
\node at (14.4,15.5) {{\bf \small $v_{N-1}$}};
\node at (12.4,14.5) {{\bf \small $v_{N-2}$}};
\node at (15.3,14.5) {{\bf \small $v_{N^2}$}};
\node at (13.45,13.5) {{\bf \small $v_{N^2-1}$}};
\node at (14.55,12.5) {{\bf \small $v_{N(N-1)}$}};
%horizontal lines top corner
\node at (16.5,15.75) {{\bf \small $h_{N^2}$}};
\node at (15.5,13.75) {{\bf \small $h_{N(N-1)}$}};
\node at (13.5,12.75) {{\bf \small $h_{N(N-1)-1}$}};
\node at (14.5,11.75) {{\bf \small $h_{N(N-3)}$}};
\node at (14.5,14.75) {{\bf \small $h_{N^2-1}$}};
\node at (12.5,13.75) {{\bf \small $h_{N^2-2}$}};
%diagonal lines top
\node at (7.2,12.25) {{\bf \small $m_{N(N-1)+1}$}};
\node at (3.6,11.25) {{\bf \small $m_{N(N-1)}$}};
\node at (3,9.75) {{\bf \small $m_{N(N-2)}$}};
%vertical lines top
\node at (4.55,10.5) {{\bf \small $v_{N(N-1)}$}};
\node at (4.75,12.5) {{\bf \small $v_{1}$}};
%horizontal lines top
\node at (3.5,10.75) {{\bf \small $h_{N^2}$}};
\node at (5.55,12.25) {{\bf \small $h_{N(N-1)}$}};
\end{tikzpicture}}}
\end{center}
where the subscript $p=\lfloor\tfrac{N}{2}\rfloor$. Notice that not all of the $3N^2$ parameters are independent due to the necessity of imposing linear constraints to guarantee the consistency of the web. Solving for these constraints, there are only $N^2+2$ independent parameters.

After cutting along the diagonal red line and re-gluing the diagram along the legs $a_i$ (for $i=1,\ldots, N$), we obtain the following equivalent web (after an appropriate $SL(2,\mathbb{Z})$-transformation that leaves the vertical lines invariant)
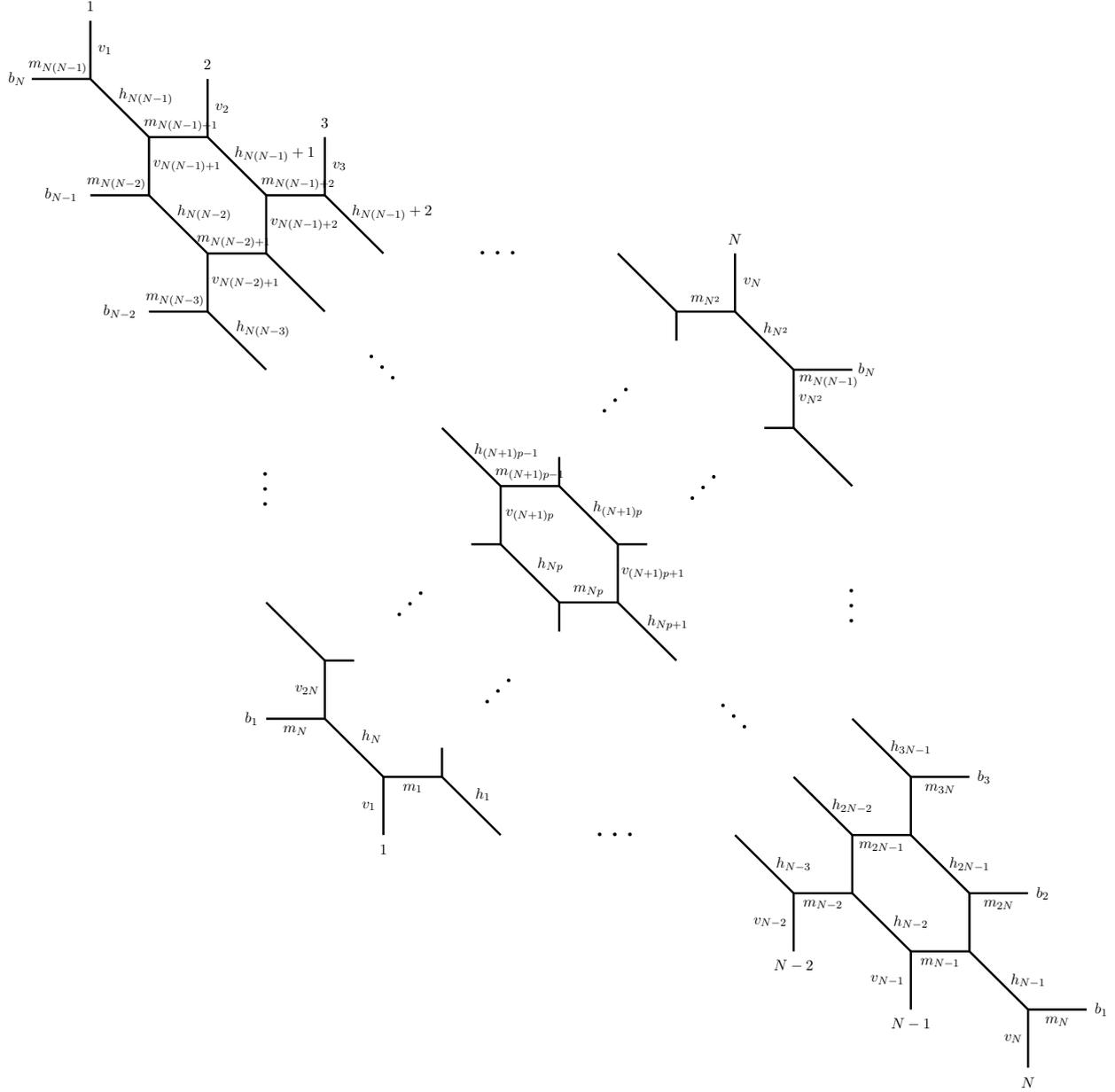
\begin{center}
\scalebox{0.58}{\parbox{28.25cm}{\begin{tikzpicture}[scale = 1.5]
\begin{scope}[rotate=90]
\draw[ultra thick] (-1,0) -- (0,0) -- (1,1) -- (2,1) -- (3,2) -- (4,2) -- (5,3);
\draw[ultra thick] (0,2) -- (1,2) -- (2,3) -- (3,3) -- (4,4);
\draw[ultra thick] (1,4) -- (2,4) -- (3,5);
%bottom vertical
\draw[ultra thick] (0,-1) -- (0,0);
\draw[ultra thick] (2,0) -- (2,1);
\draw[ultra thick] (4,1) -- (4,2);
%first row
\draw[ultra thick] (1,1) -- (1,2);
\draw[ultra thick] (3,2) -- (3,3);
%second row
\draw[ultra thick] (2,3) -- (2,4);
%dots for the cutting
\node[rotate=0] at (3,7) {\Huge $\cdots$};
\node[rotate=-45] at (5,5) {\Huge $\cdots$};
\node[rotate=90] at (7,3) {\Huge $\cdots$};
%Cells to be cut
\draw[ultra thick] (7,8) -- (8,9) -- (9,9) -- (10,10);
\draw[ultra thick] (6,6) -- (7,7) -- (8,7) -- (9,8) -- (9.5,8);
\draw[ultra thick] (8,9) -- (8,9.5);
\draw[ultra thick] (8,6.5) -- (8,7);
\draw[ultra thick] (7,8) -- (6.5,8);
%\draw[ultra thick] (10,7.5) -- (10,8);
\draw[ultra thick] (7,7) -- (7,8);
\draw[ultra thick] (9,8) -- (9,9);
%Cells bottom to be cut
\begin{scope}[xshift=1cm,yshift=-1cm]
\draw[ultra thick] (8,4) -- (9,5) -- (10,5) -- (11,6) -- (12,6);
\draw[ultra thick] (10,5) -- (10,4);
\draw[ultra thick] (11,6) -- (11,7) -- (12,8);
\draw[ultra thick] (11,7) -- (10.5,7);
\draw[ultra thick] (9,5) -- (9,5.5);
\node[rotate=45] at (8,6.5) {\Huge $\cdots$};
\node[rotate=45] at (9.5,8) {\Huge $\cdots$};
\end{scope}
%Cells top to be cut
\begin{scope}[xshift=-1cm,yshift=1cm]
\draw[ultra thick] (5.5,9) -- (5,9) -- (5,10) -- (4,10);
\draw[ultra thick] (4,8) -- (5,9);
\draw[ultra thick] (5,10) -- (6,11) -- (7,11) -- (8,12);
\draw[ultra thick] (6,11) -- (6,12);
\draw[ultra thick] (7,11) -- (7,10.5);
\node[rotate=45] at (8,9.5) {\Huge $\cdots$};
\node[rotate=45] at (6.5,8) {\Huge $\cdots$};
\end{scope}
\begin{scope}[rotate=180,xshift=-16cm,yshift=-16cm]
\draw[ultra thick] (-1,0) -- (0,0) -- (1,1) -- (2,1) -- (3,2) -- (4,2) -- (5,3);
\draw[ultra thick] (0,2) -- (1,2) -- (2,3) -- (3,3) -- (4,4);
\draw[ultra thick] (1,4) -- (2,4) -- (3,5);
%bottom vertical
\draw[ultra thick] (0,-1) -- (0,0);
\draw[ultra thick] (2,0) -- (2,1);
\draw[ultra thick] (4,1) -- (4,2);
%first row
\draw[ultra thick] (1,1) -- (1,2);
\draw[ultra thick] (3,2) -- (3,3);
%second row
\draw[ultra thick] (2,3) -- (2,4);
\node[rotate=0] at (3,7) {\Huge $\cdots$};
\node[rotate=-45] at (5,5) {\Huge $\cdots$};
\node[rotate=90] at (7,3) {\Huge $\cdots$};
\end{scope}
\end{scope}
%vertical labels
\node at (0,-1.25) {{\bf \small $N$}};
\node at (-2,-0.25) {{\bf \small $N-1$}};
\node at (-4,0.75) {{\bf \small $N-2$}};
\node at (-11,2.75) {{\bf \small $1$}};
\node at (-16,17.25) {{\bf \small $1$}};
\node at (-14,16.25) {{\bf \small $2$}};
\node at (-12,15.25) {{\bf \small $3$}};
\node at (-5,13.25) {{\bf \small $N$}};
%horizontal labels
\node at (-17.25,16) {{\bf \small $b_N$}};
\node at (-16.5,14) {{\bf \small $b_{N-1}$}};
\node at (-15.5,12) {{\bf \small $b_{N-2}$}};
\node at (-13.25,5) {{\bf \small $b_{1}$}};
\node at (1.25,0) {{\bf \small $b_{1}$}};
\node at (0.25,2) {{\bf \small $b_{2}$}};
\node at (-0.75,4) {{\bf \small $b_{3}$}};
\node at (-2.75,11) {{\bf \small $b_{N}$}};
%labels bottom left
\node at (-11.25,3.5) {{\small $v_{1}$}};
\node at (-4.4,1.5) {{\small $v_{N-2}$}};
\node at (-2.4,0.5) {{\small $v_{N-1}$}};
\node at (-0.25,-0.5) {{\small $v_{N}$}};
\node at (-12.3,5.5) {{\small $v_{2N}$}};
\node at (-11.2,4.7) {{\small $h_{N}$}};
\node at (-9.3,3.7) {{\small $h_{1}$}};
\node at (-12.5,4.8) {{\small $m_{N}$}};
\node at (-10.5,3.8) {{\small $m_{1}$}};
%labels bottom right
\node at (0,0.5) {{\small $h_{N-1}$}};
\node at (-2,1.5) {{\small $h_{N-2}$}};
\node at (-4,2.5) {{\small $h_{N-3}$}};
\node at (0.5,-0.2) {{\small $m_{N}$}};
\node at (-1.5,0.8) {{\small $m_{N-1}$}};
\node at (-3.5,1.8) {{\small $m_{N-2}$}};
\node at (-0.5,1.8) {{\small $m_{2N}$}};
\node at (-2.5,2.8) {{\small $m_{2N-1}$}};
\node at (-1.5,3.8) {{\small $m_{3N}$}};
\node at (-1,2.5) {{\small $h_{2N-1}$}};
\node at (-3,3.5) {{\small $h_{2N-2}$}};
\node at (-2,4.5) {{\small $h_{3N-1}$}};
%labels center
\node at (-8.9,9.6) {{\small $h_{(N+1)p-1}$}};
\node at (-8.5,9.2) {{\small $m_{(N+1)p-1}$}};
\node at (-7.5,7.2) {{\small $m_{Np}$}};
\node at (-8.5,8.5) {{\small $v_{(N+1)p}$}};
\node at (-6.4,7.5) {{\small $v_{(N+1)p+1}$}};
\node at (-7,8.6) {{\small $h_{(N+1)p}$}};
\node at (-8.15,7.65) {{\small $h_{Np}$}};
\node at (-6.15,6.65) {{\small $h_{Np+1}$}};
%labels top left
\node at (-16.6,16.2) {{ \small $m_{N(N-1)}$}};
\node at (-14.5,15.2) {{ \small $m_{N(N-1)+1}$}};
\node at (-12.5,14.2) {{ \small $m_{N(N-1)+2}$}};
\node at (-15.1,15.7) {{ \small $h_{N(N-1)}$}};
\node at (-14.1,13.7) {{ \small $h_{N(N-2)}$}};
\node at (-15.6,14.2) {{ \small $m_{N(N-2)}$}};
\node at (-13.6,13.2) {{ \small $m_{N(N-2)+1}$}};
\node at (-12.9,14.7) {{ \small $h_{N(N-1)}+1$}};
\node at (-14.6,12.2) {{ \small $m_{N(N-3)}$}};
\node at (-10.9,13.7) {{ \small $h_{N(N-1)}+2$}};
\node at (-13.1,11.7) {{ \small $h_{N(N-3)}$}};
\node at (-15.75,16.5) {{\small $v_1$}};
\node at (-14.35,14.5) {{\small $v_{N(N-1)+1}$}};
\node at (-13.35,12.5) {{\small $v_{N(N-2)+1}$}};
\node at (-13.75,15.5) {{\small $v_2$}};
\node at (-12.35,13.5) {{\small $v_{N(N-1)+2}$}};
\node at (-11.75,14.5) {{\small $v_3$}};
%labels top right
\node at (-3.4,10.8) {{\small $m_{N(N-1)}$}};
\node at (-5.5,12.2) {{\small $m_{N^2}$}};
\node at (-4.7,12.5) {{\small $v_{N}$}};
\node at (-3.7,10.5) {{\small $v_{N^2}$}};
\node at (-4.3,11.7) {{\small $h_{N^2}$}};
\end{tikzpicture}}}
\end{center}
Here, the nodes $b_i$ indicate the gluing of the lines that have been cut in the previous step along the red line. As we can see, the diagram is again of the form of a $(N,N)$ web, except that the parameters $\{\mathbf{m}\}$ and $\{\mathbf{h}\}$ have been exchanged. This indicates, that the diagonal expansion also leads to a gauge theory with gauge group $U(N)^N$ where the mass parameters (from the perspective of the horizontal or vertical description) $\{ \bf m \}$ are now related to the coupling constants.

Finally, due to the known $SL(2,\mathbb{Z})$ duality $(N,M)\longleftrightarrow (M,N)$ of the compact brane configurations (\emph{i.e.} those compactified on a torus),  we can find a similar transformation that exchanges the parameters $\{\mathbf{m}\}$ and $\{\mathbf{v}\}$. As a result, we find a self-triality for the configurations $(N,N)$, given by the exchange of any of the three sets of parameters $\{\mathbf{h}\}$, $\{\mathbf{v}\}$ or $\{\mathbf{m}\}$
\begin{align}
\{\mathbf{m}\}\longleftrightarrow \{\mathbf{h}\}\longleftrightarrow\{\mathbf{v}\}\,.\label{TrialityNN}
\end{align}
Since the transformation described above transforms a $(N,N)$ web diagram into a $(N,N)$ web diagram, it is clear that (\ref{TrialityNN}) is a symmetry of the partition function ${\cal Z}_{N, N}({\bf h}, {\bf v}, {\bf m}, \epsilon_1, \epsilon_2)$ Eq.(\ref{PartitionFunctionXNM}). As seen from the above figures, the precise map is somewhat complicated in the generic case. However, it can be made very explicit for specific examples: in the simplest case $(2,2)$, the transformation is shown in detail in the following figure
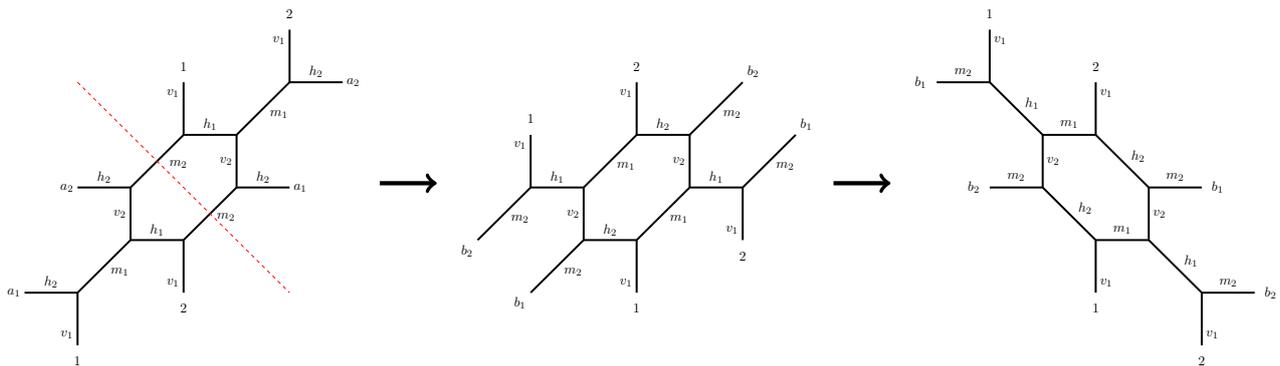
\begin{center}
\scalebox{0.465}{\parbox{10cm}{\begin{tikzpicture}[scale = 1.5]
\draw[ultra thick] (-6,0) -- (-5,0);
\draw[ultra thick] (-5,-1) -- (-5,0);
\draw[ultra thick] (-5,0) -- (-4,1);
\draw[ultra thick] (-4,1) -- (-3,1);
\draw[ultra thick] (-4,1) -- (-4,2);
\draw[ultra thick] (-3,1) -- (-3,0);
\draw[ultra thick] (-3,1) -- (-2,2);
\draw[ultra thick] (-4,2) -- (-3,3);
\draw[ultra thick] (-5,2) -- (-4,2);
\draw[ultra thick] (-3,3) -- (-3,4);
\draw[ultra thick] (-3,3) -- (-2,3);
\draw[ultra thick] (-2,2) -- (-2,3);
\draw[ultra thick] (-2,2) -- (-1,2);
\draw[ultra thick] (-2,3) -- (-1,4);
\draw[ultra thick] (-1,4) -- (0,4);
%
%\draw[ultra thick] (0,4) -- (1,5);
\draw[ultra thick] (-1,4) -- (-1,5);
\node at (-6.2,0) {{\small \bf $a_1$}};
\node at (-5.2,2) {{\small \bf $a_2$}};
\node at (-0.8,2) {{\small \bf $a_1$}};
\node at (0.2,4) {{\small \bf $a_2$}};
\node at (-5,-1.3) {{\small \bf $1$}};
\node at (-3,-0.3) {{\small \bf $2$}};
\node at (-3,4.3) {{\small \bf $1$}};
\node at (-1,5.3) {{\small \bf $2$}};
\node at (-4.2,0.4) {{\small $m_1$}};
\node at (-2.2,1.45) {{\small $m_2$}};
\node at (-3.1,2.45) {{\small $m_2$}};
\node at (-1.2,3.4) {{\small $m_1$}};
\node at (-3.5,1.2) {{\small $h_1$}};
\node at (-1.5,2.2) {{\small $h_2$}};
\node at (-5.5,0.2) {{\small $h_2$}};
\node at (-2.5,3.2) {{\small $h_1$}};
\node at (-0.5,4.2) {{\small $h_2$}};
\node at (-4.5,2.2) {{\small $h_2$}};
\node at (-5.2,-0.8) {{\small $v_1$}};
\node at (-3.2,3.8) {{\small $v_1$}};
\node at (-3.2,0.2) {{\small $v_1$}};
\node at (-1.2,4.8) {{\small $v_1$}};
\node at (-4.2,1.5) {{\small $v_2$}};
\node at (-2.2,2.5) {{\small $v_2$}};
\draw[dashed,red] (-5,4) -- (-1,0);
\end{tikzpicture}}}
%%%%%%%%%%%%%%
\hspace{0.01cm}
\parbox{0.75cm}{\begin{tikzpicture}
\draw[ultra thick,->] (0,0) -- (0.75,0);
\end{tikzpicture}}
\hspace{0.01cm}
%%%%%%%%%%%%%%
\scalebox{0.465}{\parbox{10cm}{\begin{tikzpicture}[scale = 1.5]
%diagonal lines
\draw[ultra thick] (0,0) -- (1,1);
\draw[ultra thick] (-1,1) -- (0,2);
\draw[ultra thick] (1,2) -- (2,3);
\draw[ultra thick] (2,1) -- (3,2);
\draw[ultra thick] (4,2) -- (5,3);
\draw[ultra thick] (3,3) -- (4,4);
%horizontal lines
\draw[ultra thick] (0,2) -- (1,2);
\draw[ultra thick] (1,1) -- (2,1);
\draw[ultra thick] (2,3) -- (3,3);
\draw[ultra thick] (3,2) -- (4,2);
%vertical lines
\draw[ultra thick] (1,1) -- (1,2);
\draw[ultra thick] (0,2) -- (0,3);
\draw[ultra thick] (2,1) -- (2,0);
\draw[ultra thick] (2,3) -- (2,4);
\draw[ultra thick] (3,2) -- (3,3);
\draw[ultra thick] (4,1) -- (4,2);
%connect vertical
\node at (2,-0.3) {{\small \bf $1$}};
\node at (0,3.3) {{\small \bf $1$}};
\node at (4,0.7) {{\small \bf $2$}};
\node at (2,4.3) {{\small \bf $2$}};
%connect diagonal
\node at (-0.2,-0.2) {{\small \bf $b_1$}};
\node at (5.2,3.2) {{\small \bf $b_1$}};
\node at (-1.2,0.8) {{\small \bf $b_2$}};
\node at (4.2,4.2) {{\small \bf $b_2$}};
%diagonal lines
\node at (0.8,0.4) {{\small $m_2$}};
\node at (-0.2,1.4) {{\small $m_2$}};
\node at (2.8,1.4) {{\small $m_1$}};
\node at (1.8,2.4) {{\small $m_1$}};
\node at (4.8,2.4) {{\small $m_2$}};
\node at (3.8,3.4) {{\small $m_2$}};
%vertical lines
\node at (1.8,0.2) {{\small $v_1$}};
\node at (3.8,1.2) {{\small $v_1$}};
\node at (-0.2,2.8) {{\small $v_1$}};
\node at (1.8,3.8) {{\small $v_1$}};
\node at (0.8,1.5) {{\small $v_2$}};
\node at (2.8,2.5) {{\small $v_2$}};
%horizontal lines
\node at (1.5,1.2) {{\small $h_2$}};
\node at (3.5,2.2) {{\small $h_1$}};
\node at (0.5,2.2) {{\small $h_1$}};
\node at (2.5,3.2) {{\small $h_2$}};
\end{tikzpicture}}}
%%%%%%%%%%%%%%
\hspace{0.01cm}
\parbox{0.75cm}{\begin{tikzpicture}
\draw[ultra thick,->] (0,0) -- (0.75,0);
\end{tikzpicture}}
\hspace{0.01cm}
%%%%%%%%%%%%%%%%
\scalebox{0.465}{\parbox{10cm}{\begin{tikzpicture}[scale = 1.5]
%vertical lines
\draw[ultra thick] (0,1) -- (0,0);
\draw[ultra thick] (1,-1) -- (1,-2);
\draw[ultra thick] (2,0) -- (2,-1);
\draw[ultra thick] (2,-3) -- (2,-4);
\draw[ultra thick] (3,-2) -- (3,-3);
\draw[ultra thick] (4,-4) -- (4,-5);
%diagonal lines
\draw[ultra thick] (0,0) -- (1,-1);
\draw[ultra thick] (1,-2) -- (2,-3);
\draw[ultra thick] (2,-1) -- (3,-2);
\draw[ultra thick] (3,-3) -- (4,-4);
%horizontal lines
\draw[ultra thick] (-1,0) -- (0,0);
\draw[ultra thick] (0,-2) -- (1,-2);
\draw[ultra thick] (1,-1) -- (2,-1);
\draw[ultra thick] (2,-3) -- (3,-3);
\draw[ultra thick] (3,-2) -- (4,-2);
\draw[ultra thick] (4,-4) -- (5,-4);
%connect vertical
\node at (0,1.3) {{\small \bf $1$}};
\node at (2,-4.3) {{\small \bf $1$}};
\node at (2,0.3) {{\small \bf $2$}};
\node at (4,-5.3) {{\small \bf $2$}};
%connect horizontal
\node at (-1.3,0) {{\small \bf $b_1$}};
\node at (4.3,-2) {{\small \bf $b_1$}};
\node at (-0.3,-2) {{\small \bf $b_2$}};
\node at (5.3,-4) {{\small \bf $b_2$}};
%diagonal lines
\node at (0.8,-0.4) {{\small $h_1$}};
\node at (1.8,-2.4) {{\small $h_2$}};
\node at (2.8,-1.4) {{\small $h_2$}};
\node at (3.8,-3.4) {{\small $h_1$}};
%horizontal lines
\node at (-0.5,0.2) {{\small $m_2$}};
\node at (1.5,-0.8) {{\small $m_1$}};
\node at (0.5,-1.8) {{\small $m_2$}};
\node at (2.5,-2.8) {{\small $m_1$}};
\node at (3.5,-1.8) {{\small $m_2$}};
\node at (4.5,-3.8) {{\small $m_2$}};
%vertical lines
\node at (0.2,0.8) {{\small $v_1$}};
\node at (2.2,-0.2) {{\small $v_1$}};
\node at (1.2,-1.5) {{\small $v_2$}};
\node at (3.2,-2.5) {{\small $v_2$}};
\node at (2.2,-3.8) {{\small $v_1$}};
\node at (4.2,-4.8) {{\small $v_1$}};
\end{tikzpicture}}}
%%%%%%%%%%%%%%%%%%%%%%%%%%%%%%
\end{center}
\noindent
We see that the cutting and re-gluing procedure simply amounts to an exchange\footnote{Notice that here the 6 parameters $(h_{1,2},v_{1,2},m_{1,2})$ are all independent and satisfy the consistency conditions of the diagram.}
\begin{align}
&m_1\longrightarrow h_1\,,&&m_2\longrightarrow h_2\,,&&h_1\longrightarrow m_1\,,
&&h_2\longrightarrow m_2\,.
\end{align}
this indicates that the $(2,2)$ web indeed enjoys the triality symmetry
\begin{align}
\left(\begin{array}{c}m_1 \\ m_2\end{array}\right)\longleftrightarrow \left(\begin{array}{c}h_1 \\ h_2\end{array}\right)\longleftrightarrow \left(\begin{array}{c}v_1 \\ v_2\end{array}\right)\,.
\end{align}
%%%%%%%%%%%%%%%%%%%%%%%%%%%%%%
%%%%%%%%%%%%%%%%%%%%%%%%%%%%%%
\subsection{Configurations $(nM,M)$ with $n\in\mathbb{N}$}
The argument of the previous subsection for self-triality among the descriptions associated with the horizontal, vertical and diagonal expansions of $(N,N)$ configuration does not directly generalise to other generic $(M, N)$ configurations. However, for configurations $(N,M)$ with $N=nM$ for $n\in\mathbb{N}$ (and $n>1$), \emph{i.e.} in the case that $N$ is an integer multiple of $M$, we find that the partition function is still invariant under the exchange of the parameters $\{\mathbf{v}\}\longleftrightarrow \{\bf{m}\}$. While, in this case, there is no single cut like in the previous section that allows us to argue for this duality, the latter is still apparent from the Newton polygon of $X_{nM,M}$ given below. As is well known the Newton polygon is the dual graph of the web diagram of the toric Calabi-Yau threefold \cite{Leung:1997tw}. Indeed, for $N=n M$, the two fundamental domains in the latter, associated with the vertical and diagonal description, are as follows:
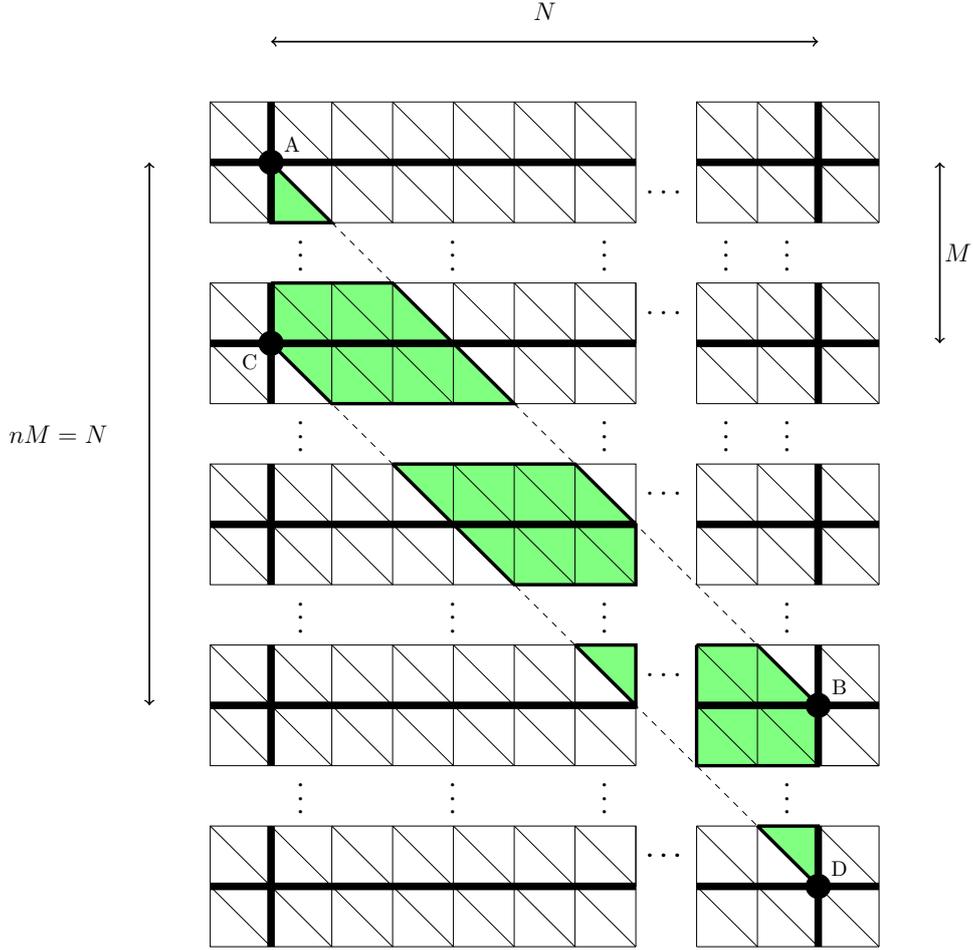
\begin{center}
\scalebox{0.8}{\parbox{16cm}{\begin{tikzpicture}[scale = 1.0]
%green
\draw[ultra thick,fill=green!50!white] (0,0) -- (1,-1) -- (0,-1) -- (0,0);
\draw[ultra thick,fill=green!50!white] (0,-2) -- (2,-2) -- (4,-4) -- (1,-4) -- (0,-3) --  (0,-2);
\draw[ultra thick,fill=green!50!white] (2,-5) -- (5,-5) -- (6,-6) -- (6,-7) -- (4,-7) --  (2,-5);
\draw[ultra thick,fill=green!50!white] (5,-8) -- (6,-8) -- (6,-9) -- (5,-8);
\draw[ultra thick,fill=green!50!white] (7,-8) -- (8,-8) -- (9,-9) -- (9,-10) -- (7,-10) -- (7,-8);
\draw[ultra thick,fill=green!50!white] (8,-11) -- (9,-11) -- (9,-12) -- (8,-11);
%%%%%%%%
\draw[-,line width=1.25mm] (-1,0) -- (0,0);
\draw[-,line width=1.25mm] (0,1) -- (0,0);
\draw[-,line width=1.25mm] (9,1) -- (9,0);
\draw[-,line width=1.25mm] (-1,-3) -- (0,-3);
\draw[-,line width=1.25mm] (0,-1) -- (0,0) -- (6,0);
\draw[-,line width=1.25mm] (7,0) -- (10,0);
\draw[-,line width=1.25mm] (9,0) -- (9,-1);
\draw[-,line width=1.25mm] (0,-2) -- (0,-3) -- (6,-3);
\draw[-,line width=1.25mm] (9,0) -- (9,-1);
\draw[-,line width=1.25mm] (7,-3) -- (10,-3);
\draw[-,line width=1.25mm] (9,-3) -- (9,-2);
\draw[-,line width=1.25mm] (9,-3) -- (9,-4);
\draw[-,line width=1.25mm] (0,-3) -- (0,-4);
%internal lines
\draw[-] (6,0) -- (6,-1) -- (-1,-1) -- (-1,1) -- (6,1) -- (6,0);
\draw[-] (0,-4) -- (-1,-4) -- (-1,-2) -- (6,-2) -- (6,-3);
\draw[-] (1,1) -- (1,-1);
\draw[-] (2,1) -- (2,-1);
\draw[-] (3,1) -- (3,-1);
\draw[-] (4,1) -- (4,-1);
\draw[-] (5,1) -- (5,-1);
\draw[-] (1,-2) -- (1,-4);
\draw[-] (2,-2) -- (2,-4);
\draw[-] (3,-2) -- (3,-4);
\draw[-] (4,-2) -- (4,-4);
\draw[-] (5,-2) -- (5,-4);
\draw[-] (6,-2) -- (6,-4);
\draw[-] (0,-4) -- (6,-4);
\draw[-] (7,-1) -- (10,-1);
\draw[-] (7,-2) -- (10,-2);
\draw[-] (7,-4) -- (10,-4);
\draw[-] (7,0) -- (7,-1);
\draw[-] (8,1) -- (8,-1);
\draw[-] (10,0) -- (10,-1);
\draw[-] (7,-2) -- (7,-4);
\draw[-] (8,-2) -- (8,-4);
\draw[-] (10,-2) -- (10,-4);
\draw[-] (7,0) -- (7,1) -- (10,1) -- (10,0);
%internal diagonal lines
\draw[-] (-1,-2) -- (0,-3);
\draw[-] (-1,-3) -- (0,-4);
\draw[-] (-1,0) -- (0,-1);
\draw[-] (-1,1) -- (1,-1);
\draw[-] (0,1) -- (2,-1);
\draw[-] (1,1) -- (3,-1);
\draw[-] (2,1) -- (4,-1);
\draw[-] (3,1) -- (5,-1);
\draw[-] (4,1) -- (6,-1);
\draw[-] (5,1) -- (6,0);
\draw[-] (0,-3) -- (1,-4);
\draw[-] (0,-2) -- (2,-4);
\draw[-] (1,-2) -- (3,-4);
\draw[-] (2,-2) -- (4,-4);
\draw[-] (3,-2) -- (5,-4);
\draw[-] (4,-2) -- (6,-4);
\draw[-] (5,-2) -- (6,-3);
\draw[-] (7,1) -- (9,-1);
\draw[-] (8,1) -- (10,-1);
\draw[-] (9,1) -- (10,0);
\draw[-] (7,0) -- (8,-1);
\draw[-] (7,-3) -- (8,-4);
\draw[-] (7,-2) -- (9,-4);
\draw[-] (8,-2) -- (10,-4);
\draw[-] (9,-2) -- (10,-3);
%lower part
\begin{scope}[yshift=-6cm]
\draw[-,line width=1.25mm] (-1,0) -- (0,0);
\draw[-,line width=1.25mm] (0,1) -- (0,0);
\draw[-,line width=1.25mm] (9,1) -- (9,0);
\draw[-,line width=1.25mm] (-1,-3) -- (0,-3);
\draw[-,line width=1.25mm] (0,-1) -- (0,0) -- (6,0);
\draw[-,line width=1.25mm] (7,0) -- (10,0);
\draw[-,line width=1.25mm] (9,0) -- (9,-1);
\draw[-,line width=1.25mm] (0,-2) -- (0,-3) -- (6,-3);
\draw[-,line width=1.25mm] (9,0) -- (9,-1);
\draw[-,line width=1.25mm] (7,-3) -- (10,-3);
\draw[-,line width=1.25mm] (9,-3) -- (9,-2);
\draw[-,line width=1.25mm] (9,-3) -- (9,-4);
\draw[-,line width=1.25mm] (0,-3) -- (0,-4);
%internal lines
\draw[-] (6,0) -- (6,-1) -- (-1,-1) -- (-1,1) -- (6,1) -- (6,0);
\draw[-] (0,-4) -- (-1,-4) -- (-1,-2) -- (6,-2) -- (6,-3);
\draw[-] (1,1) -- (1,-1);
\draw[-] (2,1) -- (2,-1);
\draw[-] (3,1) -- (3,-1);
\draw[-] (4,1) -- (4,-1);
\draw[-] (5,1) -- (5,-1);
\draw[-] (1,-2) -- (1,-4);
\draw[-] (2,-2) -- (2,-4);
\draw[-] (3,-2) -- (3,-4);
\draw[-] (4,-2) -- (4,-4);
\draw[-] (5,-2) -- (5,-4);
\draw[-] (6,-2) -- (6,-4);
\draw[-] (0,-4) -- (6,-4);
\draw[-] (7,-1) -- (10,-1);
\draw[-] (7,-2) -- (10,-2);
\draw[-] (7,-4) -- (10,-4);
\draw[-] (7,0) -- (7,-1);
\draw[-] (8,1) -- (8,-1);
\draw[-] (10,0) -- (10,-1);
\draw[-] (7,-2) -- (7,-4);
\draw[-] (8,-2) -- (8,-4);
\draw[-] (10,-2) -- (10,-4);
\draw[-] (7,0) -- (7,1) -- (10,1) -- (10,0);
%internal diagonal lines
\draw[-] (-1,-2) -- (0,-3);
\draw[-] (-1,-3) -- (0,-4);
\draw[-] (-1,0) -- (0,-1);
\draw[-] (-1,1) -- (1,-1);
\draw[-] (0,1) -- (2,-1);
\draw[-] (1,1) -- (3,-1);
\draw[-] (2,1) -- (4,-1);
\draw[-] (3,1) -- (5,-1);
\draw[-] (4,1) -- (6,-1);
\draw[-] (5,1) -- (6,0);
\draw[-] (0,-3) -- (1,-4);
\draw[-] (0,-2) -- (2,-4);
\draw[-] (1,-2) -- (3,-4);
\draw[-] (2,-2) -- (4,-4);
\draw[-] (3,-2) -- (5,-4);
\draw[-] (4,-2) -- (6,-4);
\draw[-] (5,-2) -- (6,-3);
\draw[-] (7,1) -- (9,-1);
\draw[-] (8,1) -- (10,-1);
\draw[-] (9,1) -- (10,0);
\draw[-] (7,0) -- (8,-1);
\draw[-] (7,-3) -- (8,-4);
\draw[-] (7,-2) -- (9,-4);
\draw[-] (8,-2) -- (10,-4);
\draw[-] (9,-2) -- (10,-3);
\node[rotate=90] at (3,-1.5) {\large $\cdots$};
\node at (6.5,-2.5) {\large $\cdots$};
\node[rotate=90] at (8.5,-1.5) {\large $\cdots$};
\node[rotate=90] at (0.5,-1.5) {\large $\cdots$};
\node[rotate=90] at (5.5,-1.5) {\large $\cdots$};
\end{scope}
\begin{scope}[yshift=-9cm]
\draw[-,line width=1.25mm] (-1,-3) -- (0,-3);
\draw[-,line width=1.25mm] (0,-2) -- (0,-3) -- (6,-3);
\draw[-,line width=1.25mm] (7,-3) -- (10,-3);
\draw[-,line width=1.25mm] (9,-3) -- (9,-2);
\draw[-,line width=1.25mm] (9,-3) -- (9,-4);
\draw[-,line width=1.25mm] (0,-3) -- (0,-4);
%internal lines
\draw[-] (0,-4) -- (-1,-4) -- (-1,-2) -- (6,-2) -- (6,-3);
\draw[-] (1,-2) -- (1,-4);
\draw[-] (2,-2) -- (2,-4);
\draw[-] (3,-2) -- (3,-4);
\draw[-] (4,-2) -- (4,-4);
\draw[-] (5,-2) -- (5,-4);
\draw[-] (6,-2) -- (6,-4);
\draw[-] (0,-4) -- (6,-4);
\draw[-] (7,-1) -- (10,-1);
\draw[-] (7,-2) -- (10,-2);
\draw[-] (7,-4) -- (10,-4);
\draw[-] (7,-2) -- (7,-4);
\draw[-] (8,-2) -- (8,-4);
\draw[-] (10,-2) -- (10,-4);
%internal diagonal lines
\draw[-] (-1,-2) -- (0,-3);
\draw[-] (-1,-3) -- (0,-4);
\draw[-] (0,-3) -- (1,-4);
\draw[-] (0,-2) -- (2,-4);
\draw[-] (1,-2) -- (3,-4);
\draw[-] (2,-2) -- (4,-4);
\draw[-] (3,-2) -- (5,-4);
\draw[-] (4,-2) -- (6,-4);
\draw[-] (5,-2) -- (6,-3);
\draw[-] (7,-3) -- (8,-4);
\draw[-] (7,-2) -- (9,-4);
\draw[-] (8,-2) -- (10,-4);
\draw[-] (9,-2) -- (10,-3);
\node[rotate=90] at (3,-1.5) {\large $\cdots$};
\node at (6.5,-2.5) {\large $\cdots$};
\node[rotate=90] at (8.5,-1.5) {\large $\cdots$};
\node[rotate=90] at (0.5,-1.5) {\large $\cdots$};
\node[rotate=90] at (5.5,-1.5) {\large $\cdots$};
\end{scope}
%dots
\node[rotate=90] at (3,-1.5) {\large $\cdots$};
\node at (6.5,-0.5) {\large $\cdots$};
\node at (6.5,-2.5) {\large $\cdots$};
\node at (6.5,-5.5) {\large $\cdots$};
\node at (6.5,-11.5) {\large $\cdots$};
\node[rotate=90] at (8.5,-1.5) {\large $\cdots$};
\node[rotate=90] at (7.5,-1.5) {\large $\cdots$};
\node[rotate=90] at (0.5,-1.5) {\large $\cdots$};
\node[rotate=90] at (5.5,-1.5) {\large $\cdots$};
\node[rotate=90] at (8.5,-4.5) {\large $\cdots$};
\node[rotate=90] at (7.5,-4.5) {\large $\cdots$};
\node[rotate=90] at (0.5,-4.5) {\large $\cdots$};
\node[rotate=90] at (5.5,-4.5) {\large $\cdots$};
%dashed lines
\draw[dashed] (1,-1) -- (2,-2);
\draw[dashed] (4,-4) -- (9,-9);
\draw[dashed] (0,-3) -- (9,-12);
\draw[fill=black] (0,0) circle (0.2cm);
\node at (0.35,0.3) {\footnotesize A};
\draw[fill=black] (9,-9) circle (0.2cm);
\node at (9.35,-8.7) {\footnotesize B};
\draw[fill=black] (0,-3) circle (0.2cm);
\node at (-0.35,-3.3) {\footnotesize C};
\draw[fill=black] (9,-12) circle (0.2cm);
\node at (9.35,-11.7) {\footnotesize D};
%arrows
\draw[thick,<->] (-2,0) -- (-2,-9);
\node at (-3.5,-4.5) {$nM=N$};
\draw[thick,<->] (11,0) -- (11,-3);
\node at (11.3,-1.5) {$M$};
\draw[thick,<->] (0,2) -- (9,2);
\node at (4.5,2.5) {$N$};
\end{tikzpicture}}}
\end{center}
Here, the four points $A$, $B$, $C$ and $D$ are all equivalent, such that the diagonal lines $\overline{\text{AB}}$ and $\overline{\text{CD}}$ are identified with each other. Therefore, the web-diagram dual to the green fundamental domain is again of the type $(nM,M)$, except that the diagonal parameters $\{\mathbf{m}\}$ are exchanged with the vertical ones $\{\mathbf{v}\}$. Thus, the theories $(nM,M)$ are self-dual under the exchange
\begin{align}
\{\mathbf{m}\}\longleftrightarrow \{\mathbf{v}\}\,.\label{SelfDuality}
\end{align}
For example, we can make this map more precise in the simplest case $M=2$ and $n=2$, \emph{i.e.} for the configuration $(N,M)=(4,2)$, whose web diagram and Newton polygon are given by
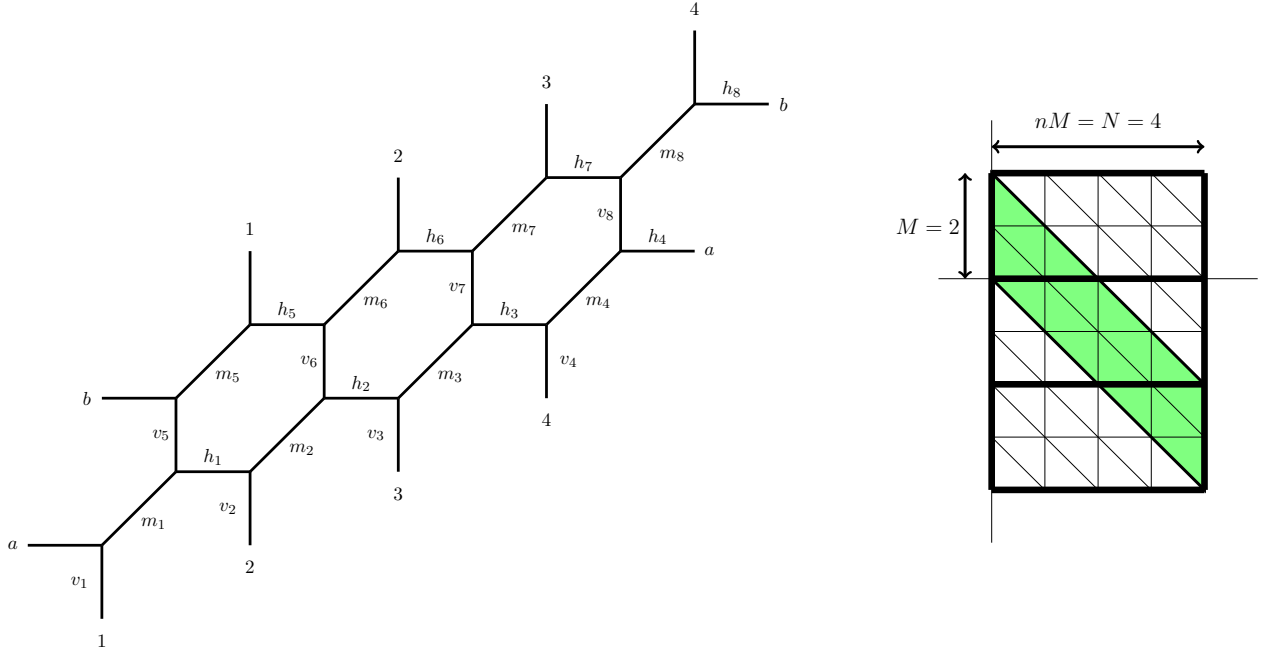
\begin{center}
%\hspace*{-2cm}%
\scalebox{0.65}{\parbox{16cm}{\begin{tikzpicture}[scale = 1.50]
\draw[ultra thick] (-6,0) -- (-5,0);
\draw[ultra thick] (-5,-1) -- (-5,0);
\draw[ultra thick] (-5,0) -- (-4,1);
\draw[ultra thick] (-4,1) -- (-3,1);
\draw[ultra thick] (-4,1) -- (-4,2);
\draw[ultra thick] (-3,1) -- (-3,0);
\draw[ultra thick] (-3,1) -- (-2,2);
\draw[ultra thick] (-4,2) -- (-3,3);
\draw[ultra thick] (-5,2) -- (-4,2);
\draw[ultra thick] (-3,3) -- (-3,4);
\draw[ultra thick] (-3,3) -- (-2,3);
\draw[ultra thick] (-2,2) -- (-2,3);
\draw[ultra thick] (-2,2) -- (-1,2);
\draw[ultra thick] (-2,3) -- (-1,4);
\draw[ultra thick] (-1,2) -- (0,3);
\draw[ultra thick] (-1,2) -- (-1,1);
\draw[ultra thick] (-1,4) -- (0,4);
\draw[ultra thick] (0,3) -- (1,3);
\draw[ultra thick] (0,3) -- (0,4);
\draw[ultra thick] (0,4) -- (1,5);
\draw[ultra thick] (1,5) -- (2,5);
\draw[ultra thick] (1,5) -- (1,6);
\draw[ultra thick] (-1,4) -- (-1,5);
\draw[ultra thick] (2,5) -- (3,6);
\draw[ultra thick] (3,6) -- (4,6);
\draw[ultra thick] (3,6) -- (3,7);
\draw[ultra thick] (2,5) -- (2,4);
\draw[ultra thick] (1,3) -- (2,4);
\draw[ultra thick] (1,2) -- (1,3);
\draw[ultra thick] (2,4) -- (3,4);
%
%connection labels vertical
\node at (-5,-1.3) {{\small \bf $1$}};
\node at (-3,-0.3) {{\small \bf $2$}};
\node at (-1,0.7) {{\small \bf $3$}};
\node at (1,1.7) {{\small \bf $4$}};
\node at (-3,4.3) {{\small \bf $1$}};
\node at (-1,5.3) {{\small \bf $2$}};
\node at (1,6.3) {{\small \bf $3$}};
\node at (3,7.3) {{\small \bf $4$}};
%
%connection labels horizontal
\node at (-6.2,0) {{\small \bf $a$}};
\node at (-5.2,2) {{\small \bf $b$}};
\node at (4.2,6) {{\small \bf $b$}};
\node at (3.2,4) {{\small \bf $a$}};
\node at (-4.3,0.3) {{\small $m_1$}};
\node at (-2.3,1.3) {{\small $m_2$}};
\node at (-0.3,2.3) {{\small $m_3$}};
\node at (1.7,3.3) {{\small $m_4$}};
\node at (-3.3,2.3) {{\small $m_5$}};
\node at (-1.3,3.3) {{\small $m_6$}};
\node at (0.7,4.3) {{\small $m_7$}};
\node at (2.7,5.3) {{\small $m_8$}};
\node at (-3.5,1.2) {{\small $h_1$}};
\node at (-1.5,2.2) {{\small $h_2$}};
\node at (0.5,3.2) {{\small $h_3$}};
\node at (2.5,4.2) {{\small $h_4$}};
\node at (-2.5,3.2) {{\small $h_5$}};
\node at (-0.5,4.2) {{\small $h_6$}};
\node at (1.5,5.2) {{\small $h_7$}};
\node at (3.5,6.2) {{\small $h_8$}};
\node at (-5.3,-0.5) {{\small $v_1$}};
\node at (-3.3,0.5) {{\small $v_2$}};
\node at (-1.3,1.5) {{\small $v_3$}};
\node at (1.3,2.5) {{\small $v_4$}};
\node at (-4.2,1.5) {{\small $v_5$}};
\node at (-2.2,2.5) {{\small $v_6$}};
\node at (-0.2,3.5) {{\small $v_7$}};
\node at (1.8,4.5) {{\small $v_8$}};
\end{tikzpicture}}}
%%%%%%%%%%%%%%
\hspace{1cm}
%%%%%%%%%%%%%%
\scalebox{0.7}{\parbox{7cm}{\begin{tikzpicture}[scale = 1.0]
\draw[ultra thick,fill=green!50!white] (0,10) -- (4,6) -- (4,8) -- (0,12) -- (0,10);
\draw[-] (-1,10) -- (5,10);
\draw[-] (0,5) -- (0,13);
\draw[-,line width=1.25mm] (0,6) -- (4,6);
\draw[-] (0,7) -- (4,7);
\draw[-,line width=1.25mm] (0,8) -- (4,8);
\draw[-] (0,9) -- (4,9);
\draw[-,line width=1.25mm] (0,10) -- (4,10);
\draw[-] (0,11) -- (4,11);
\draw[-,line width=1.25mm] (0,12) -- (4,12);
\draw[-,line width=1.25mm] (0,6) -- (0,12);
\draw[-] (1,6) -- (1,12);
\draw[-] (2,6) -- (2,12);
\draw[-] (3,6) -- (3,12);
\draw[-,line width=1.25mm] (4,6) -- (4,12);
%diagonal
\draw[-] (0,7) -- (1,6);
\draw[-] (0,8) -- (2,6);
\draw[-] (0,9) -- (3,6);
\draw[-] (0,10) -- (4,6);
\draw[-] (0,11) -- (4,7);
\draw[-] (0,12) -- (4,8);
\draw[-] (1,12) -- (4,9);
\draw[-] (2,12) -- (4,10);
\draw[-] (3,12) -- (4,11);
\draw[ultra thick,<->] (-0.5,10) -- (-0.5,12);
\draw[ultra thick,<->] (0,12.5) -- (4,12.5);
\node at (-1.2,11) {$M=2$};
\node at (2,13) {$nM=N=4$};
\end{tikzpicture}}}
\end{center}
This web-diagram can also be presented in the form
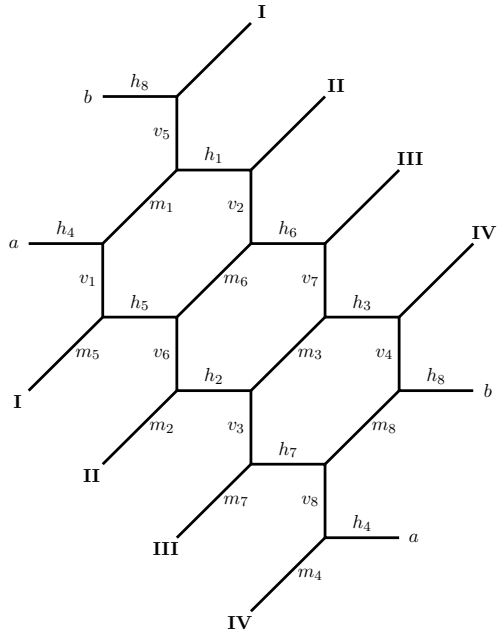
\begin{center}
\scalebox{0.65}{\parbox{10cm}{\begin{tikzpicture}[scale = 1.50]
\draw[ultra thick] (0,0) -- (1,0) -- (1,-1) -- (2,-1) -- (2,-2) -- (3,-2) -- (3,-3) -- (4,-3) -- (4,-4) -- (5,-4);
%diagonals up
\draw[ultra thick] (1,0) -- (2,1);
\draw[ultra thick] (2,-1) -- (3,0);
\draw[ultra thick] (3,-2) -- (4,-1);
\draw[ultra thick] (4,-3) -- (5,-2);
%diagonals down
\draw[ultra thick] (1,-1) -- (0,-2);
\draw[ultra thick] (2,-2) -- (1,-3);
\draw[ultra thick] (3,-3) -- (2,-4);
\draw[ultra thick] (4,-4) -- (3,-5);
\node at (-0.15,-2.15) {{\small \bf I}};
\node at (0.85,-3.15) {{\small \bf II}};
\node at (1.85,-4.15) {{\small \bf III}};
\node at (2.85,-5.15) {{\small \bf IV}};
%labels horizontal
\node at (-0.2,0) {{\small \bf $a$}};
\node at (5.2,-4) {{\small \bf $a$}};
%labels horizontal
\node at (0.5,0.2) {{\small $h_4$}};
\node at (1.5,-0.8) {{\small $h_5$}};
\node at (2.5,-1.8) {{\small $h_2$}};
\node at (3.5,-2.8) {{\small $h_7$}};
\node at (4.5,-3.8) {{\small $h_4$}};
%labels vertical
\node at (0.8,-0.5) {{\small $v_1$}};
\node at (1.8,-1.5) {{\small $v_6$}};
\node at (2.8,-2.5) {{\small $v_3$}};
\node at (3.8,-3.5) {{\small $v_8$}};
\node at (0.8,-1.5) {{\small $m_5$}};
\node at (1.8,-2.5) {{\small $m_2$}};
\node at (2.8,-3.5) {{\small $m_7$}};
\node at (3.8,-4.5) {{\small $m_4$}};
%%%%%%%
\begin{scope}[yshift=2cm,xshift=1cm]
\draw[ultra thick] (0,0) -- (1,0) -- (1,-1) -- (2,-1) -- (2,-2) -- (3,-2) -- (3,-3) -- (4,-3) -- (4,-4) -- (5,-4);
%diagonals up
\draw[ultra thick] (1,0) -- (2,1);
\draw[ultra thick] (2,-1) -- (3,0);
\draw[ultra thick] (3,-2) -- (4,-1);
\draw[ultra thick] (4,-3) -- (5,-2);
%labels up
\node at (2.15,1.15) {{\small \bf I}};
\node at (3.15,0.15) {{\small \bf II}};
\node at (4.15,-0.85) {{\small \bf III}};
\node at (5.15,-1.85) {{\small \bf IV}};
%labels down
%labels horizontal
\node at (-0.2,0) {{\small \bf $b$}};
\node at (5.2,-4) {{\small \bf $b$}};
%labels horizontal
\node at (0.5,0.2) {{\small $h_8$}};
\node at (1.5,-0.8) {{\small $h_1$}};
\node at (2.5,-1.8) {{\small $h_6$}};
\node at (3.5,-2.8) {{\small $h_3$}};
\node at (4.5,-3.8) {{\small $h_8$}};
%labels vertical
\node at (0.8,-0.5) {{\small $v_5$}};
\node at (1.8,-1.5) {{\small $v_2$}};
\node at (2.8,-2.5) {{\small $v_7$}};
\node at (3.8,-3.5) {{\small $v_4$}};
\node at (0.8,-1.5) {{\small $m_1$}};
\node at (1.8,-2.5) {{\small $m_6$}};
\node at (2.8,-3.5) {{\small $m_3$}};
\node at (3.8,-4.5) {{\small $m_8$}};
\end{scope}
\end{tikzpicture}}}
\end{center}
which is equivalent to the original diagram under the following exchange of parameters
\begin{align}
&v_1\longleftrightarrow m_4\,,&&v_2\longleftrightarrow m_7\,,&&v_3\longleftrightarrow m_2\,,&&v_4\longleftrightarrow m_5\,,\nonumber\\
&v_5\longleftrightarrow m_8\,,&&v_6\longleftrightarrow m_3\,,&&v_7\longleftrightarrow m_6\,,&&v_8\longleftrightarrow m_1\,,\nonumber\\
&h_1\longleftrightarrow h_7\,,&&h_3\longleftrightarrow h_5\,,\nonumber
\end{align}
with $h_{2,4,6,8}$ remaining invariant. This makes the self-duality relation (\ref{SelfDuality}) precise for the case $(4,2)$. Note, however, that these parameters are not independent one another, but instead the consistency conditions associated with the web diagram need to be imposed.

%%%%%%%%%%%%%%%%%%%%%%%%%%%%%%
%%%%%%%%%%%%%%%%%%%%%%%%%%%%%%
%%%%%%%%%%%%%%%%%%%%%%%%%%%%%%
\subsection{Non-Compact Web Diagrams}
Up to this point, we have studied $(N,M)$ web configuration that are defined on a torus (\emph{i.e.} both the horizontal and vertical directions are periodic with radii $\rho$ and $\tau$, respectively). This setup defines the corresponding little string theories, as outlined in the introduction. 
It is also of interest to examine if various dualities we identified so far also hold for non-compact web configurations, where one of the two directions is decompactified by sending \emph{e.g.} $\rho\to i\infty$. These brane configurations, at a particular point in the moduli space, give rise to superconformal field theories. See \cite{Heckman:2013pva,Heckman:2014qba,Ohmori:2014kda, Haghighat:2014vxa,DelZotto:2014hpa,Heckman:2015bfa,Bertolini:2015bwa,DelZotto:2017pti} for a recent discussion of six dimensional SCFTs.

It turns out that we can equally compute the equivalent of the diagonal expansion Eq.(\ref{TrialitySchemat}) while the $SL(2,\mathbb{Z})$ transformation maps between the vertical and diagonal expansion. 

\begin{figure}[htbp]
\begin{center}
%\hspace*{-2cm}%
\scalebox{0.6}{\parbox{27.5cm}{\begin{tikzpicture}[scale = 1.1]
%left bottom
\draw[ultra thick] (-1,0) -- (0,0) -- (1,1) -- (2,1) -- (3,2) -- (4,2);
\draw[ultra thick] (0,2) -- (1,2) -- (2,3) -- (3,3) -- (4,4) -- (5,4);
\draw[ultra thick] (0,0) -- (0,-1);
\draw[ultra thick] (2,1) -- (2,0);
\draw[ultra thick] (1,1) -- (1,2);
\draw[ultra thick] (3,2) -- (3,3);
\draw[ultra thick] (2,3) -- (2,4);
\draw[ultra thick] (4,4) -- (4,5);
%left mid
\draw[ultra thick] (2,7) -- (3,8) -- (4,8) -- (5,9) -- (6,9);
\draw[ultra thick] (3,8) -- (3,9);
\draw[ultra thick] (5,9) -- (5,10);
%\draw[ultra thick] (2,6) -- (2,7);
\draw[ultra thick] (4,7) -- (4,8);
%left top
\draw[ultra thick] (2,12) -- (3,12) -- (4,13) -- (5,13) -- (6,14) -- (7,14);
\draw[ultra thick] (3,14) -- (4,14) -- (5,15) -- (6,15) -- (7,16) -- (8,16);
\draw[ultra thick] (4,13) -- (4,14);
\draw[ultra thick] (6,14) -- (6,15);
\draw[ultra thick] (3,11) -- (3,12);
\draw[ultra thick] (5,12) -- (5,13);
\draw[ultra thick] (5,15) -- (5,16);
\draw[ultra thick] (7,16) -- (7,17);
%mid bottom
\draw[ultra thick] (6,2) -- (7,2) -- (8,3);
\draw[ultra thick] (9,3)-- (10,4) -- (11,4);
\draw[ultra thick] (8,4) -- (9,5) -- (10,5) -- (11,6) -- (12,6);
\draw[ultra thick] (7,1) -- (7,2);
\draw[ultra thick] (9,2) -- (9,3);
\draw[ultra thick] (8,3) -- (8,4);
\draw[ultra thick] (10,4) -- (10,5);
\draw[ultra thick] (9,5) -- (9,6);
\draw[ultra thick] (11,6) -- (11,7);
%mid mid
\draw[ultra thick] (8,9) -- (9,9) -- (10,10) -- (11,10) -- (12,11) -- (13,11);
\draw[ultra thick] (9,8) -- (9,9);
\draw[ultra thick] (11,9) -- (11,10);
\draw[ultra thick] (10,10) -- (10,11);
\draw[ultra thick] (12,11) -- (12,12);
%mid top
\draw[ultra thick] (9,14) -- (10,14) -- (11,15) -- (12,15) -- (13,16);% -- (14,16);
\draw[ultra thick] (10,16) -- (11,16) -- (12,17);
\draw[ultra thick] (13,17) -- (14,18) -- (15,18);
\draw[ultra thick] (10,13) -- (10,14);
\draw[ultra thick] (12,14) -- (12,15);
\draw[ultra thick] (11,15) -- (11,16);
%\draw[ultra thick] (13,16) -- (13,17);
%\draw[ultra thick] (12,17) -- (12,18);
\draw[ultra thick] (14,18) -- (14,19);
%right bottom
\draw[ultra thick] (13,4) -- (14,4) -- (15,5) -- (16,5) -- (17,6) -- (18,6);
\draw[ultra thick] (14,6) -- (15,6) -- (16,7) -- (17,7) -- (18,8) -- (19,8);
\draw[ultra thick] (14,3) -- (14,4);
\draw[ultra thick] (16,4) -- (16,5);
\draw[ultra thick] (15,5) -- (15,6);
\draw[ultra thick] (17,6) -- (17,7);
\draw[ultra thick] (16,7) -- (16,8);
\draw[ultra thick] (18,8) -- (18,9);
%right mid
\draw[ultra thick] (15,11) -- (16,11) -- (17,12) -- (18,12) -- (19,13);% -- (20,13);
\draw[ultra thick] (16,10) -- (16,11);
\draw[ultra thick] (18,11) -- (18,12);
\draw[ultra thick] (17,12) -- (17,13);
%\draw[ultra thick] (19,13) -- (19,14);
%right top
\draw[ultra thick] (16,16) -- (17,16) -- (18,17) -- (19,17) -- (20,18) -- (21,18);
\draw[ultra thick] (17,18) -- (18,18) -- (19,19) -- (20,19) -- (21,20) -- (22,20);
\draw[ultra thick] (17,15) -- (17,16);
\draw[ultra thick] (19,16) -- (19,17);
\draw[ultra thick] (18,17) -- (18,18);
\draw[ultra thick] (20,18) -- (20,19);
\draw[ultra thick] (19,19) -- (19,20);
\draw[ultra thick] (21,20) -- (21,21);
%dots
\node[rotate=90] at (2,5) {\Large $\cdots$};
\node[rotate=90] at (4,6) {\Large $\cdots$};
\node[rotate=90] at (3,10) {\Large $\cdots$};
\node[rotate=90] at (5,11) {\Large $\cdots$};
\node at (5,2) {\Large $\cdots$};
\node at (6,4) {\Large $\cdots$};
\node[rotate=90] at (9,7) {\Large $\cdots$};
\node[rotate=90] at (11,8) {\Large $\cdots$};
\node at (7,9) {\Large $\cdots$};
\node[rotate=90] at (10,12) {\Large $\cdots$};
\node[rotate=90] at (12,13) {\Large $\cdots$};
\node at (8,14) {\Large $\cdots$};
\node at (9,16) {\Large $\cdots$};
\node at (12,4) {\Large $\cdots$};
\node at (13,6) {\Large $\cdots$};
\node[rotate=90] at (16,9) {\Large $\cdots$};
\node[rotate=90] at (18,10) {\Large $\cdots$};
\node at (14,11) {\Large $\cdots$};
\node[rotate=90] at (17,14) {\Large $\cdots$};
\node[rotate=90] at (19,15) {\Large $\cdots$};
\node at (15,16) {\Large $\cdots$};
\node at (16,18) {\Large $\cdots$};
%path
\draw[ultra thick,red] (1,7) -- (2,7) -- (2,6);
\draw[ultra thick,red] (7,4) -- (8,4) -- (8,3) -- (9,3) -- (9,2);
\draw[ultra thick,red] (12,18) -- (12,17) -- (13,17) -- (13,16) -- (14,16);
\draw[ultra thick,red] (19,14) -- (19,13) -- (20,13);
%nodes bottom
\node at (0,-1.3) {{\small $1$}};
\node at (2,-0.3) {{\small $2$}};
\node at (7,0.7) {{\small $n+rM$}};
\node at (9,1.7) {{\small $n+rM+1$}};
\node at (14,2.7) {{\small $N-1$}};
\node at (16,3.7) {{\small $N$}};
%nodes top
\node at (5,16.3) {{\small $1$}};
\node at (7,17.3) {{\small $2$}};
\node at (12,18.3) {{\small $n+rM+1$}};
\node at (14,19.3) {{\small $n+rM+2$}};
\node at (19,20.3) {{\small $N-1$}};
\node at (21,21.3) {{\small $N$}};
%nodes left
\node at (-1.2,0) {{\small $\emptyset$}};
\node at (-0.2,2) {{\small $\emptyset$}};
\node at (0.8,7) {{\small $\emptyset$}};
\node at (1.8,12) {{\small $\emptyset$}};
\node at (2.8,14) {{\small $\emptyset$}};
%nodes right
\node at (18.2,6) {{\small $\emptyset$}};
\node at (19.2,8) {{\small $\emptyset$}};
\node at (20.2,13) {{\small $\emptyset$}};
\node at (21.2,18) {{\small $\emptyset$}};
\node at (22.2,20) {{\small $\emptyset$}};
%arrows
\draw[thick,<->] (-2,0) -- (-2,7);
\node at (-2.3,3.5) {$n$};
\draw[thick,<->] (21,6) -- (21,13);
\node[rotate=-90] at (21.4,9.5) {$M(r_{\text{max}}+1)-N+n\text{ mod }M$};
%labels path
\node[red] at (2.4,6.5) {{\small $v^{(n)}_{1}$}};
\node[red] at (7.4,4.4) {{\small $h^{(n)}_{n+rM-1}$}};
\node[red] at (7.4,3.5) {{\small $v^{(n)}_{n+rM}$}};
\node[rotate=45,red] at (8.7,3.6) {{\small $h^{(n)}_{n+rM}$}};
\node[red] at (9.6,2.5) {{\small $v^{(n)}_{n+rM+1}$}};
\node[red] at (11.3,17.5) {{\small $v^{(n)}_{n+rM+1}$}};
\node[rotate=45,red] at (12.7,17.7) {{\small $h^{(n)}_{n+rM+1}$}};
\node[red] at (13.8,16.5) {{\small $v^{(n)}_{n+rM+2}$}};
\node[red] at (13.9,15.7) {{\small $h^{(n)}_{n+rM+2}$}};
\node[red] at (18.6,13.5) {{\small $v^{(n)}_{N}$}};
%labels diagonal
\node at (2.3,7.9) {{\small $m^{(n)}_{1}$}};
\node[rotate=45] at (8.4,4.9) {{\small $m^{(n)}_{n+rM}$}};
\node at (10.2,3.3) {{\small $m^{(n)}_{n+rM+1}$}};
\node at (6.8,2.6) {{\small $\tilde{m}^{(n)}_{n+rM}$}};
\node at (10.8,16.9) {{\small $\tilde{m}^{(n)}_{n+rM+1}$}};
\node at (14.1,17.2) {{\small $m^{(n)}_{n+rM+2}$}};
\node at (13,15) {{\small $\tilde{m}^{(n)}_{n+rM+2}$}};
\node at (18.7,12.2) {{\small $\tilde{m}^{(n)}_{N}$}};
%strip main line
\draw[red,ultra thick] (3,-3) -- (4,-3) -- (4,-4) -- (5,-4);
\draw[red,ultra thick] (7,-4) -- (8,-4) -- (8,-5) -- (9,-5) -- (9,-6) -- (10,-6) -- (10,-7) -- (11,-7);
\draw[red,ultra thick] (13,-7) -- (14,-7) -- (14,-8) -- (15,-8);
\node[red] at (6,-4) {\Large $\cdots$};
\node[red] at (12,-7) {\Large $\cdots$};
\node at (2.8,-3) {{\small $\emptyset$}};
\node at (15.2,-8) {{\small $\emptyset$}};
%strip diagonal lines up
\draw[ultra thick] (4,-3) -- (5,-2);
\draw[ultra thick] (8,-4) -- (9,-3);
\draw[ultra thick] (9,-5) -- (10,-4);
\draw[ultra thick] (10,-6) -- (11,-5);
\draw[ultra thick] (14,-7) -- (15,-6);
%strip diagonal lines down
\draw[ultra thick] (4,-4) -- (3,-5);
\draw[ultra thick] (8,-5) -- (7,-6);
\draw[ultra thick] (9,-6) -- (8,-7);
\draw[ultra thick] (10,-7) -- (9,-8);
\draw[ultra thick] (14,-8) -- (13,-9);
%labels diagonal up
\node at (5.3,-1.7) {{\small $m^{(n)}_{1}$}};
\node at (9.3,-2.7) {{\small $m^{(n)}_{n+rM}$}};
\node at (10.3,-3.7) {{\small $m^{(n)}_{n+rM+1}$}};
\node at (11.3,-4.7) {{\small $m^{(n)}_{n+rM+2}$}};
\node at (15.3,-5.7) {{\small $m^{(n)}_{N}$}};
%labels diagonal down
\node at (2.7,-5.3) {{\small $\tilde{m}^{(n)}_{1}$}};
\node at (6.7,-6.3) {{\small $\tilde{m}^{(n)}_{n+rM}$}};
\node at (7.7,-7.3) {{\small $\tilde{m}^{(n)}_{n+rM+1}$}};
\node at (8.7,-8.3) {{\small $\tilde{m}^{(n)}_{n+rM+2}$}};
\node at (12.7,-9.3) {{\small $\tilde{m}^{(n)}_{N}$}};
%labels vertical
\node[red] at (3.5,-3.5) {{\small $v^{(n)}_{1}$}};
\node[red] at (7.5,-4.5) {{\small $v^{(n)}_{n+rM}$}};
\node[red] at (8.3,-5.5) {{\small $v^{(n)}_{n+rM+1}$}};
\node[red] at (10.9,-6.5) {{\small $v^{(n)}_{n+rM+2}$}};
\node[red] at (13.6,-7.5) {{\small $v^{(n)}_{N}$}};
%labels horizontal
\node[red] at (4.5,-4.4) {{\small $h^{(n)}_{1}$}};
\node[red] at (7.3,-3.6) {{\small $h^{(n)}_{n+rM-1}$}};
\node[rotate=45,red] at (8.7,-4.4) {{\small $h^{(n)}_{n+rM}$}};
\node[rotate=45,red] at (9.7,-5.3) {{\small $h^{(n)}_{n+rM+1}$}};
\node[red] at (10.7,-7.5) {{\small $h^{(n)}_{n+rM+2}$}};
\node[red] at (13.5,-6.6) {{\small $h^{(n)}_{N-1}$}};
\end{tikzpicture}}}
\end{center}
\caption{\emph{Non-compact web diagram (top) with a generic strip (bottom) that runs between two lines stretching to infinity and is only composed of horizontal and vertical lines.}}
\label{Fig:OpenWebDiagram}
\end{figure}
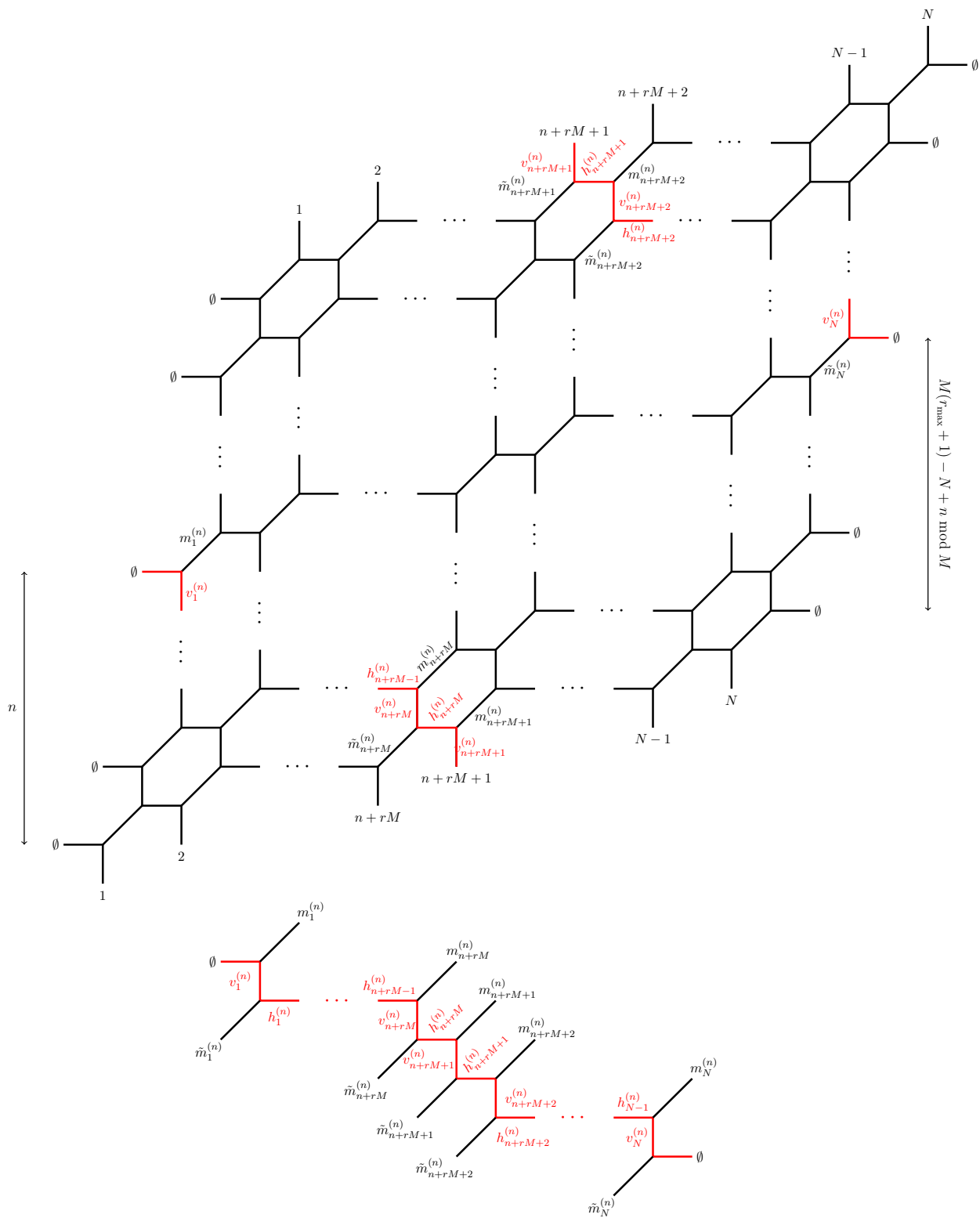

To show this, we consider a generic configuration of the type $(N,M)$ whose horizontal direction is decompactified and decompose it into $M$ strips of length $N$ that are glued together along the diagonal lines. A generic configuration of such strips (labelled by $n\in\{0,\ldots,M-1\}$) along with a labelling of the parameters involved is highlighted in \figref{Fig:OpenWebDiagram}, 
%%%%%%%%%%%%%%%%%%%%%%%%%%%%%%
\noindent
where the $\emptyset$ indicates that the horizontal lines stretch to infinity. Furthermore, for a given $n$, $r$ runs over all elements $\mathcal{S}_n=\{r\in\mathbb{N}\cup\{0\}|n+rM+1\leq N\}$ and
\begin{align}
r_{\text{max}}=\left\{\begin{array}{lcl}\text{max}[\{r\in\mathbb{N}\cup\{0\}|n+rM+1\leq N\}] & \text{if} & \mathcal{S}_n\neq\{\} \\ -1 & \text{it} &\mathcal{S}_n=\{\} \end{array} \right.
\end{align}
Note that, for $\mathcal{S}_n=\{\}$, the strip never reaches the bottom of the diagram.

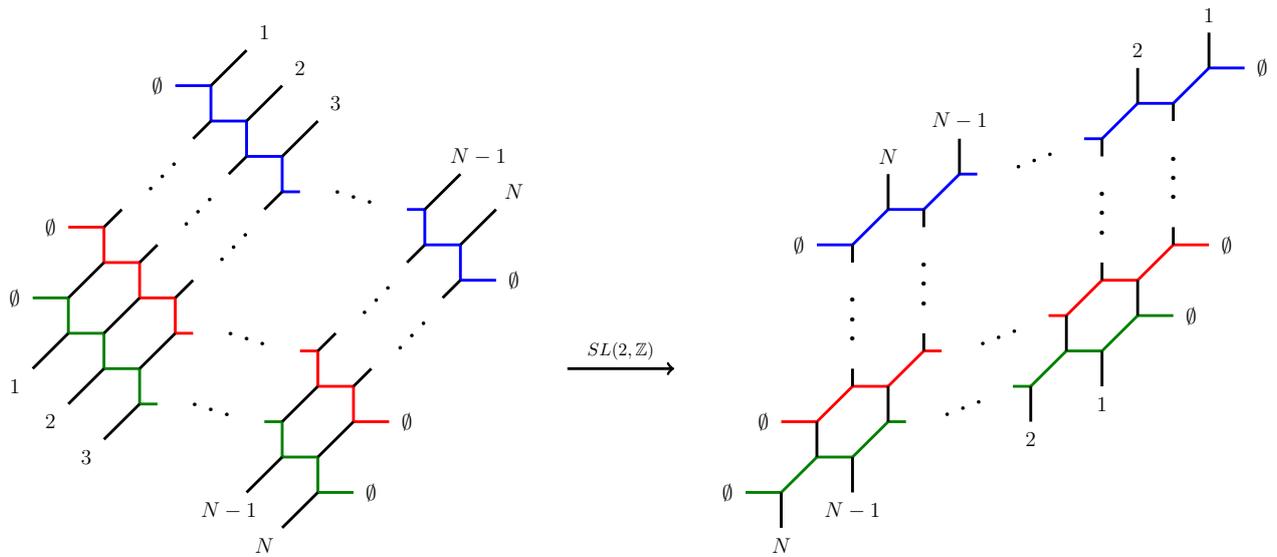
\begin{figure}[htbp]
\begin{center}
\scalebox{0.67}{\parbox{25cm}{\begin{tikzpicture}[scale = 0.7]
%paths
\draw[green!50!black, ultra thick] (0,0) -- (1,0) -- (1,-1) -- (2,-1) -- (2,-2) -- (3,-2) -- (3,-3) -- (3.5,-3);
\draw[green!50!black, ultra thick] (6.5,-3.5) -- (7,-3.5) -- (7,-4.5) -- (8,-4.5) -- (8,-5.5) -- (9,-5.5);
\draw[red, ultra thick] (1,2) -- (2,2) -- (2,1) -- (3,1) -- (3,0) -- (4,0) -- (4,-1) -- (4.5,-1);
\draw[red, ultra thick] (7.5,-1.5) -- (8,-1.5) -- (8,-2.5) -- (9,-2.5) -- (9,-3.5) -- (10,-3.5);
\draw[blue, ultra thick] (4,6) -- (5,6) -- (5,5) -- (6,5) -- (6,4) -- (7,4) -- (7,3) -- (7.5,3);
\draw[blue, ultra thick] (10.5,2.5) -- (11,2.5) -- (11,1.5) -- (12,1.5) -- (12,0.5) -- (13,0.5);
%dots horizontal
\node at (5,-3)  {\Huge $\ddots$};
\node at (6,-1)  {\Huge $\ddots$};
\node at (9,3)  {\Huge $\ddots$};
%diagonals (start bottom left)
\draw[ultra thick] (1,-1) -- (0,-2);
\draw[ultra thick] (2,-2) -- (1,-3);
\draw[ultra thick] (3,-3) -- (2,-4);
\draw[ultra thick] (7,-4.5) -- (6,-5.5);
\draw[ultra thick] (8,-5.5) -- (7,-6.5);
\draw[ultra thick] (1,0) -- (2,1);
\draw[ultra thick] (2,-1) -- (3,0);
\draw[ultra thick] (3,-2) -- (4,-1);
\draw[ultra thick] (7,-3.5) -- (8,-2.5);
\draw[ultra thick] (8,-4.5) -- (9,-3.5);
\draw[ultra thick] (2,2) -- (2.5,2.5);
\draw[ultra thick] (3,1) -- (3.5,1.5);
\draw[ultra thick] (4,0) -- (4.5,0.5);
\draw[ultra thick] (8,-1.5) -- (8.5,-1);
\draw[ultra thick] (9,-2.5) -- (9.5,-2);
%diagonal dots
\node[rotate=65] at (3.5,3.5) {\Huge $\ddots$};
\node[rotate=65] at (4.5,2.5) {\Huge $\ddots$};
\node[rotate=65] at (5.5,1.5) {\Huge $\ddots$};
\node[rotate=65] at (9.5,0) {\Huge $\ddots$};
\node[rotate=65] at (10.5,-1) {\Huge $\ddots$};
\draw[ultra thick] (4.5,4.5) -- (5,5);
\draw[ultra thick] (5.5,3.5) -- (6,4);
\draw[ultra thick] (6.5,2.5) -- (7,3);
\draw[ultra thick] (10.5,1) -- (11,1.5);
\draw[ultra thick] (11.5,0) -- (12,0.5);
\draw[ultra thick] (5,6) -- (6,7);
\draw[ultra thick] (6,5) -- (7,6);
\draw[ultra thick] (7,4) -- (8,5);
\draw[ultra thick] (11,2.5) -- (12,3.5);
\draw[ultra thick] (12,1.5) -- (13,2.5);
% Empty
\node at (-0.5,0)  {$\emptyset$};
\node at (0.5,2)  {$\emptyset$};
\node at (3.5,6)  {$\emptyset$};
\node at (9.5,-5.5)  {$\emptyset$};
\node at (10.5,-3.5)  {$\emptyset$};
\node at (13.5,0.5)  {$\emptyset$};
% External legs bottom
\node at (-0.5,-2.5)  {$1$};
\node at (0.5,-3.5)  {$2$};
\node at (1.5,-4.5)  {$3$};
\node at (5.5,-6)  {$N-1$};
\node at (6.5,-7)  {$N$};
% External legs top
\node at (6.5,7.5)  {$1$};
\node at (7.5,6.5)  {$2$};
\node at (8.5,5.5)  {$3$};
\node at (12.5,4)  {$N-1$};
\node at (13.5,3)  {$N$};
%%%%% right side
%paths
\draw[ultra thick, green!50!black] (20,-5.5) -- (21,-5.5) -- (22,-4.5) -- (23,-4.5) -- (24,-3.5) --(24.5,-3.5);
\draw[ultra thick, green!50!black] (27.5,-2.5) -- (28,-2.5) -- (29,-1.5) -- (30,-1.5) -- (31,-0.5) --(32,-0.5);
\draw[ultra thick, red] (21,-3.5) -- (22,-3.5) -- (23,-2.5) -- (24,-2.5) -- (25,-1.5) --(25.5,-1.5);
\draw[ultra thick, red] (28.5,-0.5) -- (29,-0.5) -- (30,0.5) -- (31,0.5) -- (32,1.5) --(33,1.5);
\draw[ultra thick, blue] (22,1.5) -- (23,1.5) -- (24,2.5) -- (25,2.5) -- (26,3.5) --(26.5,3.5);
\draw[ultra thick, blue] (29.5,4.5) -- (30,4.5) -- (31,5.5) -- (32,5.5) -- (33,6.5) --(34,6.5);
%external legs bottom
\draw[ultra thick] (21,-5.5) -- (21,-6.5);
\draw[ultra thick] (23,-4.5) -- (23,-5.5);
\draw[ultra thick] (28,-2.5) -- (28,-3.5);
\draw[ultra thick] (30,-1.5) -- (30,-2.5);
\draw[ultra thick] (22,-4.5) -- (22,-3.5);
\draw[ultra thick] (24,-3.5) -- (24,-2.5);
\draw[ultra thick] (29,-1.5) -- (29,-0.5);
\draw[ultra thick] (31,-0.5) -- (31,0.5);
\draw[ultra thick] (23,-2.5) -- (23,-2);
\draw[ultra thick] (25,-1.5) -- (25,-1);
\draw[ultra thick] (30,0.5) -- (30,1);
\draw[ultra thick] (32,1.5) -- (32,2);
\draw[ultra thick] (23,1) -- (23,1.5);
\draw[ultra thick] (25,2) -- (25,2.5);
\draw[ultra thick] (30,4) -- (30,4.5);
\draw[ultra thick] (32,5) -- (32,5.5);
% external legs top
\draw[ultra thick] (24,2.5) -- (24,3.5);
\draw[ultra thick] (26,3.5) -- (26,4.5);
\draw[ultra thick] (31,5.5) -- (31,6.5);
\draw[ultra thick] (33,6.5) -- (33,7.5);
%dots
\node[rotate=40] at (26,-3) {\Huge $\ddots$};
\node[rotate=40] at (27,-1) {\Huge $\ddots$};
\node[rotate=90] at (23,-0.5) {\Huge $\dots$};
\node[rotate=90] at (25,0.5) {\Huge $\dots$};
\node[rotate=90] at (30,2.5) {\Huge $\dots$};
\node[rotate=90] at (32,3.5) {\Huge $\dots$};
\node[rotate=40] at (28,4) {\Huge $\ddots$};
% Empty
\node at (19.5,-5.5) {$\emptyset$};
\node at (20.5,-3.5) {$\emptyset$};
\node at (21.5,1.5) {$\emptyset$};
\node at (32.5,-0.5) {$\emptyset$};
\node at (33.5,1.5) {$\emptyset$};
\node at (34.5,6.5) {$\emptyset$};
% Identifications
\node at (21,-7) {$N$};
\node at (23,-6) {$N-1$};
\node at (24,4) {$N$};
\node at (26,5) {$N-1$};
\node at (28,-4) {$2$};
\node at (30,-3) {$1$};
\node at (31,7) {$2$};
\node at (33,8) {$1$};
%sl2
\draw[ultra thick, ->] (15,-2) -- (18,-2);
\node at (16.5,-1.5) {\footnotesize{$SL(2,\mathbb{Z})$}};
\end{tikzpicture}}}
\end{center}
\caption{{\it Gluing the $M$ individual strips constructed in \figref{Fig:OpenWebDiagram} yields another diagram of the type $(N,M)$. For better visibility, three different strips have been coloured.}}
\label{Fig:GluedStrips}
\end{figure}

\begin{figure}[htbp]
\begin{center}
%\hspace*{-2cm}%
\scalebox{0.8}{\parbox{13cm}{\begin{tikzpicture}[scale = 1.50]
\draw[ultra thick] (-6,0) -- (-5,0);
\draw[ultra thick] (-5,-1) -- (-5,0);
\draw[ultra thick] (-5,0) -- (-4,1);
\draw[ultra thick] (-4,1) -- (-3,1);
\draw[ultra thick] (-4,1) -- (-4,2);
\draw[ultra thick] (-3,1) -- (-3,0);
\draw[ultra thick] (-3,1) -- (-2,2);
\draw[ultra thick] (-4,2) -- (-3,3);
\draw[ultra thick] (-5,2) -- (-4,2);
\draw[ultra thick] (-3,3) -- (-3,4);
\draw[ultra thick] (-3,3) -- (-2,3);
\draw[ultra thick] (-2,2) -- (-2,3);
\draw[ultra thick] (-2,2) -- (-1,2);
\draw[ultra thick] (-2,3) -- (-1,4);
\draw[ultra thick] (-1,2) -- (0,3);
\draw[ultra thick] (-1,2) -- (-1,1);
\draw[ultra thick] (-1,4) -- (0,4);
\draw[ultra thick] (0,3) -- (1,3);
\draw[ultra thick] (0,3) -- (0,4);
\draw[ultra thick] (0,4) -- (1,5);
\draw[ultra thick] (1,5) -- (2,5);
\draw[ultra thick] (1,5) -- (1,6);
\draw[ultra thick] (-1,4) -- (-1,5);
\node at (-4.3,0.3) {{\small $m_1$}};
\node at (-2.3,1.3) {{\small $m_2$}};
\node at (-0.3,2.3) {{\small $m_3$}};
\node at (-3.3,2.3) {{\small $m_4$}};
\node at (-1.3,3.3) {{\small $m_5$}};
\node at (0.7,4.3) {{\small $m_6$}};
\node at (-3.5,1.2) {{\small $h_1$}};
\node at (-1.5,2.2) {{\small $h_2$}};
\node at (-2.5,3.2) {{\small $h_3$}};
\node at (-0.5,4.2) {{\small $h_4$}};
\node at (-5.2,-0.5) {{\small $v_1$}};
\node at (-3.2,0.5) {{\small $v_2$}};
\node at (-1.2,1.5) {{\small $v_3$}};
\node at (-4.2,1.5) {{\small $v_4$}};
\node at (-2.2,2.5) {{\small $v_5$}};
\node at (-0.2,3.5) {{\small $v_6$}};
\node at (-6.2,0) {{\small \bf $\emptyset$}};
\node at (-5.2,2) {{\small \bf $\emptyset$}};
\node at (1.2,3) {{\small \bf $\emptyset$}};
\node at (2.2,5) {{\small \bf $\emptyset$}};
\node at (-5,-1.3) {{\small \bf $1$}};
\node at (-3,-0.3) {{\small \bf $2$}};
\node at (-1,0.7) {{\small \bf $3$}};
\node at (-3,4.3) {{\small \bf $1$}};
\node at (-1,5.3) {{\small \bf $2$}};
\node at (1,6.3) {{\small \bf $3$}};
\end{tikzpicture}}}
\label{noncompactconf}
\end{center}
\caption{\it Non-compact $(3,2)$ web diagram with a labelling of the K\"ahler parameters. Notice that the consistency conditions (\ref{ConsRel1}) -- (\ref{ConsRel3}) still need to be imposed.}
\end{figure}
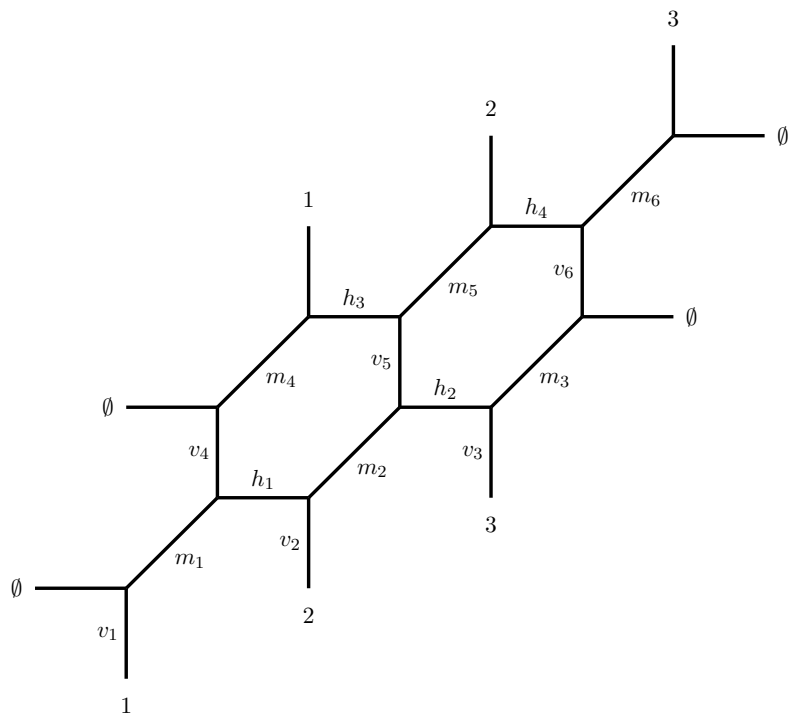

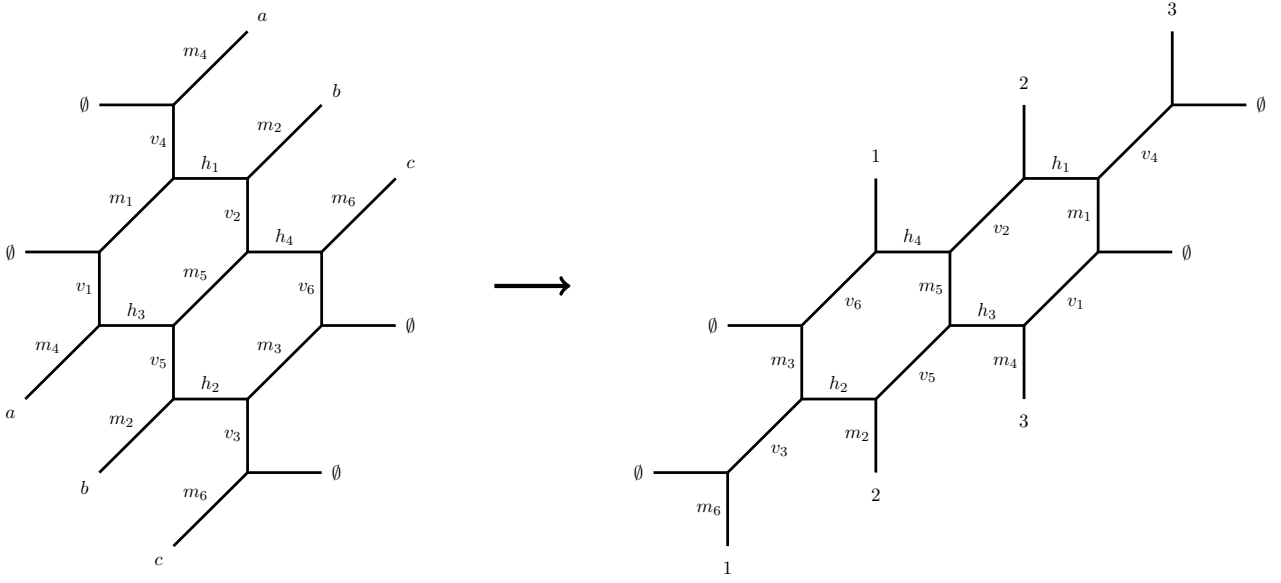
\begin{figure}[htbp]
\begin{center}
%\hspace*{-2cm}%
\scalebox{0.65}{\parbox{8.8cm}{\begin{tikzpicture}[scale = 1.50]
\draw[ultra thick] (-1,0) -- (0,0) -- (0,-1) -- (1,-1) -- (1,-2) -- (2,-2) -- (2,-3) -- (3,-3);
\draw[ultra thick] (0,2) -- (1,2) -- (1,1) -- (2,1) -- (2,0) -- (3,0) -- (3,-1) -- (4,-1);
\draw[ultra thick] (-1,-2) -- (0,-1);
\draw[ultra thick] (0,-3) -- (1,-2);
\draw[ultra thick] (1,-4) -- (2,-3);
\draw[ultra thick] (0,0) -- (1,1);
\draw[ultra thick] (1,-1) -- (2,0);
\draw[ultra thick] (2,-2) -- (3,-1);
\draw[ultra thick] (1,2) -- (2,3);
\draw[ultra thick] (2,1) -- (3,2);
\draw[ultra thick] (3,0) -- (4,1);
\node at (-1.2,-2.2) {{\small \bf $a$}};
\node at (-0.2,-3.2) {{\small \bf $b$}};
\node at (0.8,-4.2) {{\small \bf $c$}};
\node at (2.2,3.2) {{\small \bf $a$}};
\node at (3.2,2.2) {{\small \bf $b$}};
\node at (4.2,1.2) {{\small \bf $c$}};
\node at (-1.2,0) {{\small \bf $\emptyset$}};
\node at (-0.2,2) {{\small \bf $\emptyset$}};
\node at (3.2,-3) {{\small \bf $\emptyset$}};
\node at (4.2,-1) {{\small \bf $\emptyset$}};
\node at (1.3,2.7) {{\small $m_4$}};
\node at (2.3,1.7) {{\small $m_2$}};
\node at (3.3,0.7) {{\small $m_6$}};
\node at (0.3,0.7) {{\small $m_1$}};
\node at (1.3,-0.3) {{\small $m_5$}};
\node at (2.3,-1.3) {{\small $m_3$}};
\node at (-0.7,-1.3) {{\small $m_4$}};
\node at (0.3,-2.3) {{\small $m_2$}};
\node at (1.3,-3.3) {{\small $m_6$}};
\node at (-0.2,-0.5) {{\small $v_1$}};
\node at (0.8,-1.5) {{\small $v_5$}};
\node at (1.8,-2.5) {{\small $v_3$}};
\node at (0.8,1.5) {{\small $v_4$}};
\node at (1.8,0.5) {{\small $v_2$}};
\node at (2.8,-0.5) {{\small $v_6$}};
\node at (0.5,-0.8) {{\small $h_3$}};
\node at (1.5,-1.8) {{\small $h_2$}};
\node at (1.5,1.2) {{\small $h_1$}};
\node at (2.5,0.2) {{\small $h_4$}};
\end{tikzpicture}}}
%%%%%%%%%
\hspace{0.5cm}
\parbox{1cm}{\begin{tikzpicture}
\draw[ultra thick,->] (0,0) -- (1,0);
\end{tikzpicture}}
\hspace{0.5cm}
%%%%%%%%%
\scalebox{0.65}{\parbox{13cm}{\begin{tikzpicture}[scale = 1.50]
\draw[ultra thick] (-6,0) -- (-5,0);
\draw[ultra thick] (-5,-1) -- (-5,0);
\draw[ultra thick] (-5,0) -- (-4,1);
\draw[ultra thick] (-4,1) -- (-3,1);
\draw[ultra thick] (-4,1) -- (-4,2);
\draw[ultra thick] (-3,1) -- (-3,0);
\draw[ultra thick] (-3,1) -- (-2,2);
\draw[ultra thick] (-4,2) -- (-3,3);
\draw[ultra thick] (-5,2) -- (-4,2);
\draw[ultra thick] (-3,3) -- (-3,4);
\draw[ultra thick] (-3,3) -- (-2,3);
\draw[ultra thick] (-2,2) -- (-2,3);
\draw[ultra thick] (-2,2) -- (-1,2);
\draw[ultra thick] (-2,3) -- (-1,4);
\draw[ultra thick] (-1,2) -- (0,3);
\draw[ultra thick] (-1,2) -- (-1,1);
\draw[ultra thick] (-1,4) -- (0,4);
\draw[ultra thick] (0,3) -- (1,3);
\draw[ultra thick] (0,3) -- (0,4);
\draw[ultra thick] (0,4) -- (1,5);
\draw[ultra thick] (1,5) -- (2,5);
\draw[ultra thick] (1,5) -- (1,6);
\draw[ultra thick] (-1,4) -- (-1,5);
\node at (-4.3,0.3) {{\small $v_3$}};
\node at (-2.3,1.3) {{\small $v_5$}};
\node at (-0.3,2.3) {{\small $v_1$}};
\node at (-3.3,2.3) {{\small $v_6$}};
\node at (-1.3,3.3) {{\small $v_2$}};
\node at (0.7,4.3) {{\small $v_4$}};
\node at (-3.5,1.2) {{\small $h_2$}};
\node at (-1.5,2.2) {{\small $h_3$}};
\node at (-2.5,3.2) {{\small $h_4$}};
\node at (-0.5,4.2) {{\small $h_1$}};
\node at (-5.25,-0.5) {{\small $m_6$}};
\node at (-3.25,0.5) {{\small $m_2$}};
\node at (-1.25,1.5) {{\small $m_4$}};
\node at (-4.25,1.5) {{\small $m_3$}};
\node at (-2.25,2.5) {{\small $m_5$}};
\node at (-0.25,3.5) {{\small $m_1$}};
\node at (-6.2,0) {{\small \bf $\emptyset$}};
\node at (-5.2,2) {{\small \bf $\emptyset$}};
\node at (1.2,3) {{\small \bf $\emptyset$}};
\node at (2.2,5) {{\small \bf $\emptyset$}};
\node at (-5,-1.3) {{\small \bf $1$}};
\node at (-3,-0.3) {{\small \bf $2$}};
\node at (-1,0.7) {{\small \bf $3$}};
\node at (-3,4.3) {{\small \bf $1$}};
\node at (-1,5.3) {{\small \bf $2$}};
\node at (1,6.3) {{\small \bf $3$}};
\end{tikzpicture}}}
\caption{\emph{Different representation of the $(3,2)$ diagram (left) and dual diagram after an $SL(2,\mathbb{Z})$ transformation (right).}}
\label{Fig:DualityNonCom32}
\end{center}
\end{figure}

For two strips $\tilde{m}^{(n+1)}_i=m^{(n)}_i$ for $i=1,\ldots,N$ and $n=0,\ldots, M-1$, the $M$ strips are glued together to form another diagram that (with the help of an $SL(2,\mathbb{Z})$ transformation) can be brought into the form of a $(N,M)$ web, as shown in \figref{Fig:GluedStrips}.
\noindent
However, comparing the two diagrams, we see that the role of diagonal and vertical lines is exchanged, leading to the self-duality
\begin{align}
\{\mathbf{m}\}\longleftrightarrow\{\mathbf{v}\}\,.
\end{align}
The precise duality map requires also to impose the consistency conditions, which can be worked out explicitly for specific examples. As an example, we may consider the non-compact $(N,M)=(3,2)$ configuration, whose web-diagram (along with a labelling of the various lines) is given by \figref{noncompactconf}. 
In this configuration, the consistency conditions for the various K\"ahler parameters are
\begin{align}
&m_1+h_1=h_3+m_5\,,&&v_1+m_1=m_5+v_2\,,&&m_2+h_2=h_4+m_6\,,&&v_2+m_2=v_3+m_6\,,\label{ConsRel1}\\
&h_1+m_2=m_4+h_3\,,&&v_4+m_4=m_2+v_5\,,&&h_2+m_3=m_5+h_4\,,&&v_5+m_5=m_3+v_6\,,\label{ConsRel2}
\end{align}
\begin{align}
v_1+m_1+v_4+m_4=v_2+m_2+v_5+m_5=v_3+m_3+m_6+v_6\,.\label{ConsRel3}
\end{align}
This web-diagram can represented in an equivalent fashion as shown in \figref{Fig:DualityNonCom32}.
Under an $SL(2,\mathbb{Z})$ transformation, this web can again be presented in the form of a $(3,2)$ configuration which is dual to the original configuration upon the change of variables
\begin{align}
&m_1\longrightarrow v_3\,,&&m_1\longrightarrow v_3\,,&&m_3\longrightarrow v_1\,,&&m_4\longrightarrow v_6\,,&&m_5\longrightarrow v_2\,,&&m_6\longrightarrow v_4\,,\nonumber\\
&v_1\longrightarrow m_6\,,&&v_2\longrightarrow m_2\,,&&v_3\longrightarrow m_4\,,&&v_4\longrightarrow m_3\,,&&v_5\longrightarrow m_5\,,&&v_6\longrightarrow m_1\,,\nonumber
\end{align}
\begin{align}
&h_1\longrightarrow h_2\,,&&h_2\longrightarrow h_3\,,&&h_3\longrightarrow h_4\,,&&h_4\longrightarrow h_1\,.
\end{align}
%%%%%%%%%%%%%%%%%%%%%%%%%%%%%%%%%%%%%%%%%%%%%
%%%%%%%%%%%%%%%%%%%%%%%%%%%%%%%%%%%%%%%%%%%%%

%%%%%%%%%%%%%%%%%%%%%%%%%%%%%%
%%%%%%%%%%%%%%%%%%%%%%%%%%%%%%§§
%%%%%%%%%%%%%%%%%%%%%%%%%%%%%%

\end{document}